\documentclass{aa}

\usepackage{graphicx}
\usepackage{xcolor}
\usepackage{ltablex}
\usepackage{color}
\usepackage{float}
\usepackage[maxfloats=88]{morefloats}
\usepackage{txfonts}
\usepackage{lscape}
\usepackage{longtable}
\usepackage{natbib,twoopt}
\usepackage[breaklinks=true, colorlinks=true, linkcolor=blue, citecolor=blue, urlcolor=blue]{hyperref}
\bibpunct{(}{)}{;}{a}{}{,}             

\newcommand{\Msun}{$\rm M_{\odot}$}

\newcommand{\teff}{\mbox{$T_{\rm eff}$}}

\newcommand{\kms}{km s$^{-1}$}

\begin{document}

   \title{Spectroscopic characterization of the known O-star population in Cygnus OB2 }

  \subtitle{Evidence of multiple star-forming bursts\thanks{Tables A.1 to A.4 are available in electronic form at the CDS via anonymous ftp to cdsarc.u-strasbg.fr (130.79.128.5) or via \url{http://cdsweb.u-strasbg.fr/cgi-bin/qcat?J/A+A/}}}

   \author{S.~R.~Berlanas\inst{1,2,3}
         \and
          A.~Herrero\inst{2,3}
          \and
          F.~Comerón\inst{4}
         \and          
          S.~Simón-Díaz\inst{2,3} 
          \and
          D.~J.~Lennon\inst{2}
          \and
          A.~Pasquali\inst{5}
         \and       
          J.~Maíz Apellániz \inst{6}   
          \and          
          A.~Sota \inst{7}
          \and          
          A.~Pellerín \inst{8}
   }

   \institute{Departamento de Física Aplicada, Universidad de Alicante, 03690 San Vicente del Raspeig, Alicante, Spain
   \and Instituto de Astrofísica de Canarias, 38200 La Laguna, Tenerife, Spain\          
         \and Departamento de Astrofísica, Universidad de La Laguna, 38205 La Laguna, Tenerife, Spain\        \and ESO, Karl-Schwarzschild-Strasse 2, 85748 Garching bei München, Germany\ 
        \and Astronomisches Rechen-Institut, Zentrum für Astronomie der Universität Heidelberg, 69120 Heidelberg,  Germany\  
         \and Centro de Astrobiología, CSIC-INTA, Campus ESAC, 28692 Villanueva de la Cañada, Madrid, Spain\  
         \and Instituto de Astrofísica de Andalucía-CSIC, 18008 Granada, Spain\  
         \and Department of Physics and Astronomy, State University of New York at Geneseo, 1 College Circle, Geneseo, NY 14 454, USA\
         }

   \date{Received month day, year; accepted month day, year}

 
  \abstract
   {Cygnus OB2 provides a unique insight into the high-mass stellar content in one of the largest groups of young massive stars in our Galaxy. Although several studies of its massive population have been carried out over the last decades, an extensive spectroscopic study of the whole known O-star population in the association is still lacking. }
   {We aim to carry out a spectroscopic characterization of all the currently known O stars in Cygnus OB2, determining the distribution of rotational velocities and accurate stellar parameters to obtain an improved view of the evolutionary status of the region. }
   {Based on existing and new optical spectroscopy, we performed a detailed quantitative spectroscopic analysis of all the known O-type stars identified in the association. For this purpose, we used the user-friendly \texttt{iacob-broad} and \texttt{iacob-gbat} automatized tools, FASTWIND stellar models, and astrometry provided by the $Gaia$ second data release.}
   {We created the most complete spectroscopic census of O stars carried out so far in Cygnus OB2 using already existing and new spectroscopy. We present the spectra for 78 O-type stars, from which we identify new binary systems, obtain the distribution of rotational velocities, and  determine the main stellar parameters for all the stars in the region that have not been detected as double-line spectroscopic binaries. We also derive radii, luminosities, and masses for those stars with reliable $Gaia$ astrometry, in addition to creating the Hertzsprung-Russell Diagram to interpret the evolutionary status of the association. Finally, we inspect the dynamical state of the population and identify runaway candidates.}
   {Our spectroscopic analysis of the O-star population in Cygnus OB2 has led to the discovery of two new binary systems and the determination of the main stellar parameters, including rotational velocities, luminosities, masses, and radii for all identified stars. This work has shown the improvement reached when using accurate spectroscopic parameters and astrometry for the interpretation of the evolutionary status of a population, revealing, in the case of Cygnus OB2, at least two star-forming bursts at $\sim$3 and $\sim$5 Myr. We find an apparent deficit of very fast rotators in the distribution of rotational velocities. The inspection of the dynamical distribution of the sample has allowed us to identify nine O stars with peculiar proper motions and discuss a possible dynamical ejection scenario or past supernova (SN) explosions in the region.}


\titlerunning{Spectroscopic characterization of the known O stars in Cygnus OB2}

   \maketitle
%

\section{Introduction}

A complete and reliable understanding of the formation and evolution of massive stars has a direct impact on many fields of modern Astrophysics, thus serving an essential  role in achieving an accurate interpretation the evolution of the Universe \citep[see e.g.,][]{woosley02,langer12,abbott16,prantzos18}.
In spite of this, important uncertainties and limits persist in our understanding of these objects, even in their early evolutionary phases. These uncertainties are related to our ability to determine accurate stellar parameters, particularly with regard to mass and effective temperatures, and to understand the implications of rotation, multiplicity, internal mixing, magnetic fields, and large scale motions for their structure and evolution. The mass and effective temperature scales of O-stars are still a matter of debate \citep{ssimon14b, herrero16, holgado18} and the actual role of rotation and binarity in the N and He enhancement of O stars are still controversial \citep{hunter08,rivero12, martins15}. Uncertainties in mass-loss and binary interactions propagate to subsequent phases of massive star evolution, modifying them and leading to inaccuracies within the model predictions  \citep{langer12}. To improve this situation, large observational databases and the analysis of extensive samples of high quality OB-star spectra are required.

Galactic intense star-forming regions, due to their relatively close distance, allow us to perform a detailed observation of their massive stellar population. 
The Cygnus-X complex represents the most powerful star-forming region at less than 2 kpc from us \citep{rygl12} that is conspicuous at all wavelengths and encompasses several rich very young OB associations. Hosting the largest number of near massive stars and an intense star-forming activity \citep{reipurth08}, this region is an ideal testbed for our theories about star formation, structure, and evolution, as well as the interplay between interstellar medium and stars and the dynamics and kinematics of OB associations and stellar groups.

Its richest association Cygnus OB2 represents the most obvious example of recent star formation. It harbors hundreds of OB stars that can be analyzed \citep{com12, wright15,berlanas18a} and although the optical extinction to Cygnus OB2 is high (A$_V$ $\sim$ 6 mag), it is not enough to prevent us from obtaining spectra of the bright massive members. It has been studied at all wavelengths with different spatial coverage \citep[e.g.,][]{mt91,hanson03,herrero02,com02,neg08,wright10,sota11,com12,wright15,rauw15,maiz16,schneider16,morford16,berlanas18a,berlanas18b} as the actual extension of the association is not known with certainty \citep[see, e.g.,][]{com08,  com12}. Recent extensive studies of its massive population have been presented by \cite{wright15} and \cite{berlanas18a}, who used a Hertzsprung-Russell Diagram (HRD) to discuss the evolutionary status of the association. However, their analysis was based on calibrations of stellar parameters versus spectral type, which introduces additional uncertainties, as has been shown by \cite{ssimon14a}. 
Thus, an extensive spectroscopic study of the Cygnus OB2 early-type population is requisite for an accurate characterization of its massive stellar content.
Moreover, since the $Gaia$ satellite has provided accurate astrometry for the whole region \citep{arenou18} and the distance to Cygnus OB2 has been re-estimated \citep{berlanas19}, we are poised to perform an accurate characterization of their massive stellar content, allowing us to create a much more precise HRD to interpret its evolutionary status.

This paper is organized as follows. In Sect.~\ref{obs}, we introduce our observing strategy. In Sect.~\ref{sample}, we describe the sample and selection criteria. In Sect.~\ref{methods}, we present the methods and tools used for the quantitative spectroscopic analysis of our sample and the results obtained. Then, in Sect.~\ref{discussion}, we discuss and interpret all the results by creating the HRD and its spectroscopic version (sHRD). We also analyze the spatial and dynamical distribution of the sample from $Gaia$ data. In Sect.~\ref{conclusions}, we summarize the conclusions of this work.

\section{New observations and data reduction}\label{obs}

We compiled  new intermediate-high resolution and high signal-to-noise ratio (S/N) spectra to build up the most complete spectroscopic census of the currently known O-type stars in Cygnus OB2 \cite[including the recently newly classified O-type members in the association by][]{berlanas18a}. These new data have been complemented  with already available spectra from previous observations: our own observations and others drawn from our collaborators (see Sect.~\ref{sample} for further details). We included in this sample all the stars identified as O-type stars within 1 deg radius centred on Galactic coordinates $l$ = 79.8$^{o}$ and $b$ = +0.8$^{o}$. 
Since our goal is to perform accurate spectroscopic analysis for the whole sample, we need spectral data in the 4000 -- 5000~\AA ~ wavelength range to access all the main diagnostic lines. Besides this, we also need to cover the H$\alpha$ line in order to properly characterize the stellar wind. 
 
To complete the database we performed four different campaigns (see Table~\ref{obs_int} for specific dates) at the Isaac Newton Telescope (INT) in La Palma.
We obtained intermediate-high resolution, high S/N (100 to 200) spectra for a total of 66 O-type stars of Cygnus OB2.  Since we had already  blue spectra for some of them from previous observations (see Sect.~\ref{sample}), we observed fewer stars in the blue than in the red, that is, 27 in the blue (3900 -- 5100~\AA) and 66 in the red (6100 -- 6800~\AA) wavelength regions.
We used the Intermediate Dispersion Spectrograph (IDS) and the EEV10 detector, and chose the  H1800V grating for the red wavelength region and the R1200B grating for the blue one.
In the first case, the instrument configuration provides a resolving power of R $\sim$10000 at 6500~\AA, which allows us to perform an accurate analysis of the H$\alpha$ profile. 
Due to bad weather conditions during the 16A and 16B observing runs (dust in the air) the exposure times were increased by a factor of two with respect to the INT Exposure Time Calculator (ETC), resulting in times of around one hour to reach S/N $>$ 100 per pixel for a star of magnitude $B =$ 14.0 in the red.
In the second case, the instrument configuration provides a resolving power of R $\sim$5000 at 4500\AA, which is adequate for a proper determination of the spectroscopic stellar parameters. 
In the blue wavelength region, the dust has a higher impact on the exposure times and, consequently, the ETC times were increased by a factor of three. Therefore, in the blue, we needed $\sim$7000 seconds to reach S/N $>$ 100 per pixel for a star of magnitude $B =$ 14.0. In all cases, exposure times above 30 minutes were divided into ranges of 20 minutes to reduce the effect of cosmic rays.

 \begin{table*}[t!]
        \begin{center}
        \caption{\small{Summary of the observational schedule at the INT.}}     
                \label{obs_int}
                \begin{tabular}{lcccccccc}
                \hline 
                \hline  \\[-1.5ex] 
                \textbf{}& \small{Semester}& \small{Dates}&\small{$\lambda$ coverage}&\small{$B$ (mag.)}&\small{$\#$ Stars}\\
   
                \hline \\[-1.5ex] 

                \small{Phase 1} & \small{16A} & \small{ July 2016} &\small{Blue} & \small{$B <$ 14.5}& \small{13 O-type} \\
                \small{} & \small{16B} & \small{ August 2016} &\small{Red} & \small{$B <$ 14.5}& \small{52 O-type} \\
                \cline{2-6}\\[-1.5ex] 
                \small{Phase 2} & \small{17A} & \small{ July 2017} &\small{Blue} & \small{14.5 $< B <$ 16.0}& \small{12 O-type} \\
                \small{} & \small{17B} & \small{August 2017} &\small{Blue+Red} & \small{14.5 $< B <$ 16.0}& \small{2+14 O-type} \\
        \hline
        \multicolumn{6}{l}{\footnotesize Note: Blue and Red indicate the (3900 -- 5100\AA) \footnotesize and (6100 -- 6800\AA) \footnotesize wavelength regions.} \\
        \multicolumn{6}{l}{\footnotesize Exposure times vary between 150 to 10000 sec, depending on the magnitude of each star.} 
                \end{tabular}

        \end{center}
        
\end{table*}

The observational plan was divided into two phases, extending over four semesters, where two spectra were obtained for each star on different days (belonging to the same semester or not) to detect possible spectral or radial velocity variations. The four observing runs were held throughout the summer of 2016 and 2017, for a total of 24 nights distributed as 
indicated in Table~\ref{obs_int}.
The spectra were reduced using IRAF\footnote{IRAF is distributed by the National Optical Astronomy Observatories, which are operated by the Association of Universities for Research in Astronomy, Inc., under cooperative agreement with the National Science Foundation.} with standard routines for bias, flat-field subtraction and wavelength calibration.
We also corrected the spectra for the radial velocity using the Doppler shift of metal and He\,I spectral lines.

\begin{figure*}[t!]
\centering
\includegraphics[width=13cm]{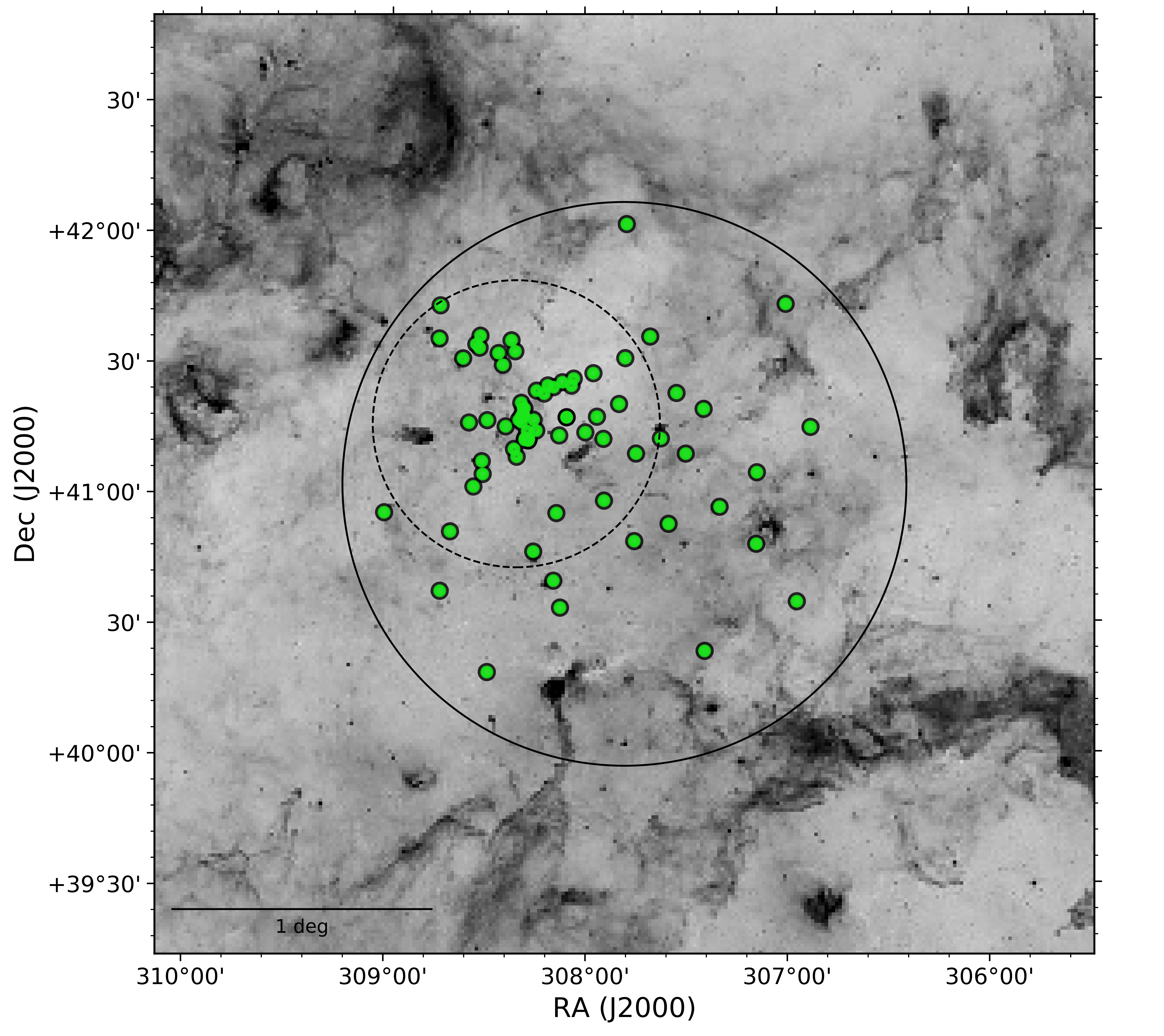}
\caption{Inverse Spitzer 8 $\mu$m image showing the location of the known O-type population in Cygnus OB2 (78 stars represented by green dots).  The solid line circle delimits the 1 degree radius area adopted for the association. For reference, the dashed-dotted line circle shows the area considered by \cite{wright15} indicating the core of the association.}
\label{fig_sample}
\end{figure*}

\section{The sample}\label{sample}

We built up the most complete spectral sample for the already known O-type population in Cygnus OB2 by complementing new observations with already available spectra (own and from collaborators) whose observational details are described as follows.

\begin{table*}[t!]
\centering
\caption{Telescopes, instruments, and settings used in this work.}       
                \label{table_inst_cyglp}
                \begin{tabular}{llcccccc}
                \hline 
                \hline \\[-1.5ex]  
                \small{Source}& \small{Instrument}& \small{Grating} &\small{Telescope}& \small{Resolving power}&\small{S/N} & \small{$\lambda$ range (\AA)}& \small{$\#$ Stars}\\ 
                \hline\\[-1.5ex]  
                \small{1}&\small{ISIS} &\small{600B} & \small{WHT} & \small{3000}&\small{$\sim$300}& \small{3900 -- 5500}& \small{20}\\
                \small{2}&\small{TWIN} &\small{1200}& \small{CAHA-3.5m} & \small{3000}&\small$\sim${300}& \small{3900 -- 5100}& \small{15}\\
                \small{3}&\small{Albireo} &\small{1800}& \small{OSN-1.5m} & \small{2500}&\small{$\sim$300}& \small{3900 -- 5100}& \small{3} \\                             
                \small{4a}&\small{HRS}&\small{600g4739} & \small{HET}  & \small{30000}&\small{$\geq$100}& \small{3800 -- 4700}& \small{12}\\
                \small{4b}&\small{HRS}&\small{600g4739} & \small{HET}  & \small{30000}&\small{$\geq$100}& \small{4700 -- 5700}& \small{12}\\             
                \small{4c}&\small{HRS}&\small{600g6302} & \small{HET}  & \small{30000}&\small{$\geq$100}& \small{5300 -- 6300}& \small{12}\\
                \small{4d}&\small{HRS}&\small{600g6302} & \small{HET}  & \small{30000}&\small{$\geq$100}& \small{6400 -- 7400}& \small{12}\\                              
                \small{5a}&\small{ISIS}&\small{H2400B} & \small{WHT}  & \small{13600}&\small{$>$100}& \small{3900 -- 5100}& \small{15}\\              
                \small{5b}&\small{ISIS}&\small{R1200R} & \small{WHT}  & \small{9300}&\small{$>$150}& \small{5500 -- 6800}& \small{15}\\              
        \small{6a}&\small{IDS}&\small{R1200B} & \small{INT} & \small{5000}&\small{$>$100}& \small{3900 -- 5100}& \small{27}\\
        \small{6b}&\small{IDS}&\small{H1800V} & \small{INT} & \small{7200}&\small{$>$150}& \small{6100 -- 6800}& \small{66}\\                  \hline\\[-1.5ex] 
        \multicolumn{8}{c}{ Total number of spectra $\rightarrow$  209 } \\
        \multicolumn{8}{c}{ Total number of stars $\rightarrow$  158 } \\         
        \multicolumn{8}{c}{ Total number of stars (not repeated) $\rightarrow$  78 } \\                                                
        \hline
        \multicolumn{8}{l}{\footnotesize Notes: Spectra for the detected binary stars are included in the table.  Spectral data from sources 1, 2, and 3 belong} \\
        \multicolumn{8}{l}{\footnotesize    to the GOSSS catalog, see \cite{sota11,maiz16}  for further observing details.}       
                \end{tabular}   

\end{table*}

Most of the available optical--blue data (see Table~\ref{table_inst_cyglp}) comes from the Galactic O-Star Spectroscopic Survey \citep[GOSSS,][]{sota11,sota14,maiz16}, which obtained a large number of Galactic O-type high S/N ($\sim$300) blue-violet spectra at R $\sim$ 2500 -- 3000. The GOSSS data used in this work were observed between 2007 and 2014 using three different instruments:  the Albireo spectrograph at the 1.5 m telescope of the Observatorio de Sierra Nevada (OSN), the TWIN spectrograph at the 3.5 m telescope of Calar Alto Observatory (CAHA, Centro Astronomico Hispano-Aleman), and the ISIS spectrograph (blue arm) at the William Herschel Telescope (WHT) of the Observatorio del Roque de los Muchachos (ORM).

We also have access to spectra from The Hobby-Eberly Telescope (HET) obtained by A. Pellerin. These data were observed throughout 2012 in four different spectral ranges (3800 -- 4700, 4700 -- 5700, 5300 -- 6300 and 6400 -- 7400~\AA) using the High-Resolution Spectrograph (HRS) at R $\sim$30000 and the 600g4739/6302 gratings. We have a total of 48 spectra for 12 stars (one in each spectral range). The S/N of these spectra is variable, ranging from 100 to 300.

Finally, we also include spectra observed by us in different runs at the William Herschel Telescope (WHT)  using the ISIS spectrograph (a total of 30 spectra, 15 in each blue and red arms). These data were acquired in two different observing runs. For 2015, we chose the H2400B grating for the blue arm, which provides a resolving power of $\sim$13600 at 4500~\AA. As the dichroic allows us to observe simultaneously with the red arm, we used the R1200R grating (R $\sim$9300 at 6500~\AA). Additionally we included stars observed by A. Herrero in 2003 also using ISIS at the WHT with the same configuration (H2400B and R1200R gratings). All ISIS spectra have S/N $>$100 (from 100 to 200, depending on the stellar magnitude).

A brief summary of the data used in this work is shown in Table~\ref{table_inst_cyglp}. The final spectral sample is composed of 78 O-type stars (12 of them previously classified as SB2) whose spatial distribution is shown in Fig.~\ref{fig_sample}. The main data and photometric information are presented in Table~\ref{table_fotom}.

\begin{figure}[t!]
\centering
\includegraphics[width=9cm]{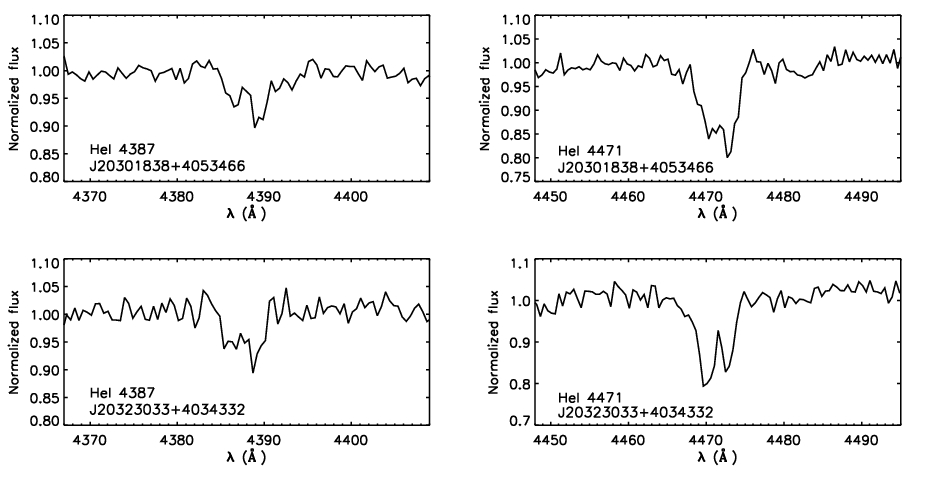}
\caption{Example of the new SB2 detections: HeI lines in stars J20301838+4053466 (top) and A32 (bottom).}
\label{fig_bin}
\end{figure}

\subsection{Binary fraction}\label{bin_fraction}

A large number of multiplicity studies among massive stars have been conducted during the last years \citep[see][for a brief overview]{sana17} where the number of detected massive O-type binaries has been continuously revised and updated. The Cygnus OB2 Radial Velocity Survey is one of the most relevant multiplicity census carried out in this association \citep{kiminki07, kiminki08, kiminki09, kiminki12, kobulnicky12, kiminki15}, where the binary properties of its massive star population have been statistically studied.  \cite{kobulnicky14} list 48 massive OB multiple systems known in Cygnus OB2, 26 of them containing at least one O-type star. In recent years, \cite{maiz16,maiz19} have provided spectral classification for most of the detected companions.

\begin{table*}[t!]
\centering
\caption{Spectroscopic binary systems identified in Cygnus OB2 and containing at least one O-type star.}
                \label{table_bins}
                \begin{tabular}{lccccc}
                \hline 
                \hline \\[-1.5ex]  
                \small{Star}& \small{Binary type}&\small{Components}& \small{SpT} &\small{SpT ref.} &\small{Bin. ref.}\\ 
                \hline\\[-1.5ex]                
                \small{Cyg OB2~$\#$1}&\small{SB1}&\small{A, B(vis)} & \small{O8 IV(n)f + B} & \small{1} & \small{13}\\          
        \small{J20311833+4121216}&\small{SB1}&\small{A, B} & \small{O9.5 IV + B} & \small{1} & \small{6}\\
        \small{Cyg OB2~$\#$20}&\small{SB1}&\small{A, B} & \small{O9 IV + mid-B} & \small{1} & \small{12}\\  
                \small{Cyg OB2~$\#$15}&\small{SB1}&\small{A, B} & \small{O8 III + B} & \small{1} & \small{12}\\             
        \small{CPR2002~A11}&\small{SB1}&\small{A, B} & \small{O7.5 Ib(f) + u} & \small{2} & \small{6}\\
        \small{Cyg OB2~$\#$17}&\small{SB1}&\small{A, B} & \small{O8 V + u} & \small{4} & \small{4}\\       
                \small{MT91-376}&\small{SB1}&\small{A, B} & \small{O9.7 III(n) + B:} & \small{2} & \small{4}\\                 
                \small{J20330292+4117431}&\small{SB1}&\small{A, B} & \small{O7.5 V((f)) + u} & \small{1} & \small{4}\\           
        \small{Cyg OB2~$\#$22B}&\small{SB1}&\small{ B1, B2} & \small{O6 V((f)) + u} & \small{9} & \small{10}\\
        \small{Cyg OB2~$\#$22C}&\small{SB1}&\small{C1, C2} & \small{O9.5 IIIn + u} & \small{9} & \small{7}\\             
                \small{Cyg OB2~$\#$9}&\small{SB2}&\small{A, B} & \small{O4 If + O5.5 III(f)} & \small{1} & \small{8}\\             
        \small{MT91-448}&\small{SB1}&\small{A, B} & \small{O6.5: V + u} & \small{1} & \small{4}\\
        \small{Cyg OB2~$\#$8A}&\small{SB2}&\small{A1, A2, A3(vis)} & \small{O6 Ib(fc) + O4.5 III(fc) + u} & \small{1} & \small{14}\\           
                \small{Cyg OB2~$\#$8D}&\small{SB1}&\small{D1,D2,D3} & \small{O8.5 V + O9 V + u} & \small{4} & \small{4}\\                 
        \small{MT91-485}&\small{SB1}&\small{A, B} & \small{O8 V + early B-O} & \small{2} & \small{4}\\
        \small{Cyg OB2~$\#$74}&\small{SB1}&\small{A, B} & \small{O8 V + B2-O} & \small{4} & \small{4}\\       
                \small{Cyg OB2~$\#$70}&\small{SB1}&\small{A, B} & \small{O9.5 IV(n) + u} & \small{1} & \small{4}\\
                \small{Cyg OB2~$\#$27}&\small{SB2}&\small{A1, A2, B(vis)} & \small{O9.7 V(n) + O9.7 V:(n) + u} & \small{2} & \small{3}\\          
        \small{MT91-720}&\small{SB2}&\small{A, B} & \small{O9.5 V + B1-2 V} & \small{4} & \small{6}\\ 
        \small{Cyg OB2~$\#$11}&\small{SB1*}&\small{A, B} & \small{O5.5 Ifc + u} & \small{9} & \small{6}\\
        \small{Cyg OB2~$\#$29}&\small{SB1}&\small{A, B} & \small{O7 V(n)((f))z + u} & \small{2} & \small{4}\\          
                \small{MT91-771}&\small{SB2}&\small{A, B} & \small{O7 V((f)) + O7 IV((f))} & \small{1} & \small{7}\\                 
        \small{Cyg OB2~$\#$3}&\small{SB2}&\small{A1, A2, B(vis)} & \small{O8.5 Ib(f) + O6 III: + B0 IV} & \small{1} & \small{13}\\
        \small{Cyg OB2~$\#$5A}&\small{SB2}&\small{A1,A2} & \small{O6.5: Iafe + O7 Iafe} & \small{1} & \small{15}\\           
                \small{Cyg OB2~$\#$ 73}&\small{SB2}&\small{A, B} & \small{O8 Vz + O8 Vz} & \small{2} & \small{12}\\          
        \small{CPR2002~B17}&\small{SB2}&\small{A, B} & \small{O6 Iaf + O9: Ia:} & \small{2} & \small{11}\\ 
        \small{CPR2002~A36}&\small{SB2}&\small{A, B} & \small{O9.5 II + O8.5 III} & \small{16} & \small{4}\\ 
        
        \small{J20301838+4053466}&\small{SB2}&\small{A, B} & \small{O9 V + u} & \small{0} & \small{0}\\          
        \small{CPR2002~A32}&\small{SB2}&\small{A, B} & \small{O9.5 IV + u} & \small{0} & \small{0}\\     
                                
        \hline
        \multicolumn{6}{l}{\footnotesize Refs.  (0) This work, (1)\cite{maiz19}, (2) \cite{maiz16}, (3) \cite{salas15},} \\
        \multicolumn{6}{l}{\footnotesize (4) \cite{kobulnicky14}, (5) \cite{sota14}, (6)\cite{kobulnicky12}, (7) \cite{kiminki12},} \\           \multicolumn{6}{l}{\footnotesize (8) \cite{naze12}, (9) \cite{sota11}, (10) \cite{maiz10} , (11) \cite{stroud10},} \\      
                \multicolumn{6}{l}{\footnotesize (12) \cite{kiminki09}, (13) \cite{kiminki08}, (14) \cite{debecker04}, (15) \cite{rauw99}. } \\  
                \multicolumn{6}{l}{\footnotesize (16) Maíz Apellániz, private comment. Notes: ($vis$) indicates that the second component is a visual companion.}\\
                \multicolumn{6}{l}{\footnotesize $u$ indicates that the spectral type is undetermined.}\\
                \multicolumn{6}{l}{\footnotesize *single and standard star in GOSSS \citep{sota11}. \cite{maiz19} indicate that this is an }\\
            \multicolumn{6}{l}{\footnotesize isolated star based on Astralux images. }\\ 
                \end{tabular}   
\end{table*}

In this work, we found two new SB2 systems within our sample of O-type members of Cygnus OB2: J20301838+4053466 and CPR2002~$\#$A32\footnote{\cite{com02} provides an extension for the original Schulte's list of members in Cygnus OB2. From now on, we refer these sources using directly the  number assigned in CPR2002 (i.e., A32 in this case). } (see Fig.~\ref{fig_bin} for an example of the binary detection). A list of the currently known binary (or multiple) systems in the association is presented in Table~\ref{table_bins} where we also show the binary type (single-lined SB1 or double-lined SB2), the components, spectral types, and bibliographic references. Taking into account the results of this work, 29 of 78 O-type stars in the region are part of binary/multiple systems (representing a fraction of 37$\%$). \cite{sana11} found that at least 45 -- 55$\%$ of the O star population in clusters and OB associations is comprised of spectroscopic binaries, making it highly likely that more binaries have gone undetected in our sample. 

We highlight some individual remarks for three of the most relevant multiple systems known in Cygnus OB2: Cyg OB2~$\#$5, Cyg OB2~$\#$22, and Cyg OB2~$\#$8.
The first is a multiple system for which \cite{maiz10} lists up to four components. Here, $\#$5A is the star that gives its name to the system, and is an eclipsing, contact binary formed by an O6.5 -- O7 and a WN9/Ofp stars \citep{rauw99} that is overluminous for its mass. The $\#$5B component, at 0.9$''$ from $\#$5A, has been classified as O7 Ib(f)p var?  by \cite{maiz19}. The GOSSS catalogue gives separated spectra for $\#$5A and $\#$5B.
For the trapezium-like system Cyg OB2~$\#$22 \cite{maiz10} has identified a total of ten components  and the GOSSS Survey has provided spectral data for the first five O components (A,B,C,D and E).
At least four O-type visual components have been detected in the Cyg OB2~$\#$8 trapezium-like system \citep{maiz12}. The first component, $\#$8A, has been identified as a SB2 star composed by two early O-type stars \citep{debecker04, maiz19}.
Both Cyg OB2~$\#$8 and $\#$22  have been traditionally treated as multiple stars but as already pointed out by \cite{bica03}, they should be considered as two clusters likely constituting the core of Cygnus OB2 and hosting the brightest stars of the area. A recent analysis by Maíz Apellániz et al. (A$\&$A submitted) has confirmed their cluster nature, and showed that both are at a distance of 1.7~kpc \cite[the same distance as obtained by][see Sect.~\ref{mrlm}]{berlanas19} and have similar proper motions. Therefore, they are a double cluster separated by 2.7~pc in the plane of the sky and the paper above proposes naming them as Villafranca O-007 and Villafranca O-008, respectively. 
As a final remark, we note that although A36 was classified as B0 Ib + B0 III by \cite{kobulnicky14}, it has been revised by \cite{kiminki15} who indicated that both components may be slightly hotter. They classified it as O9 -- O9.5 III: + O9.5 -- B0 IV: thus, we include the system in our sample.  Recently, it  was reclassified as O9.5 II + O8.5 III by J. Maíz Apellániz (in prep.) and we adopt this classification in our work.

\subsection{Selecting the final spectral sample}\label{sample_selection}

The detected SB2 stars (see Table.~\ref{table_bins}) are excluded from our final sample for the spectroscopic analysis. We also excluded the Cyg OB2~$\#$5B and Cyg OB2~$\#22$E  components since they are not suitable targets for an accurate determination of the stellar parameters. The $\#$5B component shows a peculiar variable spectra with many emission lines and the faint $\#22$E component shows a noisy spectrum.
Tentatively, we also exclude the Cyg OB2~$\#22$D component. For this star, we have spatially resolved spectroscopy from sources 1 and 6 of Table~\ref{table_inst_cyglp}, but a detailed inspection of the spectra reveals peculiar line profiles in the red spectrum that are not compatible with those in the blue. Since these spectra were obtained in different observing campaigns with a lapse of a few years, new data are required to evaluate stellar variability or even a possible target confusion. Finally, the star J20272428+4114458 was classified as O9.5 V by \cite{com12} but it has been reclassified as B0 IV in this work. As it is the only star that changes from O to B-types, we decided to keep it in the sample to see its behavior and inspect its properties.

We therefore have a final sample of 62 O-type stars (plus the star indicated above, out of a total of 78 present in our census) suitable for a quantitative spectroscopic analysis. Since we have compiled spectral data from different sources, in many cases, we have different spectra for the same star. In these cases we adopted the following strategy. To determine the distribution of rotational velocities (henceforth, sample A), we select the diagnostic lines in each spectrum following the criteria given in Sect.~\ref{broad}. Thus, we chose the spectra with the higher resolving power since it allows us to better identify spectroscopic binaries and to disentangle the macroturbulence broadening component ($vmac$), obtaining more accurate projected rotational velocity ($vsini$) determinations \citep[see][and references therein]{ssimon07,ssimon14a, ssimon17}.
 However, in order to determine the main stellar parameters, we use the spectra with the best quality in terms of S/N (henceforth, sample B) and, whenever possible, the best resolution for the H and He lines.

\begin{figure*}[t!]
\centering
\includegraphics[width=7cm]{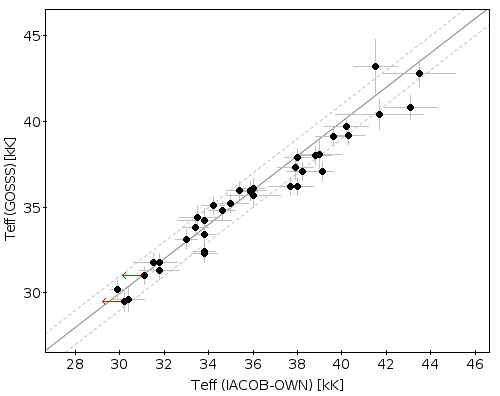}
\includegraphics[width=7cm]{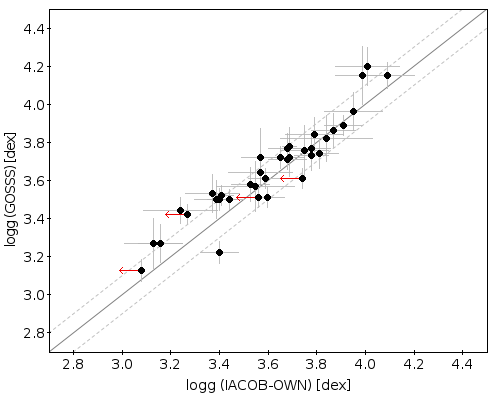}
\caption{Comparison between the stellar parameters obtained using spectra from the IACOB-OWN (R $\geq$ 25000) and the GOSSS surveys (R $\sim$ 2500) for a sample of 37 standard O-type stars in common. The same $vsini$ values and set of diagnostic lines were used for the spectroscopic analysis. Red arrows indicate upper limits.}
\label{test_R}
\end{figure*}

As an example, for the Cyg OB2~$\#$6 star, we have data from three different instruments: sources 2, 4, and 6 of Table~\ref{table_inst_cyglp}. In order to determine the rotational velocity and macroturbulence broadening component, we use the source number 4 because of the higher spectral resolution (R $\sim$30000) and the fact that the  O\,III~5592 line is included in the spectrum. On the other hand, for the stellar parameters we use sources number 2 (for the optical-blue range) and 6b (for the optical-red range) because of the higher quality of the spectra (in terms of S/N). We note that we are assuming no significant variations between individual blue and red spectral data (e.g., no strong variations in H$\alpha$); however in the cases where only one spectrum is available for each wavelength range, this assumption cannot be confirmed.
A final list of the whole O-type population of our sample, along with the new spectral classification  and the spectral data selected for each case (A or B) is shown in Table~\ref{table_stars1}.

\subsection{Implications of using an inhomogeneous spectral sample}\label{rv_effect}

Since our final sample is spectrally inhomogeneous in terms of R, S/N, and wavelength coverage, we evaluated the implications of using data with different spectral characteristics for the spectroscopic analysis.

In the case of the wavelength coverage, we know that a limited set of lines affects the parameter determination (mainly for gravity and, therefore, mass) and the errors decrease as we increase the number of diagnostic lines \citep{berlanas17}. Nevertheless, all our spectra cover the whole range where the main diagnostic lines for O-type stars are located (4000 -- 5000~\AA) and we have compensated for the lack of the H$\alpha$ and He\,I+II~6678 lines that existed for some stars by carrying out new spectroscopic observations (see Sect.~\ref{obs}).

Although all the  data used for the analysis have high S/N (exceeding 100 in all cases), the resolving power varies in a wide range (from 2500 to 30000, see Table~\ref{table_inst_cyglp}). We therefore decided to evaluate the possible effect that resolution could have in the derivation of the stellar parameters using spectra from the IACOB-OWN\footnote{See \cite{holgado18} and references therein for the characteristics of these observations.} (R $\geq$ 25000) and GOSSS (R $\sim$2500) surveys. For the test, we selected the same 37 standard O-type stars (at least one for each type and luminosity class) from both surveys. In order to avoid possible additional  uncertainties we performed the spectroscopic analysis considering only the common set of lines and the same values of $vsini$ and $vmac$.
In Fig.~\ref{test_R}, we present a comparison  between the derived effective temperature ($T_{\rm eff}$) and surface gravity (log $g$). We can see that both parameters are only slightly affected by the spectral resolution and most of the stars are located within the usual uncertainties without a global impact. Mean and standard deviation of the difference in $T_{\rm eff}$ and log $g$ from high resolution IACOB-OWN  and low resolution GOSSS  spectra is 0.4 $\pm$ 0.9 kK and -0.05 $\pm$ 0.08 dex, respectively.
We note that the pair ($vsini$, $vmac$) used in this test was obtained from the high resolution IACOB-OWN spectra since the O\,III~5592 line is included and it is one of the best lines for an accurate line-broadening characterization. We also know that resolution affects the $vsini$ determination: \cite{ssimon14a} (henceforth, SDH14) show that a low resolution could lead to an overestimate of the derived $vsini$ of $\sim$25 ($\pm$ 20) km s$^{-1}$ in stars rotating more slowly than 120 km s$^{-1}$. Sabin-Sanjulian (2014, PhD) shows that this could lead to errors in the gravity determination of up to 0.1 dex and a consequent change in \teff~ up to 1000 K (although these are upper limits and changes would tend to be smaller in most cases).

 \begin{table*}[t!]
        \begin{center}
        \caption{\small{Diagnostic lines used for the line-broadening characterization of our final sample.}}
                \begin{tabular}{cccccc}
                \hline 
                \hline \\[-1.5ex] 
                \small{Line}& \small{O\,III~5592}& \small{Si\,III~4552} & \small{He\,I$^{a}$}& \small{N\,V~4603} & \small{He\,II~4541}\\
                \hline \\[-1.5ex] 
                \small{$\#$ Stars} & \small{8} & \small{9} & \small{42} & \small{2} & \small{2}  \\
                \small{$\%$ Sample} & \small{12$\%$} & \small{14$\%$} & \small{67$\%$} & \small{3$\%$} & \small{3$\%$}  \\
        \hline
        \multicolumn{6}{l}{\footnotesize a) unweighted average of nebular free or weakly contaminated He\,4713, 4922,}\\
        \multicolumn{6}{l}{\footnotesize 4387 and/or 4471 lines (whenever present).}      
                \end{tabular}
                \label{table_lines}
        \end{center}
        
\end{table*}

\section{Methods and results}\label{methods}

\subsection{Line-broadening characterization}\label{broad}

Since  metallic lines do not suffer from strong Stark broadening nor from nebular contamination, they are best suited for obtaining accurate $vsini$ values. Unfortunately, they are not present in all our spectra. Consequently, we assumed the following criteria based on the works by \cite{herrero92}, \cite{ssimon14b} and \cite{ragudelo13}: 
(i) we  first choose the O\,III~5592 or Si\,III~4552 diagnostic lines, when either of them is available. If both are present, we prioritize the O\,III~5592 line since it is usually stronger and more isolated than Si\,III~4552 for O-type stars;
(ii)  if none of them are present, we use nebular free or weakly contaminated He\,I lines (an unweighted average of the He\,I~4387, He\,I~4471, He\,I~4713 and He\,I~4922 lines);
(iii) if any star suffers strong nebular contamination or  the He\,I lines are weak or too noisy, we then rely on He\,II~4542; 
(iv) finally, if the stellar wind is weak and the line is of photospheric origin, we also use N\,V~4603.

\begin{figure}[t!]
\centering
\includegraphics[width=8cm]{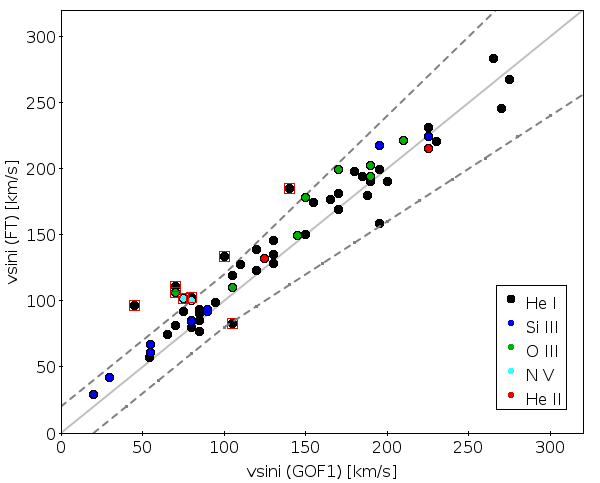}
\caption{Comparison of projected rotational velocities resulting from the Fourier Transform (FT) and the Goodness-Of-Fit (GOF$_{1}$) techniques obtained using the \texttt{iacob-broad} tool. Dashed lines represent a difference of 20 km s$^{-1}$ or 20$\%$ from the 1:1 relation, whichever is the largest. Red squares indicate those stars out of these limits. Different colors indicate the diagnostic lines used for the line-broadening characterization.}
\label{gof1_ft}
\end{figure}

As in \cite{ragudelo13}, for those cases in which different diagnostic lines are available we compared the measured $vsini$ values. We found similar results as the aforementioned authors, obtaining a good degree of agreement. In all cases differences between lines are within the limits represented in Fig.~\ref{gof1_ft} and have no impact on global or individual conclusions. 
A summary of the diagnostic lines used for the line-broadening characterization of our final sample of 63 O-type stars is shown in Table~\ref{table_lines}.

We used the \texttt{iacob-broad} tool \citep{ssimon07,ssimon14a} to obtain the projected rotational velocities and the macroturbulent broadening component for the whole stellar sample (excluding detected SB2 and problematic stars). 
This is a user-friendly IDL procedure for the line-broadening characterization of OB stars, based on a combined Fourier transform (FT) plus a goodness-of-fit (GOF) methodology. It allows the user to  determine easily the stellar projected rotational velocity and the amount of extra broadening (assuming a radial-tangential profile) from a specifically selected diagnostic line. The FT technique is based on the identification of the first zero in the Fourier transform of a given line profile \citep{gray08,ssimon07}. The GOF technique is based on a comparison between the observed and a synthetic line profile that is convolved with different values of $vsini$ and $vmac$ to obtain the best-fit by means of a $\chi^{2}$ optimization. 
The main advantage of the \texttt{iacob-broad} analysis is that we obtain two independent measurements of the $vsini$ (resulting from either the FT or the GOF analysis) whose comparison is used as a consistency check and to better understand problematic cases. The results, with uncertainties in the range 10 -- 20$\%$, are given in Table~\ref{table_param}.

 \begin{figure*}[t!]
\centering
\includegraphics[width=6cm, trim={0.3cm 6cm 8cm 0},clip]{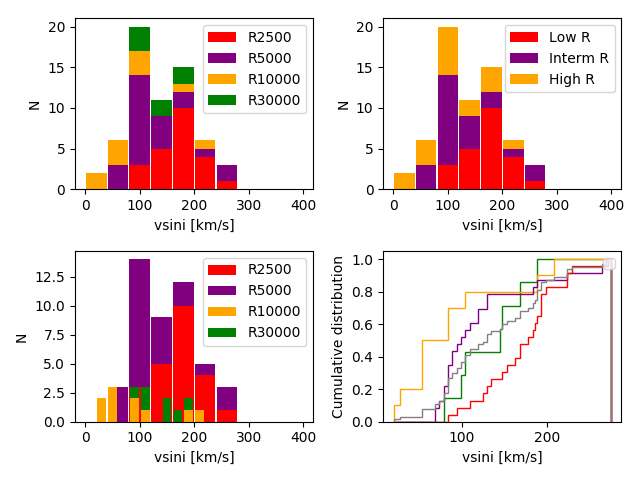}
\includegraphics[width=12cm, trim={0 0 0 6cm},clip]{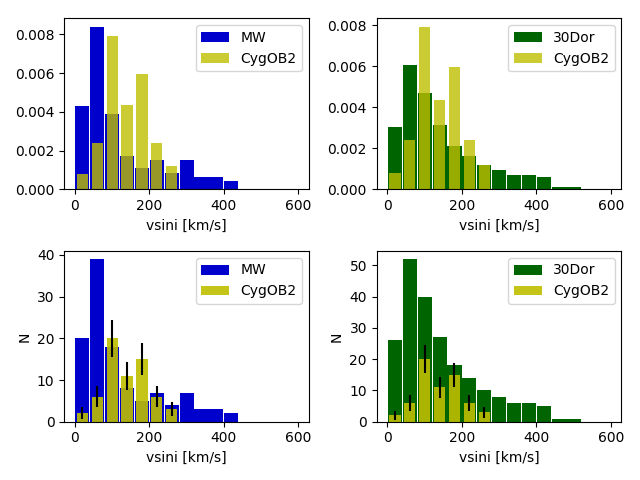}
\caption{  Distribution of rotational velocities of our final sample of 63 O-type isolated stars in Cygnus OB2. {\it Left}: Histogram color-coded on the basis of the resolving power of the spectra used for each target. {\it Middle}: $vsini$ histogram showing the comparison between the distribution of rotational velocities found for our sample of O-type stars in Cygnus OB2 and the distribution of O-type stars found in the MW. Vertical black lines indicate the associated Poisson noise. {\it Right}: As in middle panel but compared to 30 Dor. }
\label{rot_before}
\end{figure*}

\subsection{Distribution of rotational velocities}\label{vsini}

Since there are no previous studies focused on the distribution of rotational velocities for O-type stars in Cygnus OB2, we explore the information contained in our data, although we have to pay attention to their inhomogeneity in resolving power when comparing results from different spectra. 

We first compared the $vsini$ values derived from the Fourier transform technique (FT) and those derived from the goodness-of-fit technique (GOF) as an assessment of the reliability of the results from the line-broadening analysis (see Fig.\ref{gof1_ft}). We see that only a few stars have discrepancies beyond the adopted uncertainties (we adopt limits of 20 km s$^{-1}$ or 20$\%$, whichever is the largest since SH14 found that the agreement is always better than 20$\%$ but for low rotational velocities effect like the spectral resolution or the microturbulence may play an additional role) with no clear systematic trends due to the lines used for the analysis. For the stars for which we find  good agreement (below the  limit given above), we assumed $vsini$ and $vmac$ provided by the GOF technique since it is less affected by the subjectivity in the selection of the first zero of the FT  \citep[see][]{ssimon14a}. For those few cases where we find larger discrepancies (red squares), we identified  clear signatures of asymmetry in the line, perhaps due to a broad-line secondary component. In these cases, the GOF solution tries to fit the line by increasing $vmac$, hence resulting in a lower $vsini$. We then used the $vsini$ provided by the FT technique (which is not very much affected) and the associated $vmac$ (obtained from the fit of the line assuming a fixed $vsini$ value derived from the FT). 
However, we remark that nearly all discrepancies are close to the adopted limit and, therefore, our choice has little impact on the final distribution.

To derive and correctly interpret the distribution of rotational velocities of our sample, we must take into account that our spectral sample for Cygnus OB2 is inhomogeneous in terms of resolving power (see Table~\ref{table_inst_cyglp}), so we must consider the lower limiting $vsini$ of each of the spectral resolutions used in this work. Using a rough approximation (v$_{lim}$ $\sim$ c/R), values of $\sim$100, $\sim$60, $\sim$30, and $\sim$10 km s$^{-1}$ are the lower limits that can be measured at R $\sim$2500, R $\sim$5000, R $\sim$10000, and R $\sim$30000, respectively.
We thus show the distribution of rotational velocities for each group of stars observed at a given resolution separately (see Fig.~\ref{rot_before}, left). We see that the main peak of slow rotators ($vsini \leq$ 100 km s$^{-1}$) is composed mainly of stars observed at R $\sim$5000, with an associated limiting velocity of $\sim$60 km s$^{-1}$. Therefore, we are not able to detect lower velocities and the main peak of slow rotators is shifted to higher values. The second peak of stars rotating at intermediate velocities ($vsini$ $\sim$180\,--\,200~ km s$^{-1}$) is mainly composed by stars observed at R $\sim$2500, with a corresponding limiting velocity of $\sim$100 km s$^{-1}$. Such a low resolution may result in some cases in $vsini$ values overestimated by $\sim$20 km s$^{-1}$ (i.e., one bin). Thus the peaks in our distribution are probably shifted towards higher $vsini$ values (but the results from higher resolution spectra will not be affected). Finally, we remark two more points in the distribution: (a) the very slow rotators ($vsini \leq$ 60 km s$^{-1}$) bins where we see the lack of extreme slow rotators ($vsini \le$ 20 km s$^{-1}$), indicating that some other physical process is preventing the derivation of lower values (probably microturbulence, see \cite{ssimon14a}); and (b) the short tail of fast rotators that extends only up to $vsini \leq$ 280 km s$^{-1}$.

To check how representative the distribution of rotational velocities that we obtained really is and to explore possible future research directions, we compared our results to other similar works, such as the study of \cite{ssimon14a} for a general sample of O-type stars in the Milky Way (MW) and the study done by \cite{ragudelo13} in 30 Doradus in the Large Magellanic Cloud (LMC).  We applied the Kolmogorov-Smirnov two-sided test \citep[see][]{smirnov39,hodges58} to check if the difference between these distributions and ours is statistically significant. The p-values returned ($10^{-5}$ and $10^{-9}$ for the MW and 30 Doradus samples, respectively) reject the null hypothesis that the distributions are the same. The same test carried out for the distributions up to 280 \kms returns values of p= 0.911 and p= 0.904, indicating that the lack of very fast rotators in the Cyg OB2 distribution is significant. 

In Fig.~\ref{rot_before} (middle, right), we compare our results with those obtained by these aforementioned studies. In the middle panel, we compare our results in Cygnus OB2 (in yellow) and the distribution obtained for a general sample of Galactic O stars in the MW (in blue). Both are Galactic samples, the first one composed by O stars belonging to the same young star-forming region and the latter composed by bright O stars belonging to many regions within the Galaxy.  For the latter, the sample consists of 116 O-type stars\footnote{earlier than B0} from the IACOB database that have been observed at R = 46000 and 25000. Both studies have followed the same methodology for the line-broadening characterization although the wavelength coverage of the IACOB spectra allows them to use the Si\,III 4552 and O\,III 5592 lines for all their sources.
The sample of Galactic O stars presents a clear bimodal distribution, with a main peak of slow rotators in the 40 -- 80 km s$^{-1}$ bin (instead of the 80 -- 120 km s$^{-1}$ found for Cygnus OB2) and a tail of fast rotators reaching 450 km s$^{-1}$ (while in Cygnus OB2 we don't reach 300 km s$^{-1}$). Taking into account the lower $vsini$ limit that can be detected for a given resolution,  the difference in the low peak can be understood as due to the impact of the lower resolution in the Cygnus OB2 sample (note also the height of the 0 -- 40  km s$^{-1}$ bin in the MW sample). On the contrary, the lack of the fast rotator tail in the Cygnus OB2 sample is much more difficult to understand and demands a more detailed study.

In the right panel, we compare the $vsini$ distribution resulting from our study in Cygnus OB2 (in yellow) and the distribution for a sample of 216  O-type stars in 30 Doradus \citep[in green, see][]{ragudelo13}. Both are samples of stars belonging to young star-forming regions that differ in the number of targets (66 versus 216), the environment metallicity (Z$_{LMC}$ = 0.5 Z$_{\odot}$) and the resolving power of the spectral data (R $\sim$8000 for all targets in 30 Dor). Again, the same methodology was used for the line-broadening characterization, and due to similar wavelength restrictions both had to rely in many cases on He\,I instead of metallic lines. The peak of slow rotators in 30 Dor is found at 40 -- 80 km s$^{-1}$ and there is also a more evident tail of fast rotators reaching 600 km s$^{-1}$, something expected for O stars located in lower metallicity regions as they lose less angular momentum through stellar winds \citep[see, e.g.,][]{langer12}. The 30 Dor distribution is therefore very similar to the MW one.

In both cases we find a similar behavior, showing significant differences with respect the distribution derived for Cygnus OB2. The main peak of slow rotators is shifted to lower velocities, which can be easily explained by the different resolutions. In addition to this, the fact of having limited number of epochs for our spectra affects the detection of SB2 stars, overestimating the derived $vsini$ values and increasing the peak. However, the extended tail of fast rotators in the MW and 30 Dor distributions is not present in our Cygnus OB2 data, which demands further attention.

\subsection{Spectral classification}\label{sp_class}

Since we have increased the quality of the previous available spectral data for a large number of stars in our sample, we revised the spectral classification of these stars using the following criteria.

The main classification criterion used for O-type stars is the comparison of the He\,II~4542 and He\,I~4471 lines, whose ratio is unity for an O7 type star. He\,I tends to increase in strength with decreasing temperature while He\,II decreases. Spectral types earlier than O8 can be classified using these criteria \citep[see e.g., Walborn in][]{gray09}. For later O types the relative strengths of He\,II 4542/He\,I 4387 and He\,II~4200/He\,I~4144 may be used, representing the main diagnostic for types O8 -- B0 \citep{sota11}. We adopted the criteria of \cite{sota11} using the list of qualifiers for O spectral types summarized in their work. 
Regarding the luminosity class, the classification criteria for early O-type stars were introduced by \cite{walborn71, walborn73}, taking into account the emission effects in the He\,II~4686 and N\,III~4634-4640-4642 lines. \cite{sota11} provided an updated classification system for late O-type stars, which we adopted in this work.
Although the classification for our O-type stars was based on He lines as described above, we corroborated their spectral types by using the Marxist Ghost Buster tool \citep[MGB, ][]{maiz12}. It compares the observed spectra with a standard library of O stars  \citep[in this work the GOSSS library, see][]{maiz16}. This interactive software allows us to vary the spectral subtype, luminosity class, line broadening, and spectral resolving power of the standard spectrum until we obtain the  best match. In addition, it also allows us to combine two standard spectra (with different velocities and flux fractions) to fit SB2 binaries. 

The previous (from the literature) and new (from this work) spectral types are listed in Table~\ref{table_stars1}. There is quite good agreement between the previous spectral types and the new ones. In nearly all cases, the differences are not larger than half a spectral subtype and one luminosity class. In three cases, the differences reach one spectral subtype (J20291617+4057371, MT91-455 and A20). 
Only in one case we find a significant change in luminosity class (B18, reclassified from O7: Ib to O7.5 IV:(f)), indicating the interest of a follow-up for this target.

\subsection{Determination of the main spectroscopic parameters}\label{param}

We determined the fundamental stellar parameters for the final sample of 63 O-type stars in Cygnus OB2 performing a quantitative spectroscopic analysis based on synthetic FASTWIND models  \citep{santolaya97, puls05, rivero12} and the \texttt{iacob-gbat} tool \citep{ssimon11, holgado18}.

\begin{table}[t!]
 \caption{\small{Parameter ranges of the HHe FASTWIND grid at solar metallicity used in this work.}}
        \begin{center}
                \begin{tabular}{lcc}
                \hline 
                \hline 

                \small{Parameter}& \small{Range or specific values}& \small{Step}\\
                \hline
                \small{$T_{\rm eff}$ [K]} & \small{22000--55000} & \small{1000} \\
                \small{log $g$ [dex]} & \small{2.6--4.4} & \small{0.1}  \\
                \small{log $Q$ [dex]} & \small{-11.7, -11.9, -12.1, -12.3, -12.5,} & \small{-}  \\
                \small{} & \small{ -12.7, -13.0, -13.5, -14.0, -15.0} & \small{}  \\                             
                \small{Y(He) [dex]} & \small{0.06, 0.10, 0.15, 0.20, 0.25, 0.30 } & \small{-}  \\                  
                \small{$\xi$ [km s$^{-1}$]} & \small{5--20} & \small{5}  \\ 
                \small{$\beta$} & \small{0.8--1.2} & \small{0.2}  \\ 
        \hline
                \multicolumn{3}{l}{\footnotesize Note: Grid calculated using the CONDOR workload management}\\
                \multicolumn{3}{l}{\footnotesize  system (http://www.cs.wisc.edu/condor/).}\\
                \end{tabular}
                \label{ranges_grid}
        \end{center}
        
\end{table}

 The code allows us to  accurately determine the basic six spectroscopic stellar parameters for OB type stars: the effective temperature ($T_{\rm eff}$), surface gravity (log $g$), wind-strength parameter ($Q$, defined as $\dot{M}$/(v$_{\infty}$ R)$^ {1.5}$), helium abundance ($Y(He)$, defined as N(He)/N(H)), microturbulence ($\xi$), and the exponent of the wind velocity-law ($\beta$) from their H and He lines\footnote{The following optical diagnostic lines were considered for the analysis of our stellar samples (whenever present):  H$\alpha$,  H$\beta$, H$\gamma$, H$\delta$, H$\epsilon$,He\,I+II~4026, He\,I~4387, He\,I~4471, He\,I~4713, He\,I~4922, He\,I~6678, He\,II~4200, He\,II~4541, He\,II~4686, He\,II~5411 and He\,II~6682.}. Once the observed spectrum is processed, the tool compares the observed and the synthetic line profiles (from FASTWIND models, in our case) by applying a $\chi^2$ algorithm. It computes the line-by-line $\chi^2$ distributions, estimating the goodness-of-fit for each model within a subgrid of models selected from the global grid. Then it iteratively computes the global $\chi^2$ distribution, from which the final parameter values and their associated uncertainties are estimated. The given parameters are the mean values computed from the models located within the 1-$\sigma$ confidence level of the total $\chi^2$ distributions  (after each model has been weighed by its corresponding  $\chi^2$ value). Then their uncertainties are given by the standard deviation within the 1-$\sigma$ level.
Our grid of models covers the wide range of  stellar and wind parameters considered for standard OB-type stars, from early-O to early-B types and from dwarf to supergiant luminosity classes (see Table~\ref{ranges_grid} for grid details). As explained in Sect.~\ref{sample_selection}, when different spectral data were available for the same target, we chose the spectra with the best quality in terms of S/N (and highest resolving power if possible). Derived stellar parameters are shown in Table~\ref{table_param}. We complement these results with a series of figures (see Appendix~\ref{appB}) in which the best-fitting model resulting from the analysis of each star is overplotted on the observed spectrum.

\subsection{Radii, luminosities, and masses from  \textit{Gaia} DR2}\label{mrlm}

The \texttt{iacob-gbat} tool also allows us to  obtain directly physical parameters when the absolute magnitude $M_{V}$ is provided. It computes the radius ($R$), luminosity ($L$), and spectroscopic mass ($M_{sp}$) following the same methodology as above.  With this aim, the tool uses the equation~(Eq. \ref{eq_r}) introduced by  \cite{kud80}, where the \( \mathcal{V} \) parameter represents the integral in wavelength  of the emergent flux of the model \citep{ms63}:
\begin{align}\label{eq_r}
5~log (R/R_{\odot}) = 29.57 -  M_{V} +  \mathcal{V} 
\end{align}

\noindent and then we use the usual formulae to derive masses and luminosities:
\begin{equation}\label{eq_l}
log (L/L_{\odot}) = 2~(R/R_{\odot}) + 4~log (T_{eff}/T_{eff,\odot})
,\end{equation}
\begin{equation}\label{eq_m}
log (M_{sp}/M_{sp,\odot}) = 2~(R/R_{\odot}) + log (g/g_{\odot})
.\end{equation}

By using the unprecedented parallaxes provided by {\it Gaia} in its second data release \citep[DR2,][]{brown18}, we derived $M_{V}$  for all our stellar sample with reliable astrometry  \citep[RUWE $\leq$ 1.4, see][for details on this criterion]{lindegren18}. However, we should take into account that these parallaxes could be affected by systematic uncertainties of up to 0.1 mas that are not well understood \citep{luri18} and, therefore, should be taken with caution.

Since V-band photometry is not available for all stars of our sample (see Table~\ref{table_fotom}) we followed the same procedure used by \cite{com08} to derive $M_{V}$ values. With $M_V$ being the absolute magnitude we have
\begin{equation}\label{eq_mv1}
M_{V} = 5 - 5*log(d) + V - A_{V} = - DM + V - A_{V}
,\end{equation}

\noindent where DM is the distance modulus. The visual extinction $A_{V}$ can be defined in terms of the $(V - K_{s})$ color excess as
$A_{V} = E(V - K_{s}) +  A_{ K_{s}}$.  Then, $M_{V}$ can be defined as
\begin{equation}\label{eq_mv2}
M_{V} = K_{s} - DM - A_{K_{s}} + (V - K_{s})_{o}
.\end{equation}

\noindent We used distances provided by \cite{bailer-jones18} (henceforth, BJ18) to estimate individual values of the DM and adopted optical and near-IR photometry from the USNO-B and 2MASS catalogs and unreddened intrinsic colors from \cite{martins06}. The use of the alternative prior of \cite{maiz01}, which is optimized for early-type stars in the Galactic disk, provides similar distances within the uncertainties. We used the $R_{V}$ = 3.1 extinction law from \cite{rieke85} to derive $A_{Ks} = 0.092* E(B - K_{s})$.   We note that our results will be dependent both on the extinction law adopted and on the assumption of a constant average $R_{V}$. This coefficient depends on the properties of the absorbing dust grains and thus on the kind of region containing them \citep[see, e.g., ][]{maiz18}. 
Nevertheless, for the whole sample we find good agreement between the derived $M_{V}$ values and those obtained from the \cite{martins06} calibration. The
$M_{V}$ uncertainties were obtained from the distance errors provided by BJ18. Uncertainties related to the apparent magnitudes are negligible compared with those obtained from distances, that are of the order of $\pm$ 0.13 magnitudes.

The final list of targets with reliable astrometry for the O-type population in Cygnus OB2 is composed of 52 stars, whose corresponding $Gaia$ DR2 sources,  estimated distances by BJ18 and derived  $M_{V}$ values are presented in Table~\ref{table_gaia}.
 This table also includes the membership group of each star from \cite{berlanas19} (henceforth B19). Group 2 is the main Cygnus OB2 group at $\sim$1.76 kpc and Group 1 is a closer one at $\sim$1.35 kpc\footnote{ \cite{rygl12} estimates a distance to Cygnus OB2 of $\sim$1.4 kpc using four masers as astrometric targets, that would be included in this foreground group.}. Group 3 contains foreground or background contaminants. Group 0 indicates that the object lies between Groups 1 and 2 and could not be reliably assigned to any group. Luminosities, radii and spectroscopic masses from the \texttt{iacob-gbat} tool are given in Table~\ref{table_gaia} as well. Uncertainties include the \texttt{iacob-gbat} formal uncertainties for the stellar parameters and those related to $M_{V}$.

\section{Discussion}\label{discussion}

\subsection{Evolutionary status}
\label{HRD}

Using the luminosities derived from $M_{V}$ and the \texttt{iacob-gbat} tool, we placed the sample of stars that have passed the criteria for reliable astrometry (RUWE $\leq$ 1.4) in the Hertzsprung-Russell diagram (HRD).  The mean uncertainties are $\sim$0.01 dex in log $T_{\rm {eff}}$ and $\sim$0.05 dex in log(L/L$_{\odot}$). 
We decided to explore four different stellar models (two families with two initial rotational velocities) in order to assess possible uncertainties in the obtained distribution.
We used rotating and non-rotating Geneva \citep{ekstrom12} and Bonn \citep{brott11} stellar evolutionary tracks and isochrones. In addition, we  also divided the sample into the different stellar groups found by B19. 

\begin{landscape}

\begin{figure}[ht!]
\centering
\includegraphics[width=12cm, angle=180,trim={0.6cm 2cm 0cm 0.5cm},clip]{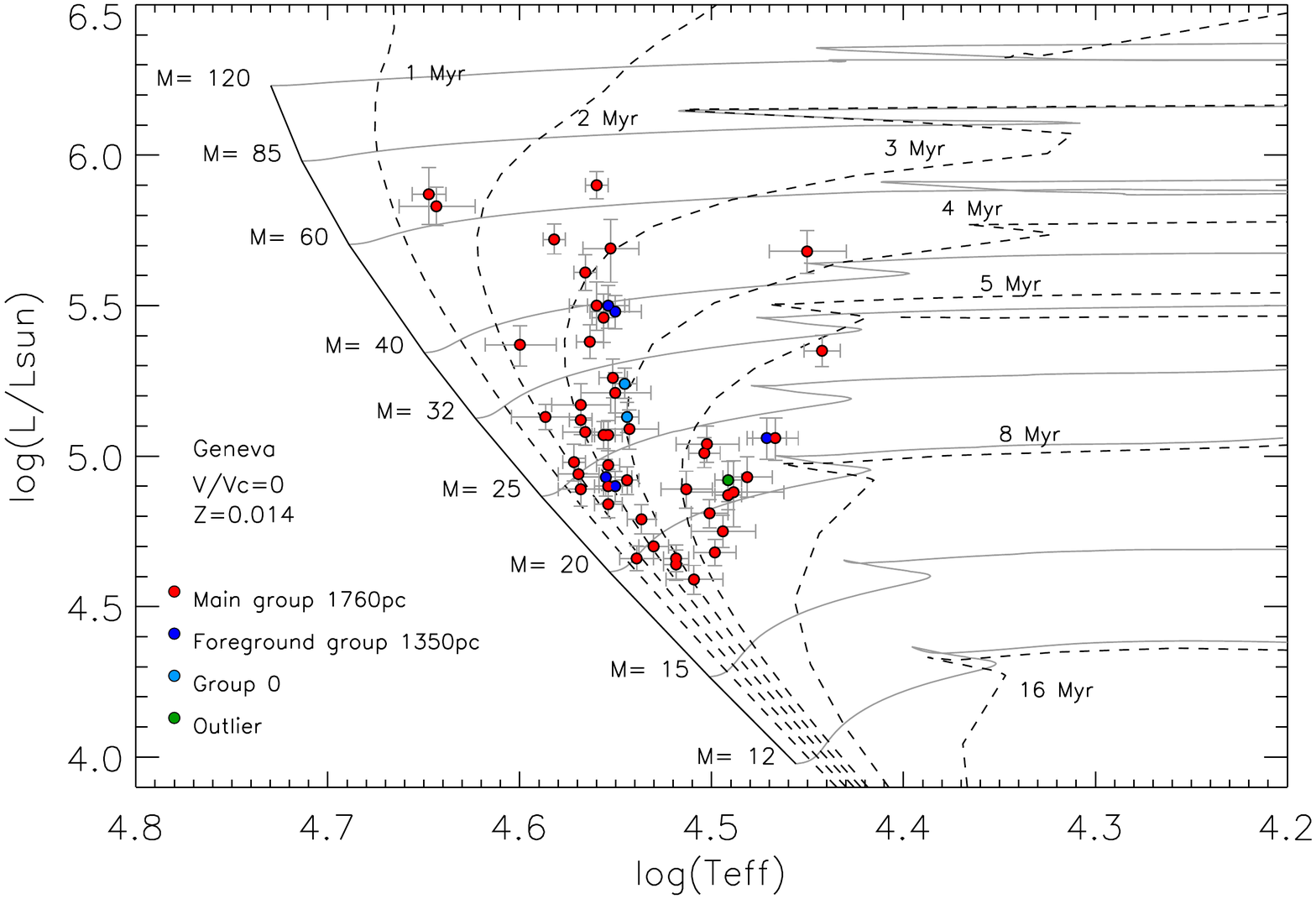}
\includegraphics[width=12cm, angle=180,trim={0.6cm 2cm 0cm 0.5cm},clip]{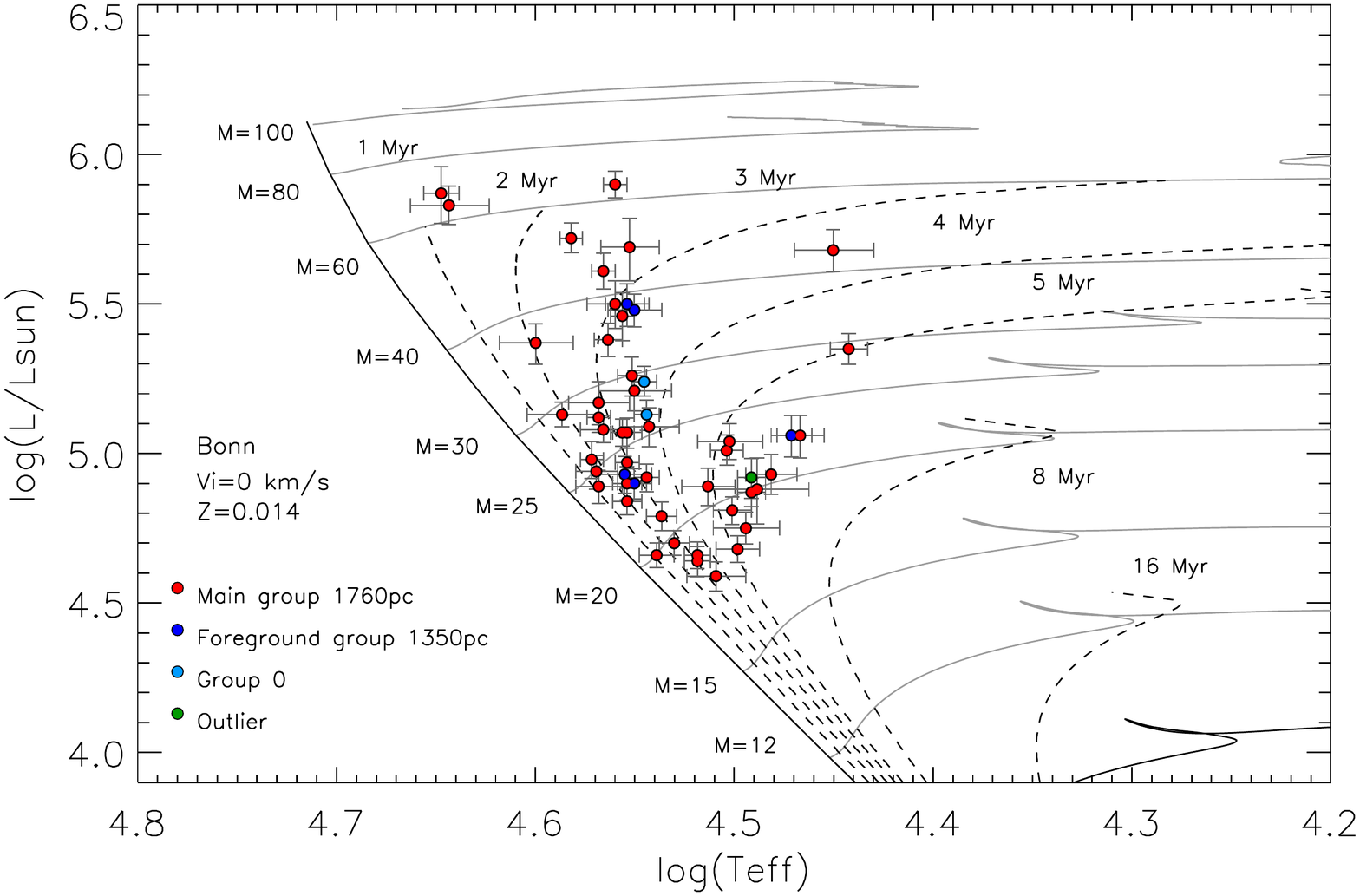}
\includegraphics[width=12cm, angle=180,trim={0.6cm 2cm 0cm 0.5cm},clip]{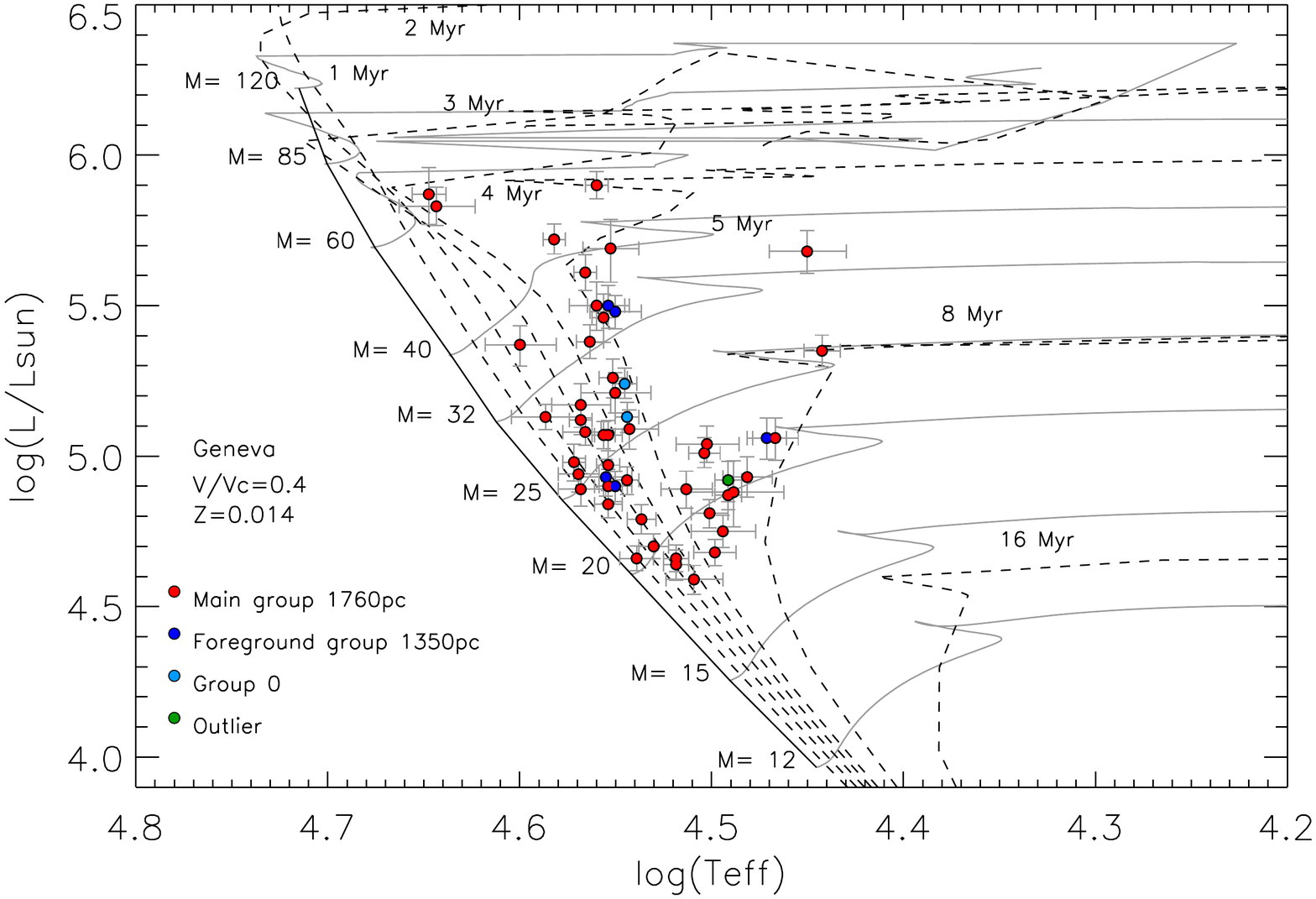} 
\includegraphics[width=12cm, angle=180,trim={0.6cm 2cm 0cm 0.5cm},clip]{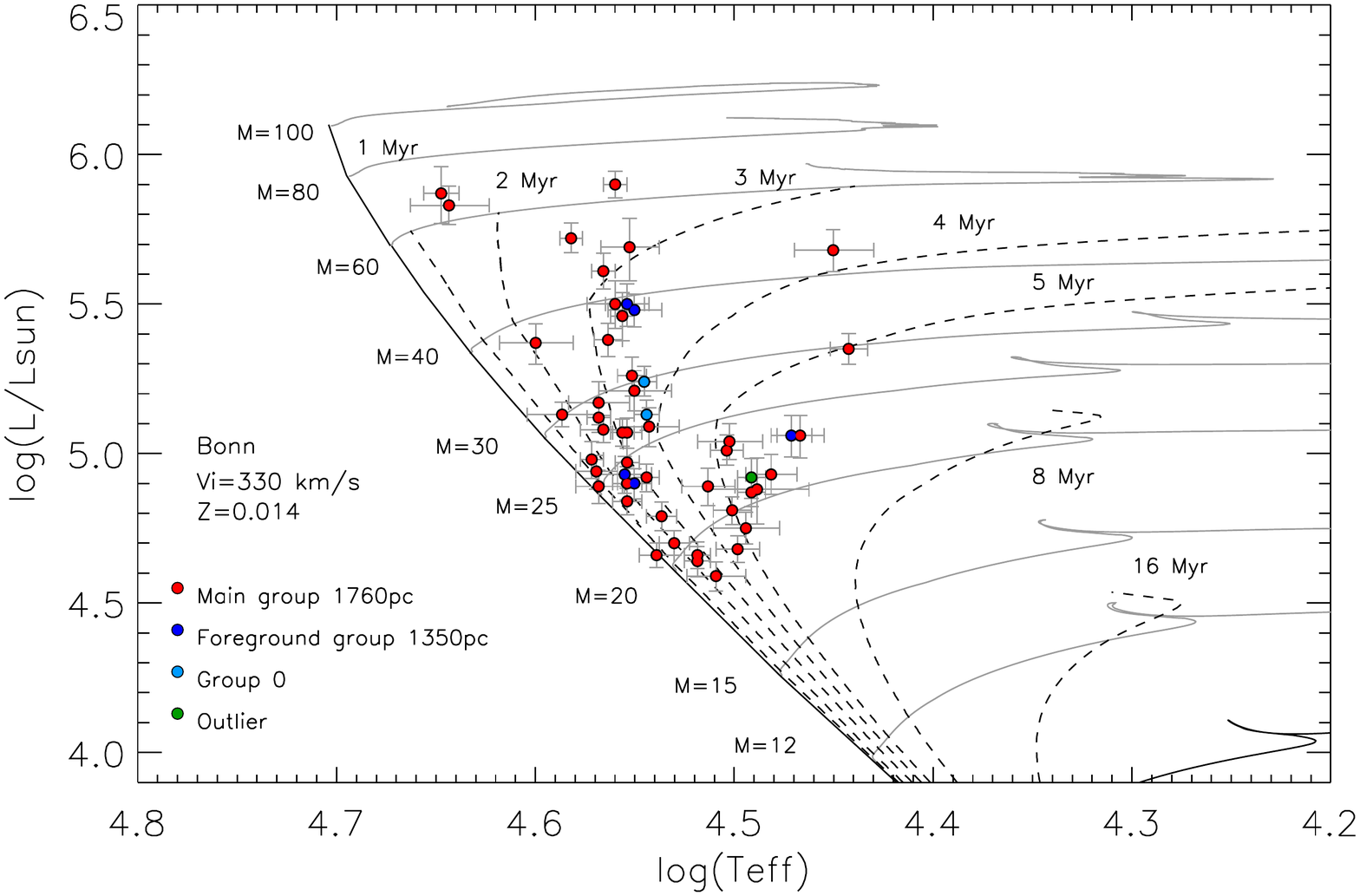} 
\caption{ HRD for the sample of O stars in Cygnus OB2 with reliable astrometry using Geneva \citep[][left]{ekstrom12} and Bonn \citep[][right]{brott11} evolutionary stellar tracks and isochrones. Non-rotating models are on the top and rotating models on the bottom. The sample is also subdivided into the spatial groups found by \cite{berlanas19}. Mean uncertainties are $\sim$0.01 dex in log $T_{eff}$ and $\sim$0.05 dex in log(L/L$_{\odot}$).}
\label{hrd}
\end{figure}

\end{landscape}

In Fig.~\ref{hrd} (top left hand panel), we show the HRD using Geneva non-rotating evolutionary models. The age of the O population is well constrained between 1 -- 6 Myr, as estimated by other authors \citep{hanson03,neg08,com12, wright15,berlanas18a}. We find the most massive star at around 75 M$_{\odot}$ (Cyg OB2~$\#22$A) and the least massive one at 18 M$_{\odot}$ (Cyg OB2~$\#23$), as expected for O-type stars.
However, different  stellar age groups can be clearly distinguished within the whole population\footnote{The same behavior is found when using global distances to the B19 groups instead individual BJ18 distances. However, if the canonical distance modulus of 10.8 \citep{rygl12} is used to derive $M_{V}$ values for all stars of our sample, some of them would appear located close or even slightly  below the ZAMS, depending on the considered stellar models. At face, they could constitute the result of a very recent star-forming burst.}.
We find the older group following the $\sim$6 Myr isochrone with initial masses between 18 -- 40 M$_{\odot}$. The group is well separated in luminosity class, with the dwarfs located along the Main Sequence (MS) and the two supergiants A29 and J20283038+4105290 close to (but before) the terminal-age main sequence (TAMS).
A more numerous and younger group is found along the $\sim$3 Myr isochrone with initial masses between 20 -- 60 M$_{\odot}$. It is more tightly concentrated than the others, containing most of the population. Interestingly, all the stars belonging to Group 1 (the foreground group at $\sim$1350 pc) and Group 0 (located between the foreground and the main group) identified by B19 seem to follow the same isochrone, except for one object in each group.
A third smaller group containing the apparently youngest stars of the sample seems to follow the $\sim$1.5 Myr isochrone with initial masses between 20 -- 75 M$_{\odot}$. The hottest stars of our sample, Cyg OB2~$\#22$A and  Cyg OB2~$\#7$  are included in this group \citep[both O3 supergiants, see][]{walborn02}. Without the presence of the most massive stars (with M$>$60 \Msun) it is possible to assign the rest of the stars to the group at $\sim$ 3 Myr. In this case, there would only be two star formation bursts (at $\sim$6 and $\sim$3 Myr) and the position of the luminous stars close to the ZAMS would be due to peculiar evolution, like merger products after binary interaction (as it was suggested e.g., in \cite{neg08}), effects of the accretion rate \citep[see][]{holgado20}, or chemically homogeneous evolution (see below). We make note of the lack of stars close to the ZAMS between 40 -- 60 M$_{\odot}$ in this group, producing an apparent gap close to the ZAMS in the global diagram. This effect has been pointed out by other authors \citep[e.g.,][]{herrero07, castro14, sabin17} and was recently revisited by \cite{holgado20}.

The same HRD is shown in the bottom left-hand panel of Fig.~\ref{hrd} but this time with a considering rotation (v/v$_{crit}$ = 0.4). As stated in previous works \citep[e.g.,][]{maeder00, wright15,berlanas18a}, Geneva rotating stellar models provide older main sequence stellar ages than their non-rotating models. 
The most massive members are more affected, and in average we find all the age groups $\sim$2 Myr older than in the non-rotating case. Interestingly, the most massive stars could also belong to an older population with stars of lower mass located further away from the ZAMS as the rotating isochrones of 3 -- 5 Myr move bluewards for the higher masses. 
The Bonn stellar models (right hand panels of Fig.~\ref{hrd}) do not exhibit large differences in stellar ages when using non-rotating or rotating models with initial rotational velocities (v$_{ini}$) as high as 330 km s$^{-1}$. This  different  behavior is most likely to be ascribed to their different treatment of angular momentum transport and magnetic fields in the stellar interior. The ages derived with the Bonn models are similar to those obtained with the non-rotating Geneva models. The most relevant difference between both families of models is that Bonn models provide a more extended TAMS, and when the models include rotation, the ZAMS is slightly displaced to higher luminosities and lower temperatures. In consequence, we find more stars close to (or on) the ZAMS.

In summary, our data indicate the presence of at least two main episodes of star formation in the main group of the Cygnus OB2 association (the one at $\sim$1.76 kpc in B19) at ages, according to Geneva non-rotating tracks, of $\sim$6 and $\sim$3 Myr. A third episode at $\sim$1.5 Myr seems possible, depending on the interpretation of the most massive stars. These results are fully in agreement with the analysis by \cite{neg08}. They assigned an age of 2.5 Myr to the association, with evidence of a slightly older population (much more evident in our results), and they explore the possibility that the most massive stars belong to a younger population, but ultimately conclude that this is not the most probable case. Their findings are also consistent with \cite{wright15}, who propose a nearly continuous star formation between 1 and 7 Myr (but our results favour a more episodic history). Although the region has currently an intense stellar formation activity \citep[see e.g.,][]{schneider06}, there is no evidence of star formation episodes more recent than 1 -- 2 Myr.

That situation resembles the one established by \cite{schneider18a, schneider18b} in 30 Dor, with a main peak of star formation 8 Myr ago, a decline in the last Myr and intense star formation in between. 
The ages derived for the stars in Cygnus OB2 become older by 1 -- 2 Myr when considering Geneva models with rotation,  except for the youngest population, which approaches the theoretical ZAMS when considering rotating Bonn models. 
For the smaller group at $\sim$1.35 kpc and the few stars that according to B19 seem to lie between the two main groups, we find that star formation took place $\sim$3 Myr ago, without substantial activity at other ages (but we note that the sample is small). Two stars might belong to an older population at again $\sim$6 Myr, but the small sample size hinders us from generalizing the result and claiming that this is another episode of massive star formation. Careful research of the associated population at lower masses would be required to confirm or reject such a possibility.

\begin{figure}[t]
\centering
\includegraphics[width= 9cm, angle=180,trim={0cm 1.7cm 0.8cm 0.9cm},clip]{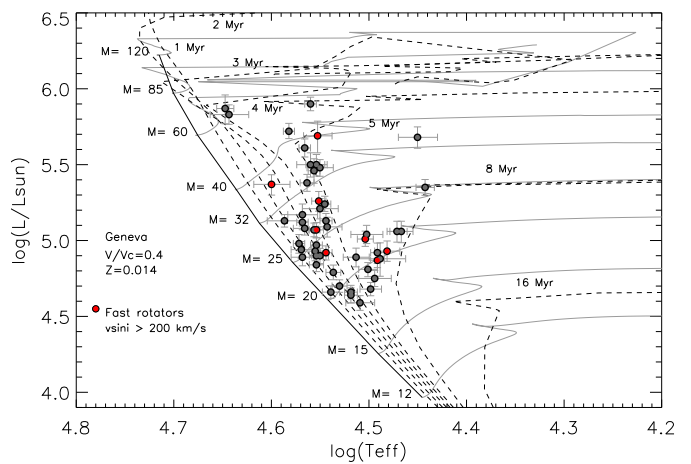}
\caption{Distribution of fast rotating stars ($vsini >$ 200 \kms, marked by red dots) in the HRD. Geneva rotating evolutionary models from \cite{ekstrom12}.}
\label{hrdfast}
\end{figure}

We also have to consider the possibility that some of the stars were born with a different initial rotational velocity than the rest or have followed peculiar (e.g., not single) evolutionary paths. For example, the two more massive stars (Cyg OB2~$\#7$ and Cyg OB2~$\#22$A) are a bit puzzling. At face, we would assign them to a 1.5 Myr group, but then they leave a gap between 25--40 M$_\odot$ (depending on the diagram we use) and 60 M$_\odot$ close to the ZAMS. Alternatively, we could assume that they belong to an older population and are the result of alternative evolutionary channels. Cyg OB2~$\#7$ and Cyg OB2~$\#22$A could be the product of fast rotation evolution, while other stars may have followed an essentially non-rotating track. In Fig.~\ref{hrdfast} we plot the stars on the HRD with Geneva rotating tracks and isochrones. We see that it would be possible to assign Cyg OB2~$\#7$ and Cyg OB2~$\#22$A to the 3--4 Myr population, with no mass gap in the population. In this scenario we would expect these high luminosity stars to show a present-day rotational velocity of the order of the observed $vsini$, but also to have a high surface Helium abundance (higher than the enhanced Helium abundance found for Cyg OB2~$\#7$), which is not supported by our analysis. However, the distribution of fast rotators ($vsini >$ 200 \kms) on the HRD shown in Fig.~\ref{hrdfast} concentrates at the lowest masses and it does not offer any support for this scenario. 
On the other hand, the stars could be interpreted as mergers after binary interaction \citep{demink14, wang20}, again corresponding to a 3 -- 4 Myr age. In that case, helium abundances would be closer to the observed ones as a consequence of the fresh hydrogen supply, but the expected rotational velocities would be very large (although the observed one would be affected by the inclination of the rotational axis).
Although it does neither contradict it, if the most massive stars belonged to a population of fast rotators following an isochrone between 2 and 4 Myr, we would expect them to be rotating fast and closely following a chemically homogeneous evolution. Yet, both stars have modest $vsini$ and only the He abundance of Cyg OB2~$\#7$ is is slightly above the solar value. 

Finally, we evaluated the possible impact of the lack of detected binary (or multiple) systems in our sample. As stated in Sect.~\ref{bin_fraction}, we may expect more binaries, as seen in other clusters \citep[see, e.g.,][]{sana11, sana12, aldoretta15}. However, only the undetected SB2 binaries would have an impact on the HRD, specifically those systems with relatively separate components. Systems with very close components would have high $vsini$ values, thus the probability of not detecting them would be small. Therefore, we would expect that only a small number of stars could be affected and our global conclusions would not vary. Moreover, we note that this small number of undetected binaries could be related to the lack of fast rotators shown in Fig.~\ref{rot_before} since binary interaction has been proposed to be the origin of the high rotational velocities \citep[][]{demink14}. 

\subsection{HRD versus sHRD}

The spectroscopic version of the HRD (henceforth, sHRD) has been widely used in the literature to interpret the evolutionary status of massive populations. It is a distance-independent diagram presented by \cite{langer14} based on the effective temperature and surface gravity where a spectroscopic luminosity is defined as log($\mathcal{L}$) = 4 $*$ log ($T_{eff}$) - log $g$. 

\begin{landscape}
\begin{figure}[ht!]
\centering
\includegraphics[width=12cm, angle=180,trim={0.6cm 2cm 0cm 0.5cm},clip]{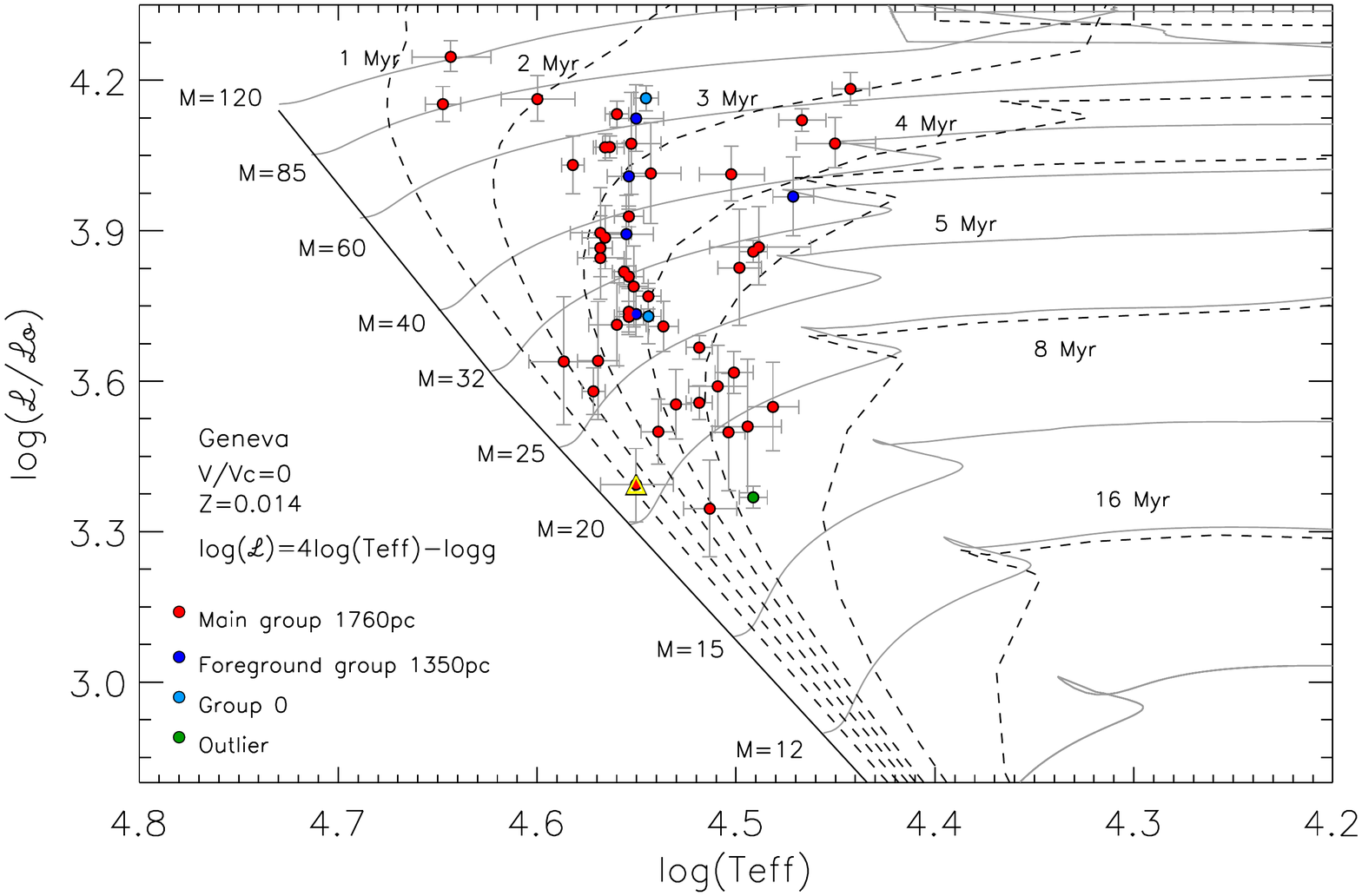}
\includegraphics[width=12cm, angle=180,trim={0.6cm 2cm 0cm 0.5cm},clip]{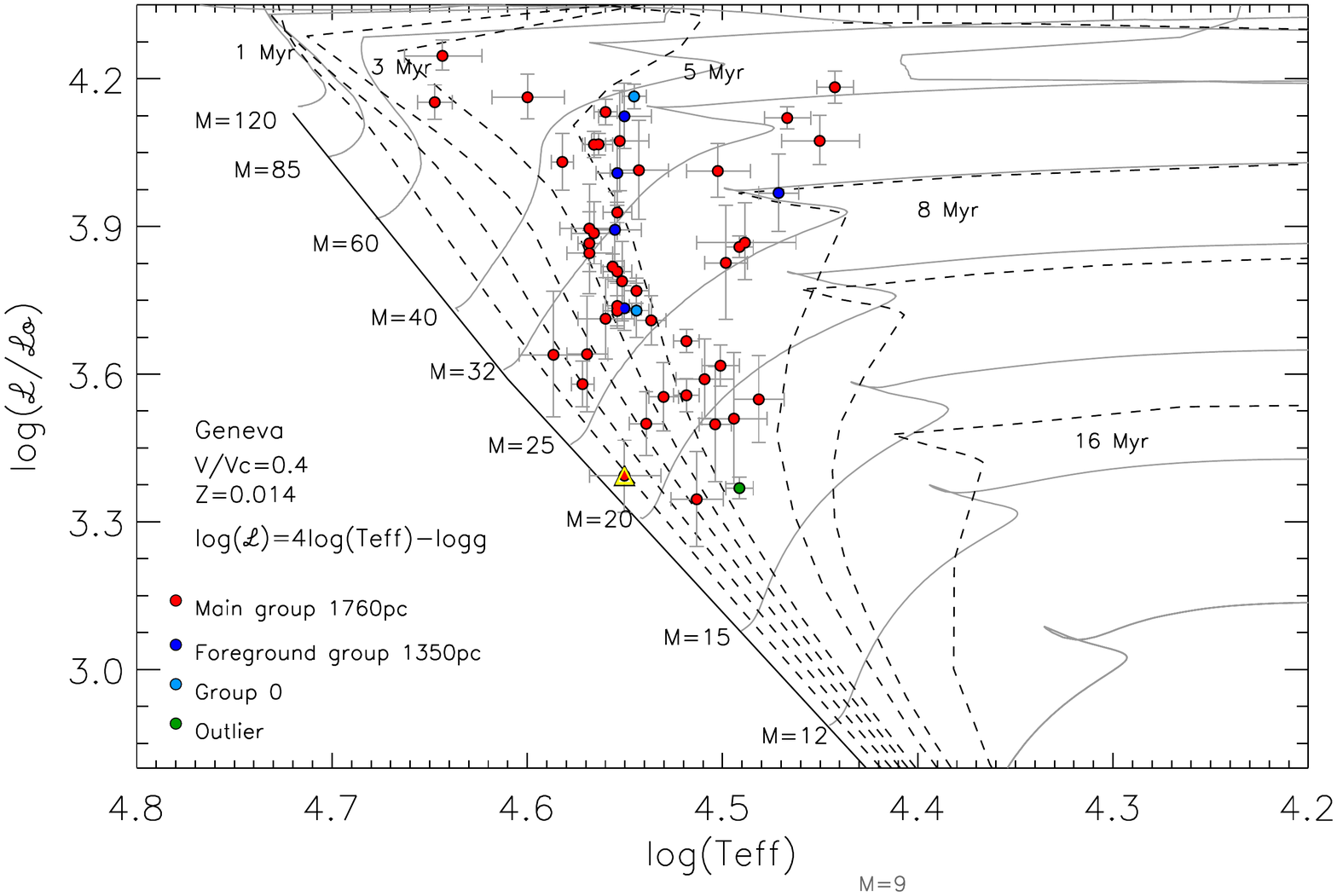}
\includegraphics[width=12cm, angle=180,trim={0.6cm 2cm 0cm 0.5cm},clip]{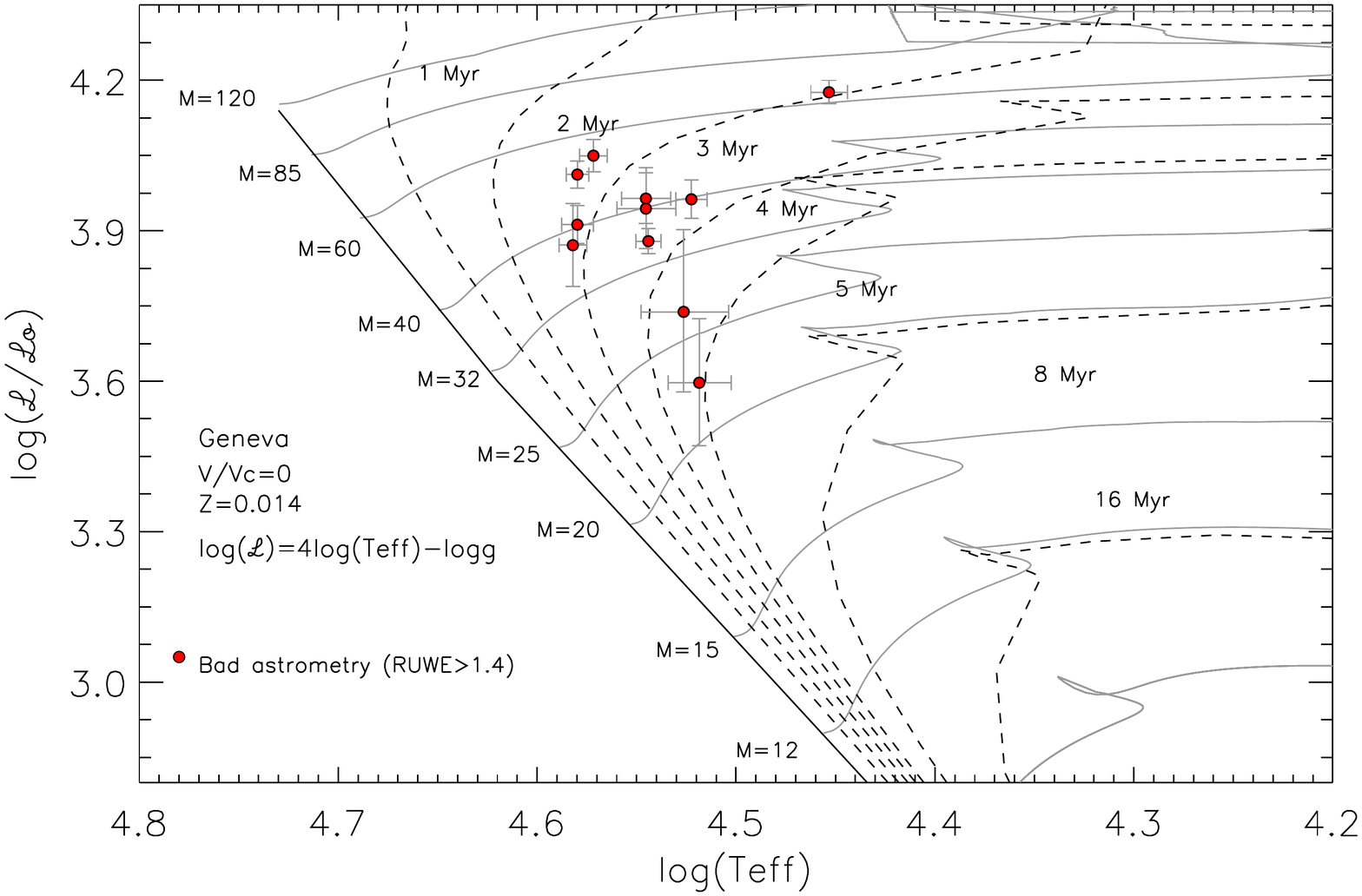}
\includegraphics[width=12cm, angle=180,trim={0.6cm 2cm 0cm 0.5cm},clip]{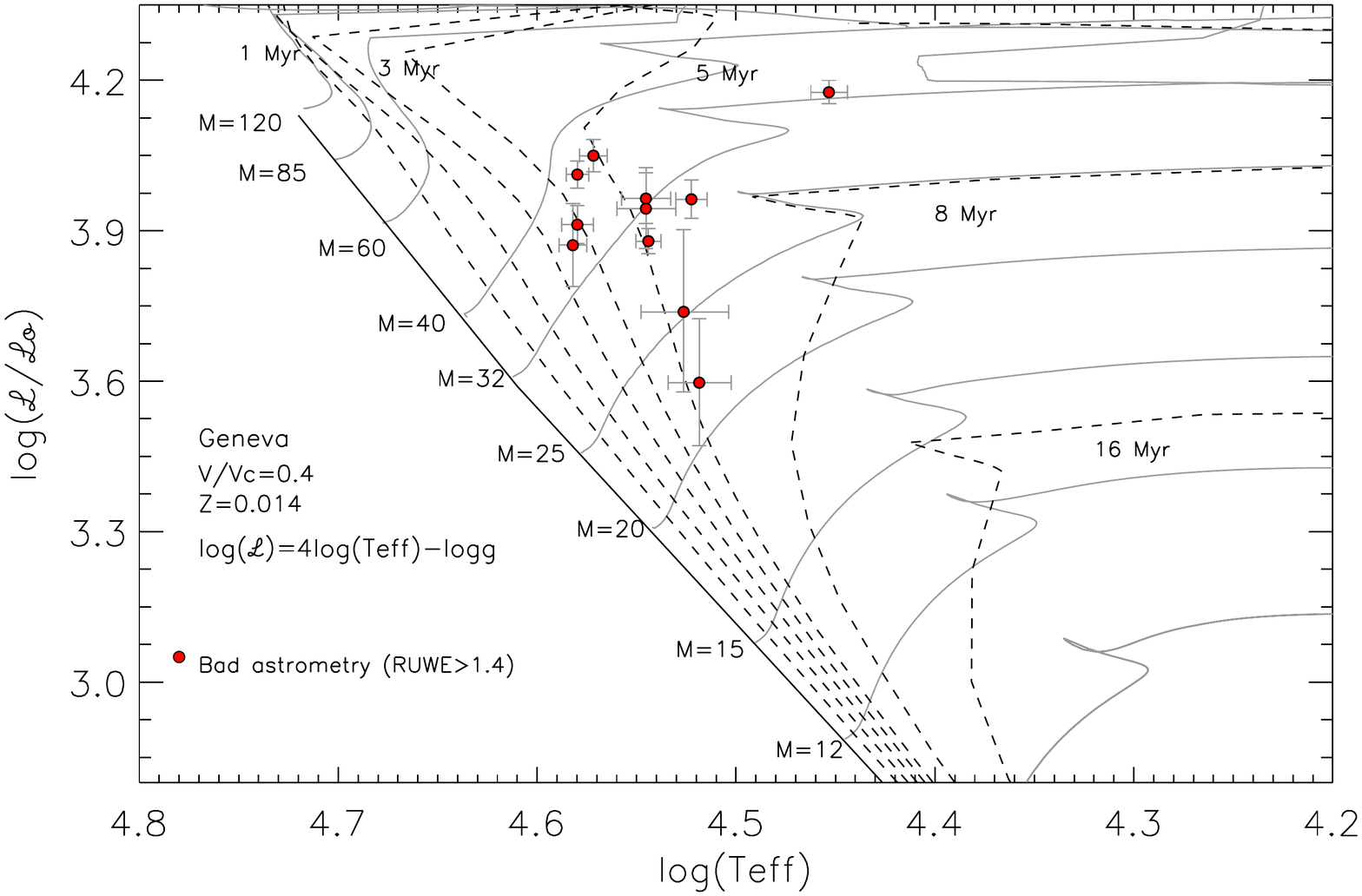}
\caption{ \textit{Top}: sHR diagrams for the sample of O stars in Cygnus OB2 with reliable astrometry using Geneva \citep[][,considering non-rotating models on the top and rotating models on the bottom, respectively]{ekstrom12} evolutionary stellar tracks and isochrones. The sample is also subdivided into the stellar groups found by \cite{berlanas18b}. \textit{Bottom}: sHR diagrams for the sample of O stars in Cygnus OB2 that have not passed the criterion for reliable astrometry considering the same set of tracks and isochrones. The yellow triangle indicates an upper limit for log $g$.}
\label{shrd}
\end{figure}

\end{landscape}

For simplicity, in Fig.~\ref{shrd} (top panels), we only show sHRDs for both non-rotating and rotating Geneva models. Comparing the HR diagrams to their respective sHR diagrams we find some notable differences.  On the one hand, the whole population seems to be displaced to higher masses in the sHRD. The most massive star is found in the sHRD at around 120 M$_{\odot}$ and the less massive one at 18 M$_{\odot}$. The age groups seem to be more dispersed although still present: the older group follows the 4 -- 5 Myr isochrones (+2 Myr for Geneva rotating models) reaching masses up to 60 M$_{\odot}$ and we find a more concentrated young group at $\sim$3 Myr (+2) with masses up to nearly 85 M$_{\odot}$. 
In addition, we now find an evident gap between the predicted ZAMS and the observed stars above 32 M$_{\odot}$, similar to that found in other studies of O stars (see Sect.~\ref{HRD} and citations there) and in our discussion of the HRD (now even more conspicuous). Again, the way in which the most massive stars have evolved determines the interpretation of such gap and the possible presence of a recent episode of star formation ($\sim$1 -- 1.5 Myr). The advantage of the sHRD is that we do not have to rely on distances. The sample of ten O stars that did not pass the selection criteria for reliable astrometry (RUWE $>$ 1.4) were placed in the sHRD (see Fig.~\ref{shrd}, bottom panels). 
The only significant difference we see with the previous results is that this small sample increases the number of stars with 3 -- 5 Myr.

In view of the above findings we conclude that we obtain a similar global information when using the HRD and the sHRD, except that we derive higher stellar masses from the sHRD than from the HRD.

The sHRD is thus useful to interpret the stellar ages and a general evolutionary status for massive stars when no reliable distances (RUWE $>$ 1.4) are available for them, although due to the significant difference in mass found here between the two types of diagrams we should be aware of a possible mass overestimation with respect to the HRD. 
To reconcile the masses from the two diagrams we would need to increase the stellar luminosity by $\sim$0.3 -- 0.4 dex in the HRD or decrease the spectroscopic luminosity by 0.2 -- 0.3 dex. But that would imply in the first case to increase the distance to Cygnus OB2 to values barely consistent with Gaia DR2 parallaxes (and incompatible with other measurements, such as that of \cite{rygl12}). 
In the second case, we would have to increase log $g$ by an amount incompatible with present day atmosphere models\footnote{Recently, \cite{markova18} proposed that inclusion of turbulent pressure in model atmospheres could increase the derived gravities; while this could help reconcile the sHRD and HRD and solve the mass discrepancy they find for stars below 30 \Msun, it would significantly increase that discrepancy for higher mass stars, as the same authors point out.}. 
 At the same time, \teff~ should decrease beyond present uncertainties. We note, however, that the determination of log $g$ depends not only on the model atmospheres, but also on a correct normalization of the observed spectrum \citep[see][]{ssimon20}.

\subsection{Spatial and dynamical distribution from $Gaia$ DR2}

In B19, the spatial substructure of the Cygnus OB2 association using parallaxes from $Gaia$ DR2 was explored. The sample used for the analysis comprised known OB members. Using a Bayesian methodology B19 found evidence of two different stellar groups superposed on the association but separated along the line-of-sight by $\sim$ 400 pc. Taking into account the spectral analyses and the two age bursts of O-type stars found in this work, we explore further the relationship between both groups and their dynamical state. We note that B19 include objects with confirmed OB spectra and RUWE $\leq$ 1.4, whereas in the present work we only include the O stars, but independently of their RUWE value. Thus, B19 include 200 objects, whereas we only consider 78.

\subsubsection{Spatial distribution}
\begin{figure*}[t!]
\centering

\includegraphics[width=7.8cm]{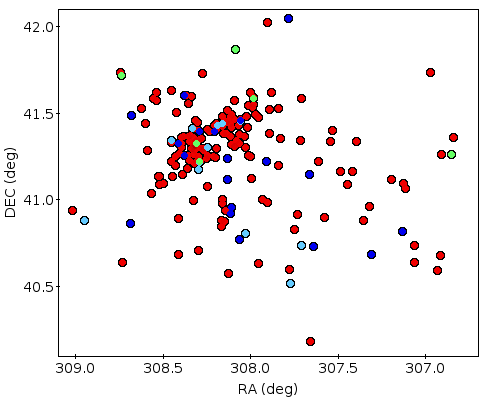}
\includegraphics[width=7.8cm]{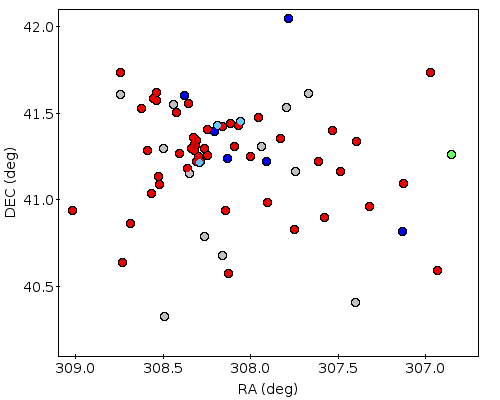}
\caption{Spatial distribution of the samples in B19 (left) and this paper (right). Coordinates are from the $Gaia$ DR2 catalog. Colors identify the different groups. Group 1 members are represented by blue symbols. Groups 2, 0, and 3 by red, cyan and green symbols. Stars with RUWE $>$ 1.4 (only in the right plot) are represented with grey symbols.}
\label{B19groups}
\end{figure*}

We plot in Fig.~\ref{B19groups} the spatial distribution of both samples, namely, B19 and ours. The left panel shows the spatial distribution of the B19 groups: the main groups at 1.35 and 1.76 kpc (Groups 1 and 2, respectively), those intermediate objects that could not be clearly assigned to either group (Group 0) and those that are foreground or background contaminants (Group 3). The distribution is dominated by the large Group 2 that appears splitted in two main concentrations: one to the left (larger right ascension) around the Cyg OB2~$\#8$ and Cyg OB2~$\#22$ systems plus Cyg OB2~$\#7$\footnote{ As indicated in Sect.~\ref{bin_fraction}, \cite{bica03} suggested that both Cyg OB2~$\#8$ and Cyg OB2~$\#22$ systems should be considered as two clusters that likely constitute the core of Cyg OB2. \cite{neg08} also divide this concentration in two clumps, one around Cyg OB2~$\#8$ and Cyg OB2~$\#7$, and another one around Cyg OB2~$\#22$. And recently, Maíz Apellániz et al. (A$\&$A submitted) named them as Villafranca O-008 and Villafranca O-007, respectively.}, and the other one, with less massive members, to the right and centered around Cyg OB2~$\#15$. Only Group 2 shows a strong stellar concentration, whereas the other groups are much more dispersed. Beyond that, it is not possible to spatially separate the groups found by B19. 

In the right panel, we present the spatial distribution of our sample (the 78 O stars) where we included the group of stars with poor parallax measurements for which we have stellar parameters (and that are not in the left plot). We see the same groups as in the full sample.
The spatial features remain the same as in the full sample, except that now a small concentration of O stars around Cyg OB2~$\#11$ (to the northeast of that around the Cyg OB2~$\#8$ complex) is more apparent. We note that there is only one representative star of the B19 Group 3 (that is composed by foreground and background objects) and it is precisely J20272428+4115458, the B0IV star that we decided to keep in our sample (see Sect.~\ref{sample_selection}). Therefore we find that all O stars that we see belong to the Cygnus star forming complex.

\subsubsection{Proper-motion distribution}
\begin{figure*}[t!]
\centering

\includegraphics[width=7.7cm]{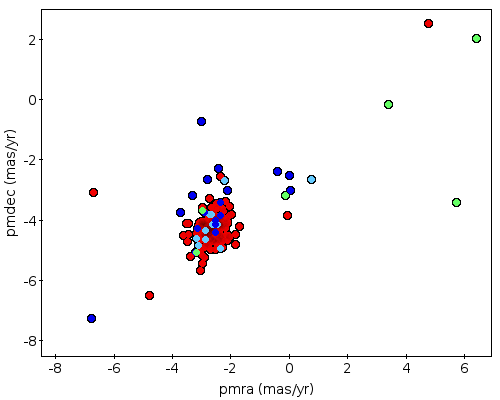}
\includegraphics[width=7.7cm]{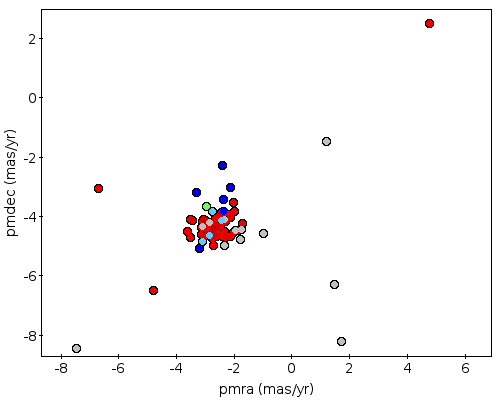}
\caption{Gaia DR2 proper motion distribution of the samples in B19 (left) and this paper (right). Colors identify the groups as in Fig.~\ref{B19groups} and light gray is used for stars with RUWE $>$ 1.4.}
\label{propmot}
\end{figure*}

 Proper motions are represented in Fig.~\ref{propmot}. Again, the B19 sample appears to the left. In that plot star J20322615+4057194 (with $\mu_\alpha$= -5.368$\pm$0.085 and $\mu_\delta$= -14.083$\pm$0.095 mas yr$^{-1}$) has been excluded to expand the scale. In the B19 sample we see that there is a small difference between Group 1 and Group 2: while their distributions overlap, Group 2 forms the lower envelope, whereas Group 1 defines the upper one. As a result, the mean proper motions of both groups are slightly different. However, statistically the difference is not significant. The core of Group 1 has ($\mu_\alpha$= -2.770$\pm$0.450, $\mu_\delta$= -3.679$\pm$0.732) mas yr$^{-1}$ and that of Group 2 shows ($\mu_\alpha$= -2.641$\pm$0.340, $\mu_\delta$= -4.301$\pm$0.380)  mas yr$^{-1}$. We note that these numbers do not take individual errors into account and assume that the errors are independent. The difference in right ascension proper motion is thus not significant (0.129$\pm$0.564) mas yr$^{-1}$ , whereas the proper motion in declination is larger, but still within the uncertainties (0.622$\pm$0.825) mas yr$^{-1}$. Nevertheless, the difference in the proper motion in declination is clearly deserving of a diligent inspection in the future $Gaia$ DR3\footnote{We note that the proper motions in right ascension of the two clusters identified within the Group 2, Villafranca O-007 and Villafranca O-008, are also very similar while in declination they differ by a little over one sigma (see Maíz Apellániz et al., A$\&$A submitted, for further details).}.  
To the right of the main concentration, we see a small number of stars with similar proper motions. They are a mixture of early B stars from different groups and although we note that three of them belong to Group 1, they lie far apart from one another on the sky and, thus, they are not physically related. The rest of the stars have proper motions departing from the average of the central objects and can be considered as runaway candidates (except those belonging to Group 3). 

On the right side of Fig.~\ref{propmot} we show the same plot, now for the stars in our sample. Star Cyg OB2~$\#$22B ($\mu_\alpha$= -4.628$\pm$0.776 and $\mu_\delta$= 8.367$\pm$0.797 mas yr$^{-1}$) has also been excluded to expand the scale (we note that it also has a very high RUWE value). Now the difference between Groups 1 and 2 is not so clear because of the low number of targets in Group 1, although they tend towards higher $\mu_\alpha$ values than the Group 2 members. Now we introduce the O-stars that could not be assigned to Group 1 or 2 and we see that six of them can be classified as runaway candidates. The runaways candidates are listed in Table~\ref{runaways}. For completeness we also list the early B-stars from the original list of B19 that may be considered runaway candidates, although they are not analyzed in this work (we exclude objects included in Group 3 as they cannot be considered runaway candidates). 
Also interesting is the fact that all O-type runaway candidates belong to Group 2 or have a poor RUWE value (and could not be assigned to any group), whereas B-type runaways belong mostly to Group 1, with only one member of Group 2.
The three stars from Group 2 that we may consider runaway candidates are A46, A15 and A37. The last two are also fast rotators. This might be seen as supporting the relation between fast rotators and runaways \citep[see, e.g.,][]{blaauw93, li20}, but other five fast rotators from Group 2 share the proper motion of the majority of the stars. In addition, most runaway candidates belong to the group of stars with poor RUWE values and none of these candidates is a fast rotator (at least, from its projected rotational velocity). Therefore, in Cygnus OB2 we cannot find a strong association between runaways and fast rotation (which is not surprising in view of the results in Sect.~\ref{vsini}).
A37 has been identified by \cite{kobulnicky16} as having a bow-shock, indicative of the star high velocity against the interstellar medium and reinforcing its status as runaway candidate. In the group of runaway candidates with poor RUWE we have Cyg OB2~$\#$10, that was also proposed as such by \cite{caballero14}, although later \cite{aldoretta15} list it as "no runaway." Interestingly, of the four objects with the highest over-luminosity in X-rays found by \cite{rauw15}\footnote{Apart from the even more extreme binary systems,Cyg OB2~$\#$5, 8A and 9, that depart from the general trend.}, three of them (A20, A26\footnote{\cite{rauw15} point out that the X-ray overluminosity of this object is highly dependent on its strong correction for X-ray absorption.} and MT91-516) appear on this list, which leads us to wonder whether the interaction of the stellar wind with the surrounding interstellar medium may play a role here. The fourth one is A11, one of the binary systems listed in Table~\ref{table_bins} with an early supergiant of type O7.5Ib(f) as primary.

To the left of Fig.~\ref{propbin}, we again show the spatial distribution of the O stars in Cygnus OB2, but now with arrows representing their residual proper motion after having substracted the mean proper motion of the corresponding group (for Group 0 we have averaged the means of Groups 1 and 2, and for stars without an assigned group, we used the mean of Group 2). We see that it is impossible to distinguish between the three groups in Cygnus (0, 1 and 2). The kinematical and dynamical status of Cygnus OB2 was carefully studied by \cite{wright16},  who found the association to be highly substructured, although their spatial scales have been recently challenged by \cite{arnold20}, who find substructure at smaller scales of 0.5 pc. A comparison of Fig. 9 in Wright et al. and on the left of our Fig.~\ref{propbin} makes it obvious that there is a similar pattern, although their sample is dominated by F-G stars. As these authors, we find no clear expansion nor contraction pattern in the O-star population of Cygnus OB2.

\subsubsection{Signatures of possible dynamical ejection events}

One of the puzzles in Cygnus OB2 is the lack of strong evidence of past supernova (SN) explosions in spite of the strong concentration of massive stars. The most compelling evidence to date is the presence of the pulsar PSR J2032+4127 \citep{albacete_colombo20} that forms a binary system with Cyg OB2~$\#$4B (a.k.a. MT91-213), a Be star \citep{salas15, Ho17}. This star has a proper motion that differs from the majority of these stars and, thus, it could be considered a runaway candidate. 
However, a monitoring of the system by \cite{cher20}  found that the periastron passage occurred in November 2017, as predicted by \cite{Ho17}. Furthermore, $Gaia$ proper motions could be affected by the orbit of Cyg OB2~$\#$4B around  PSR J2032+4127, although given the mass ratio large effects are not expected. Since we currently have no access to the $Gaia$ individual epochs, we tentatively include Cyg OB2~$\#$4B in Table~\ref{runaways}.
Other runaway candidates in the field could also be the result of ejection after a SN explosion or after a dynamical encounter and, in fact, all O-type runaway candidate stars listed in Table~\ref{runaways} have an age that is in excess of 2 Myr (as derived from the HRD and the evolutionary tracks), which makes them compatible with any scenario for a possible ejection. However, we point out that the region considered in this paper is relatively small and that evidence for runaways from Cygnus OB2 should consider a larger extension \citep[see e.g., the case of the runaway BD+43$^\circ$3654 in][]{comeron07, wright15}.

\begin{figure*}[t!]
\centering

\includegraphics[width=7.7cm]{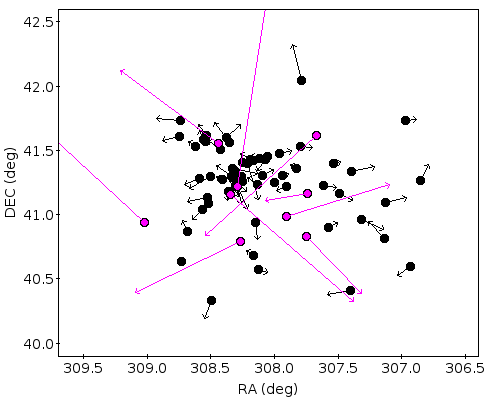}
\includegraphics[width=7.7cm]{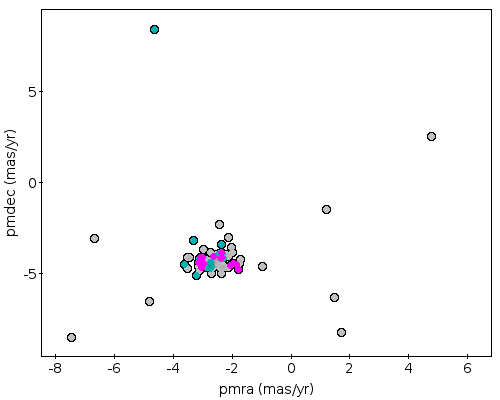}
\caption{Left: Spatial distribution and proper motion arrows for the O stars in Cygnus OB2. Runaway candidates are represented in purple. Right: Gaia DR2 proper motion distribution of our sample divided by binary status. Cyan: SB1 systems; purple: SB2 systems; light gray: single stars. We note that the ordinate scale is different as in the previous figures to include Cyg OB2~$\#$22B.}
\label{propbin}
\end{figure*}

\begin{table*}[t!]
 \caption{\small{List of O and B stars with proper motions departing from the bulk of the sample in Cygnus OB2. Proper motions are given in milliarcseconds per year. For O stars, spectral types and projected rotational velocities (in kilometers per second) come from this work. For B stars, we list spectral types given by \cite{com12,berlanas18a,berlanas18b} and $vsini$ values from the literature, whenever available. Stars with poor RUWE values have no B19 group assigned. Stars included in group 3 (foreground and background objects) are not actually runaway candidates and, thus, they are not listed.}}
        \begin{center}
                \begin{tabular}{lcccccc}
                \hline 
                \hline 
                \small{O Star}& \small{SpT}&\small{$\mu_\alpha$} & \small{$\mu_\delta$} &\small{$vsini$}& \small{B19 group}&\small{Comments} \\
                \hline
                \small{A46} &\small{O7 V((f))} & \small{-4.7975$\pm$0.0401} & \small{-6.5080$\pm$0.0445} &\small{75} & \small{2}&\small{} \\
                \small{A15} &\small{O7 Ibf}& \small{-6.6917$\pm$0.1052} & \small{-3.0674$\pm$0.1075} &\small{270}& \small{2}&\small{X-ray emission$^{a}$} \\
                \small{A37} &\small{O5 V((f))}& \small{+4.7680$\pm$0.0660} & \small{+2.5090$\pm$0.0743} &\small{225}& \small{2}&\small{Stellar bowshock nebulae candidate$^{b,c}$ } \\
                \small{MT91-5} &\small{O6.5 V((f))z}& \small{+1.7037$\pm$0.7238} & \small{-8.2103$\pm$0.6239} &\small{70}& \small{-}&\small{X-ray emission$^{a}$}\\
                \small{A26} &\small{O9.5 IV}& \small{-0.9824$\pm$1.4337} & \small{-4.5861$\pm$1.6144} &\small{85}& \small{-}&\small{X-ray emission$^{a}$}\\
                \small{A20} &\small{O7 III((f))}& \small{1.4764$\pm$0.7290} & \small{-6.2771$\pm$0.6385}&\small{80} & \small{-}&\small{X-ray emission$^{a}$}\\
                \small{Cyg OB2~$\#$22B} &\small{O6 IV((f))}& \small{-3.0826$\pm$0.1007} & \small{-4.6284$\pm$0.7757} &\small{145}& \small{-}&\small{Member of multiple system$^{d}$}\\
                \small{MT91-516} &\small{O6 IVf}& \small{-7.4619$\pm$1.5750} & \small{-8.4631$\pm$1.6608} &\small{85}& \small{-}&\small{X-ray emission$^{a}$}\\
                \small{Cyg OB2~$\#$10} &\small{O9.7 Iab}& \small{+1.2143$\pm$0.8583} & \small{-1.4617$\pm$0.7715}&\small{55} & \small{-}&\small{}\\
                \hline
                \hline                  
        \small{B Star}&\small{SpT}& \small{$\mu_\alpha$} & \small{$\mu_\delta$} &\small{$vsini$}& \small{B19 group} &\small{Comments}\\
                \hline
        \small{BD+40$^\circ$4208} &\small{B1 V}& \small{+0.765$\pm$0.055} & \small{-2.64$\pm$0.053} &\small{165}& \small{0}&\small{$vsini$ from (e)}\\
                \small{J20331029+4123446} &\small{B4 II-III}& \small{-6.761$\pm$0.043} & \small{-7.248$\pm$0.045}&\small{-} & \small{1}&\small{}\\
 
                \small{Cyg OB2~$\#$4B} &\small{B0:V ne}& \small{-2.991$\pm$0.048} & \small{+0.742$\pm$0.055} &\small{-}& \small{1}&\small{}\\
                \small{BD+40$^\circ$4193} &\small{B2 V}& \small{-0.397$\pm$0.049} & \small{-2.387$\pm$0.059} &\small{70}& \small{1}&\small{$vsini$ from (e)}\\
                \small{Schulte 30} &\small{B2 IIIe}& \small{-0.001$\pm$0.043} & \small{-2.51$\pm$0.044} &\small{182}& \small{1}&\small{$vsini$ from (f)}\\
                \small{A39} &\small{B2V}& \small{+0.048$\pm$0.054} & \small{-3.009$\pm$0.058} &\small{200}& \small{1}&\small{$vsini$ from (e)}\\
                \small{J20303833+4010538} &\small{B1 V}& \small{-0.05$\pm$0.044} & \small{-3.829$\pm$0.047}&\small{255} & \small{2}&\small{$vsini$ from (e)}\\
                \small{J20325440+4115219} &\small{B3 V}& \small{-1.4434$\pm$0.6728} & \small{+3.5137$\pm$0.7639}&\small{-} & \small{-}&\small{X-ray emission$^{a}$}\\
                \small{J20313364+4136046} &\small{B3 V}& \small{-0.2003$\pm$0.3608} & \small{-6.6582$\pm$0.3415} &\small{-}& \small{-}&\small{}\\
                \small{J20333979+4122523} &\small{B1 V}& \small{-5.2397$\pm$0.7872} & \small{-3.2279$\pm$0.6934}&\small{142} & \small{-}&\small{$vsini$ from (f)}\\
                \hline
                \multicolumn{7}{l}{\footnotesize Refs. (a) \cite{rauw15}; (b) \cite{kobulnicky16}; (c) \cite{kobul17}; (d) \cite{maiz10}; (e) \cite{berlanas18b};} \\
                \multicolumn{7}{l}{\footnotesize (f) \cite{wolff07}}\\                  
                \end{tabular}
                \label{runaways}
        \end{center}
        
\end{table*}

We also explored the distribution of binary systems in the proper motion diagram to see whether they are affected in some way in their proper motions. All of them but three show RUWE values below 1.4. Two of the three stars with RUWE $>$ 1.4 are just above this value (Cyg OB2~$\#$1 and Cyg OB2~$\#$27, with values of 1.421 and 1.521, respectively). The outlier is Cyg OB2~$\#$22B, with a RUWE value of 5.365. Interestingly, the distribution of binaries (SB1 and SB2) in the proper motion diagram (see Fig.~\ref{propbin}, right) shows that they all are consistent with that of the main concentration. The only exception is again Cyg OB2~$\#$22B\footnote{We note that MT91-516 has a companion, as reported by \cite{maiz16} and \cite{caballero14}. However, the latter indicate that it is radial velocity constant.}. This implies that most probably these systems still keep their original velocities and reinforce the possibility that peculiar proper motion values are the result of a dynamical ejection or a SN kick and are not attributed to the impact of binary motion.

\section{Conclusions}\label{conclusions}

Our new spectroscopic census of O stars in Cygnus OB2 provides an updated view of the high-mass stellar population in one of the largest groups of young stars in our Galaxy. The main results from this study are listed as follows.

\begin{enumerate}[I)]

\item We compiled the most complete spectroscopic census of O stars so far in the association, consisting of a sample of 78 O-type stars. We identified two new binary systems and performed quantitative spectroscopic analysis of the sample (excluding detected SB2 stars) to obtain the distribution of rotational velocities and main stellar parameters.\vspace{0.1cm}

\item  The distribution of rotational velocities is similar to those obtained from analogous studies of early-type stars in other Galactic (and extragalactic) regions, except that we find a short tail of fast rotators, which requires further investigation.

\vspace{0.1cm}

\item In addition, using distances from {\it Gaia} DR2 we estimated radii, luminosities, and spectroscopic masses for all isolated stars of our sample with reliable astrometry (RUWE $\leq$ 1.4). Using the Hertzsprung-Russell diagram the global evolutionary status of the association has been discussed. We find  star formation during the last 1 -- 6 Myr, with two main bursts centred roughly at 3 and 5 Myr. A third smaller group of stars at $\sim$1.5 Myr containing the hottest stars of our sample might be identified, but its veracity depends on the possible evolutionary paths followed by these stars. \vspace{0.1cm}

\item We obtained similar results when using the HRD and its spectroscopic version (sHRD), except that we derived higher stellar masses from the sHRD than from the HRD. If confirmed, this would be a cautionary remark for the use of the sHRD, although it would still be possible to use it to study the evolutionary status of a stellar population.\vspace{0.1cm}

\item We investigated the spatial and dynamical distribution of the sample from $Gaia$ astrometry. We find the association to be quite substructured with no evidence of expansion nor contraction pattern in its O-star population. Moreover, we identified nine O stars as runaway candidates, that would support a possible dynamical ejection scenario or a past SN explosion. Proper motions of binary systems also support this possibility since the sample shares the proper motions of the global population, indicating that they have kept their initial dynamical properties. \vspace{0.1cm}

\end{enumerate}

This work has demonstrated the advantages and difficulties of performing accurate spectroscopic analysis in a very active and obscured region as Cygnus OB2. It shows the precision that can be reached at using spectroscopic parameters for the interpretation of the evolutionary status of a region, that is, revealing internal star-forming bursts that otherwise would be unnoticed. It also shows the importance of the spectral resolution at performing quantitative spectroscopic analysis (mainly for obtaining an accurate distribution of rotational velocities).

The  improved  view  we  now have of Cygnus OB2  brings us closer to a complete understanding of the origin and evolution of this association, Cygnus-X, and OB associations, in general. It provides us with a solid background to interpret the results from other studies, such as the forthcoming Cygnus survey of WEAVE, which is the next multi-object spectrograph at the 4.2 m William Herschel Telescope (WHT), whose first light is planned by the end of 2020. Over the next few years, the WEAVE survey will deliver high quality ground-based spectroscopy for several hundred high-mass stars in the field. This work  will help to optimize and homogenize the sample, adding additional epochs for detecting multiplicity and serving as a template for the other OB associations in the complex. 
The procedures and results presented here will thus constrain the overall strategy for the spectroscopic analysis of a large amount of OB spectra over the whole region. Together with the upcoming improved $Gaia$ DR3 astrometry, this will allow us to make a much more accurate interpretation of the entire Cygnus-X evolutionary status.

\begin{acknowledgements}

This paper is dedicated to the memory of Rebeca Galera Rosillo and her green fireflies. We thank the referee for useful and valuable comments that helped improve this paper.  We acknowledge financial support from the Spanish Ministry of Economy and Competitiveness (MINECO) under the grants AYA2012-39364-C02-01, AYA 2015-68012-C2-01, Severo Ochoa SEV-2015-0548 and  PGC2018-093741-B-C21/C22 (MICIU/AEI/FEDER, UE).

\end{acknowledgements}

%
%

\def\bibname{References}

\bibliographystyle{aa}
\bibliography{biblio.bib}

\begin{appendix}
\onecolumn
\section{Tables of stellar parameters}\label{appA}
In this appendix we include all the tabulated data of the whole O-type population brighter than $B$ = 16 mag in Cygnus OB2 and the results obtained from the spectroscopic analysis.

\setlength\LTcapwidth{16cm}
\begin{longtable}{p{0.5cm}|cc|cc|cccc}
\caption{\label{table_fotom} Basic photometric data of the currently known O-type population of Cygnus OB2. Columns give their names, coordinates (J2000.0 epoch), $V$, $B$, $J$ and $K_{s}$ magnitudes. $B$ magnitudes are from the USNO-B and GOSC catalogs (the latter indicated with an asterisk). $K_{s}$ and $J$ magnitudes are from the 2MASS catalog. $V$ magnitudes gathered from  the literature (see table footnote).}\\

\hline
\hline
 \small{ID}&\small{Common Name}&\small{2MASS Name}& \small{RA(hhmmss)}& \small{Dec($^{\circ}$ $^\prime$ $^{\prime\prime}$)} & \small{$V$ [mag]}& \small{$B$ [mag]} & \small{$J$ [mag]} & \small{$K_{s}$ [mag]}\\  	 	 \hline\\[-1.5ex] 
\endfirsthead	

\caption{continued.} \\
\hline
\hline
  \small{ID}&\small{Common Name}&\small{2MASS Name}& \small{RA(hhmmss)}& \small{Dec($^{\circ}$ $^\prime$ $^{\prime\prime}$)} & \small{$V$ [mag]}& \small{$B$ [mag]} & \small{$J$ [mag]} & \small{$K_{s}$ [mag]}\\  	 	 \hline\\[-1.5ex] 
\endhead

\small{1}& \small{-}            &  \small{J20272428+4115458} & \small{20 27 24.28}& \small{+41 15 45.82}& \small{11.76$^{5}$}& \small{12.62}& \small{9.36}& \small{8.83}\\
\small{2}& \small{ALS 11363}    &  \small{J20274361+4035435} & \small{20 27 43.62}& \small{+40 35 43.53}& \small{9.71$^{2}$}& \small{10.11}& \small{8.42}& \small{8.26}\\
\small{3}& \small{-}            &  \small{J20275293+4144067} & \small{20 27 52.92}& \small{+41 44 06.65}& \small{-}& \small{13.33}& \small{8.14}& \small{7.28}\\
\small{4}& \small{-}            &  \small{J20283038+4105290} & \small{20 28 30.38}& \small{+41 05 29.04}& \small{12.16$^{5}$}& \small{13.36}& \small{7.09}& \small{6.12}\\
\small{5}& \small{ALS 11376}     &  \small{J20283203+4049027} & \small{20 28 32.03}& \small{+40 49 02.88}& \small{8.83$^{6}$}& \small{9.50}& \small{6.67}& \small{6.26}\\
\small{6}& \small{-}            &  \small{J20291617+4057371} & \small{20 29 16.17}& \small{+40 57 37.19}& \small{-}& \small{15.03}& \small{8.85}& \small{7.89}\\
\small{7}& \small{-}            &  \small{J20293480+4120089} & \small{20 29 34.79}& \small{+41 20 08.93}& \small{-}& \small{15.43}& \small{9.45}& \small{8.48}\\
\small{8}& \small{-}            &  \small{J20293563+4024315} & \small{20 29 35.63}& \small{+40 24 31.45}& \small{11.64$^{5}$}& \small{12.47}& \small{8.83}& \small{8.27}\\
\small{9}& \small{A42}          &  \small{J20295701+4109538} & \small{20 29 57.01}& \small{+41 09 53.84}& \small{12.21$^{5}$}& \small{13.01}& \small{9.12}& \small{8.45}\\
 \small{10}&\small{A18}          &  \small{J20300788+4123504} & \small{20 30 07.88}& \small{+41 23 50.50}& \small{-}& \small{15.64}& \small{9.39}& \small{8.36}\\
 \small{11}&\small{-}            &  \small{J20301838+4053466} & \small{20 30 18.39}& \small{+40 53 46.56}& \small{-}& \small{15.15}& \small{8.92}& \small{7.96}\\
 \small{12}&\small{B17}   &  \small{J20302730+4113253} & \small{20 30 27.30}& \small{+41 13 25.31}& \small{-}& \small{14.74}& \small{7.63}& \small{6.44}\\
 \small{13}&\small{MT91-5}     &  \small{J20303981+4136507} & \small{20 30 39.82}& \small{+41 36 50.72}& \small{12.93$^{2}$}& \small{14.57}& \small{9.10}& \small{8.31}\\    
\small{14}& \small{A26}          &  \small{J20305772+4109575} & \small{20 30 57.73}& \small{+41 09 57.60}& \small{-}& \small{14.61}& \small{9.09}& \small{8.19}\\ 
\small{15}& \small{A46}          &  \small{J20310019+4049497} & \small{20 31 00.20}& \small{+40 49 49.70}& \small{11.22$^{5}$}& \small{12.07}& \small{8.38}& \small{7.83}\\
\small{16}& \small{A41}          &  \small{J20310838+4202422} & \small{20 31 08.30}& \small{+42 02 42.00}& \small{12.29$^{5}$}& \small{13.10}& \small{7.82}& \small{7.02}\\
 \small{17}&\small{Cyg OB2~$\#$1A }    &  \small{J20311055+4131535} & \small{20 31 10.55}& \small{+41 31 53.54}& \small{11.18$^{4}$}& \small{12.48}& \small{7.96}& \small{7.36}\\
 \small{18}&\small{MT91-70}     &  \small{J20311833+4121216} & \small{20 31 18.33}& \small{+41 21 21.66}& \small{12.99$^{2}$}& \small{14.13}& \small{8.61}& \small{7.75}\\
 \small{19}&\small{A15}          &  \small{J20313690+4059092} & \small{20 31 36.91}& \small{+40 59 09.06}& \small{-}& \small{15.38}& \small{7.91}& \small{6.81}\\
 \small{20}&\small{Cyg OB2~$\#$3A}    &  \small{J20313749+4113210} & \small{20 31 37.50}& \small{+41 13 21.05}& \small{10.35$^{4}$}& \small{11.63}& \small{6.49}& \small{5.75}\\
 \small{21}&\small{MT91-138}     &  \small{J20314540+4118267} & \small{20 31 45.40}& \small{+41 18 26.75}& \small{12.26$^{2}$}& \small{13.62}& \small{8.06}& \small{7.26}\\
 \small{22}&\small{Cyg OB2~$\#$20}    &  \small{J20314965+4128265} & \small{20 31 49.65}& \small{+41 28 26.52}& \small{11.52$^{1}$}& \small{12.57}& \small{9.07}& \small{8.63}\\
\small{23}& \small{-}            &  \small{J20315961+4114504} & \small{20 31 59.61}& \small{+41 14 50.45}& \small{-}& \small{14.46}& \small{9.12}& \small{8.33}\\ 
 \small{24}&\small{Cyg OB2~$\#$4A }    &  \small{J20321383+4127120} & \small{20 32 13.82}& \small{+41 27 12.01}& \small{10.23$^{1}$}& \small{11.23}& \small{7.58}& \small{7.10}\\
 \small{25}&\small{Cyg OB2~$\#$14}    &  \small{J20321656+4125357} & \small{20 32 16.56}& \small{+41 25 35.71}& \small{11.47$^{1}$}& \small{12.25}& \small{8.71}& \small{8.18}\\
 \small{26}&\small{Cyg OB2~$\#$5A}    &  \small{J20322242+4118190} & \small{20 32 22.42}& \small{+41 18 19.00}& \small{-}& \small{10.80*}& \small{5.30}& \small{-}\\
 \small{27}&\small{Cyg OB2~$\#$5B}    &  \small{J20322248+4118190} & \small{20 32 22.49}& \small{+41 18 19.00}& \small{-}& \small{13.40*}& \small{7.80}& \small{-}\\ 
 \small{28}&\small{Cyg OB2~$\#$15}    &  \small{J20322766+4126220} & \small{20 32 27.66}& \small{+41 26 22.11}& \small{11.10$^{1}$}& \small{12.30}& \small{8.54}& \small{8.02}\\
 \small{29}&\small{Cyg OB2~$\#$A32}  &  \small{J20323033+4034332} & \small{20 32 30.33}& \small{+40 34 33.30}& \small{-}& \small{14.0}& \small{7.89}& \small{7.07}\\  
 \small{30}&\small{A11}     &  \small{J20323154+4114082} & \small{20 32 31.53}& \small{+41 14 08.18}& \small{-}& \small{14.72}& \small{7.82}& \small{6.66}\\
 \small{31}&\small{A38}          &  \small{J20323486+4056174} & \small{20 32 34.87}& \small{+40 56 17.40}& \small{-}& \small{14.93}& \small{9.38}& \small{8.56}\\
 \small{32}&\small{A25}          &  \small{J20323843+4040445} & \small{20 32 38.44}& \small{+40 40 44.5}& \small{-}& \small{15.33}& \small{8.35}& \small{7.38}\\   
 \small{33}&\small{Cyg OB2~$\#$16}    &  \small{J20323857+4125137} & \small{20 32 38.58}& \small{+41 25 13.66}& \small{10.84$^{1}$}& \small{11.86}& \small{8.19}& \small{7.72}\\
 \small{34}&\small{Cyg OB2~$\#$6 }    &  \small{J20324545+4125374} & \small{20 32 45.44}& \small{+41 25 37.51}& \small{10.68$^{1}$}& \small{11.67}& \small{7.95}& \small{7.42}\\
 \small{35}&\small{Cyg OB2~$\#$17}    &  \small{J20325002+4123446} & \small{20 32 50.02}& \small{+41 23 44.68}& \small{11.60$^{1}$}& \small{12.43}& \small{8.58}& \small{7.98}\\ 
\small{36}& \small{MT91-376}     &  \small{J20325919+4124254} & \small{20 32 59.19}& \small{+41 24 25.47}& \small{11.91$^{1}$}& \small{12.82}& \small{8.88}& \small{8.31}\\  
 \small{37}&\small{MT91-378}     &  \small{J20325964+4115146} & \small{20 32 59.64}& \small{+41 15 14.67}& \small{13.49$^{1}$}& \small{14.94}& \small{9.05}& \small{8.14}\\  
 \small{38}&\small{A20}          &  \small{J20330292+4047254} & \small{20 33 02.92}& \small{+40 47 25.40}& \small{-}& \small{14.40}& \small{7.25}& \small{6.27}\\
 \small{39}&\small{MT91-390}     &  \small{J20330292+4117431} & \small{20 33 02.92}& \small{+41 17 43.13}& \small{12.95$^{1}$}& \small{14.50}& \small{8.72}& \small{7.87}\\
 \small{40}&\small{Cyg OB2~$\#$22A}   &  \small{J20330879+4113182} & \small{20 33 08.77}& \small{+41 13 18.74}& \small{-}& \small{14.20*}& \small{7.60}& \small{6.72}\\
 \small{41}&\small{Cyg OB2~$\#$22B}   &  \small{J20330883+4113174} & \small{20 33 08.84}& \small{+41 13 17.48}& \small{-}& \small{14.80*}& \small{8.20}& \small{7.32}\\
\small{42}& \small{Cyg OB2~$\#$22E}   &  \small{J20330944+4112583} & \small{20 33 09.44}& \small{+41 12 58.30}& \small{-}& \small{16.80*}& \small{10.60}& \small{-}\\ 
 \small{43}&\small{Cyg OB2~$\#$22C}   &  \small{J20330960+4113005} & \small{20 33 09.60}& \small{+41 13 00.60}& \small{12.84$^{2}$}& \small{15.00}& \small{8.65}& \small{7.76}\\
 \small{44}&\small{Cyg OB2~$\#$22D}   &  \small{J20331011+4113101} & \small{20 33 10.11}& \small{+41 13 10.10}& \small{13.62$^{3}$}& \small{15.60}& \small{9.44}& \small{8.62}\\
 \small{45}&\small{Cyg OB2~$\#$9 }    &  \small{J20331074+4115081} & \small{20 33 10.73}& \small{+41 15 08.22}& \small{10.96$^{1}$}& \small{12.77}& \small{6.47}& \small{5.57}\\
 \small{46}&\small{MT91-448}     &  \small{J20331326+4113287} & \small{20 33 13.26}& \small{+41 13 28.67}& \small{13.61$^{1}$}& \small{15.02}& \small{8.98}& \small{8.01}\\
 \small{47}&\small{MT91-455}     &  \small{J20331369+4113057} & \small{20 33 13.69}& \small{+41 13 05.78}& \small{12.92$^{1}$}& \small{14.06}& \small{9.03}& \small{8.28}\\
 \small{48}&\small{Cyg OB2~$\#$7 }    &  \small{J20331411+4120218} & \small{20 33 14.11}& \small{+41 20 21.91}& \small{10.55$^{1}$}& \small{11.86}& \small{7.25}& \small{6.61}\\
 \small{49}&\small{Cyg OB2~$\#$8B}    &  \small{J20331476+4118416} & \small{20 33 14.76}& \small{+41 18 41.63}& \small{10.33$^{1}$}& \small{11.77}& \small{7.21}& \small{6.57}\\
\small{50}& \small{Cyg OB2~$\#$8A}    &  \small{J20331508+4118504} & \small{20 33 15.08}& \small{+41 18 50.48}& \small{9.06$^{1}$}& \small{10.36}& \small{6.12}& \small{5.50}\\

 \small{51}&\small{Cyg OB2~$\#$23} &  \small{J20331571+4120172} & \small{20 33 15.71}& \small{+41 20 17.20}& \small{12.50$^{1}$}& \small{12.92}& \small{9.33}& \small{8.72}\\
 \small{52}&\small{Cyg OB2~$\#$8D} &  \small{J20331634+4119017} & \small{20 33 16.26}& \small{+41 19 00.16}& \small{12.02$^{1}$}& \small{12.58}& \small{8.84}& \small{8.24}\\
 \small{53}&\small{Cyg OB2~$\#$24} &  \small{J20331748+4117093} & \small{20 33 17.48}& \small{+41 17 09.35}& \small{11.88$^{1}$}& \small{13.09}& \small{8.35}& \small{7.65}\\
 \small{54}&\small{Cyg OB2~$\#$8C} &  \small{J20331798+4118311} & \small{20 33 17.98}& \small{+41 18 31.19}& \small{10.19$^{1}$}& \small{11.06}& \small{7.16}& \small{6.58}\\
 \small{55}&\small{MT91-485}  &  \small{J20331803+4121366} & \small{20 33 18.03}& \small{+41 21 36.65}& \small{12.06$^{1}$}& \small{12.69}& \small{8.74}& \small{8.11}\\
\small{56}& \small{MT91-507}  &  \small{J20332101+4117401} & \small{20 33 21.02}& \small{+41 17 40.14}& \small{12.70$^{1}$}& \small{13.63}& \small{9.30}& \small{8.67}\\
 \small{57}&\small{MT91-516} &  \small{J20332346+4109130} & \small{20 33 23.47}& \small{+41 09 12.90}& \small{11.84$^{1}$}& \small{13.50}& \small{7.02}& \small{6.05}\\
 \small{58}&\small{Cyg OB2~$\#$25A}&  \small{J20332556+4133269} & \small{20 33 25.56}& \small{+41 33 27.00}& \small{11.58$^{1}$}& \small{13.15}& \small{8.17}& \small{7.52}\\
 \small{59}&\small{MT91-534}  &  \small{J20332674+4110595} & \small{20 33 26.75}& \small{+41 10 59.51}& \small{13.00$^{1}$}& \small{14.24}& \small{8.97}& \small{8.16}\\  
 \small{60}&\small{Cyg OB2~$\#$74}  &  \small{J20333030+4135578} & \small{20 33 30.31}& \small{+41 35 57.90}& \small{12.51$^{1}$}& \small{13.79}& \small{8.38}& \small{7.57}\\
 \small{61}&\small{Cyg OB2~$\#$70}  &  \small{J20333700+4116113} & \small{20 33 37.00}& \small{+41 16 11.30}& \small{12.40$^{1}$}& \small{13.53}& \small{8.68}& \small{7.93}\\   
 \small{62}&\small{MT91-611}  &  \small{J20334086+4130189} & \small{20 33 40.87}& \small{+41 30 18.98}& \small{12.77$^{2}$}& \small{13.68}& \small{9.26}& \small{8.61}\\
 \small{63}&\small{Cyg OB2~$\#$10} &  \small{J20334610+4133010} & \small{20 33 46.11}& \small{+41 33 01.05}& \small{9.88$^{1}$}& \small{11.19}& \small{6.29}& \small{5.58}\\
 \small{64}&\small{-}         &  \small{J20335842+4019411} & \small{20 33 58.42}& \small{+40 19 41.13}& \small{-}& \small{15.48}& \small{7.96}& \small{6.93}\\
 \small{65}&\small{Cyg OB2~$\#$27A} &  \small{J20335952+4117354} & \small{20 33 59.53}& \small{+41 17 35.48}& \small{12.32$^{1}$}& \small{13.18}& \small{8.53}& \small{7.89}\\
 \small{66}&\small{MT91-716}  &  \small{J20340486+4105129} & \small{20 34 04.86}& \small{+41 05 12.90}& \small{13.50$^{1}$}& \small{14.53}& \small{9.56}& \small{8.84}\\
 \small{67}&\small{MT91-720}  &  \small{J20340601+4108090} & \small{20 34 06.02}& \small{+41 08 09.00}& \small{13.59$^{1}$}& \small{14.84}& \small{9.05}& \small{8.15}\\
 \small{68}&\small{CygOB2~$\#$11} &  \small{J20340850+4136592} & \small{20 34 08.51}& \small{+41 36 59.39}& \small{10.03$^{1}$}& \small{11.22}& \small{6.65}& \small{5.99}\\
 \small{69}&\small{MT91-736}  &  \small{J20340951+4134136} & \small{20 34 09.52}& \small{+41 34 13.69}& \small{12.79$^{1}$}& \small{14.25}& \small{9.30}& \small{8.65}\\ 
 \small{70}&\small{Cyg OB2~$\#$29} &  \small{J20341350+4135027} & \small{20 34 13.51}& \small{+41 35 02.86}& \small{11.91$^{1}$}& \small{13.56}& \small{8.55}& \small{7.92}\\
 \small{71}&\small{A28}       &  \small{J20341604+4102196} & \small{20 34 16.05}& \small{+41 02 19.59}& \small{13.89$^{7}$}& \small{15.10}& \small{9.41}& \small{8.52}\\
 \small{72}&\small{S73}       &  \small{J20342193+4117016} & \small{20 34 21.93}& \small{+41 17 01.60}& \small{12.40$^{1}$}& \small{13.66}& \small{8.39}& \small{7.60}\\
 \small{73}&\small{MT91-771}  &  \small{J20342959+4131455} & \small{20 34 29.60}& \small{+41 31 45.49}& \small{12.06$^{1}$}& \small{13.37}& \small{7.56}& \small{6.71}\\
 \small{74}&\small{A24}       &  \small{J20344410+4051584} & \small{20 34 44.15}& \small{+40 51 58.67}& \small{-}& \small{14.74}& \small{8.40}& \small{7.45}\\
 \small{75}&\small{A29}  &  \small{J20345605+4038180} & \small{20 34 56.06}& \small{+40 38 17.92}& \small{-}& \small{13.54}& \small{7.44}& \small{6.54}\\
 \small{76}&\small{B18}&  \small{J20345785+4143542} & \small{20 34 57.85}& \small{+41 43 54.25}& \small{-}& \small{15.43}& \small{8.45}& \small{7.42}\\
 \small{77}&\small{Cyg OB2~$\#$A36}       &  \small{J20345878+4136174} & \small{20 34 58.78 }& \small{+41 36 17.36}& \small{11.42$^{4}$}& \small{12.81}& \small{7.19}& \small{6.36}\\  
 \small{78}&\small{A37}       &  \small{J20360451+4056129} & \small{20 36 04.50}& \small{+40 56 13.01}& \small{-}& \small{13.95}& \small{8.57}& \small{7.68}\\ 

\hline  
 \multicolumn{9}{l}{\footnotesize  Refs: (1) \cite{mt91}, (2)\cite{reed03}, (3)\cite{wright15}, (4)\cite{laur17}, (5) Tycho-2 catalog.}  \\
 \multicolumn{9}{l}{\footnotesize  (6) \cite{wess82}, (7) \cite{zach12}}.  \\
 
\end{longtable}\vspace{1cm}

\begin{longtable}{p{0.5cm}|cc|ccc|cc}
\caption{\small{List of the whole O-type population used in this work along with previous (see table footnote for references) and new spectral classification as well as the spectral data used to characterize the broadening (case A) and determine the spectroscopic parameters (case B). See Sect.~\ref{sample_selection} for further details.}}
\label{table_stars1}\\
\hline
\hline
 \small{ID}&\small{Name}&\small{2MASS Name}& \small{SpT(prev)}& \small{SpT(new)} & \small{Comments}& \small{Data (A)} & \small{Data (B)} \\  	 	 
 \hline\\[-1.5ex] 
\endfirsthead	

\caption{\small{continued.}} \\
\hline
\hline
 \small{ID}&\small{Common Name}&\small{2MASS Name}& \small{SpT(prev)}& \small{SpT(new)} & \small{Comments}& \small{Data (A)} & \small{Data (B)} \\  	 	 
 \hline\\[-1.5ex] 
\endhead

\small{1}& \small{-}         & \small{J20272428+4115458} & \small{O9.5 V$^{5}$              }& \small{B0 IV              }& \small{Group3        }& \small{5a}& \small{5a $\&$ 5b}\\
\small{2}& \small{ALS 11363} & \small{J20274361+4035435} & \small{O8 Vz$^{4}$               }& \small{O8 Vz              }& \small{-             }& \small{3}& \small{3 $\&$ 6b}\\
\small{3}& \small{-}         & \small{J20275293+4144067} & \small{O9.5 II$^{2}$             }& \small{O9.5 II            }& \small{-             }& \small{6a}& \small{6a $\&$ 6b}\\
\small{4}& \small{-}         & \small{J20283038+4105290} & \small{OC9.7 Ia$^{5}$            }& \small{OC9.7 Iab          }& \small{-             }& \small{6a}& \small{6a $\&$ 6b}\\
\small{5}& \small{ALS 11376}  & \small{J20283203+4049027} & \small{O7$^{11}$                 }& \small{O7 Ib(f)           }& \small{-             }& \small{3}& \small{3 $\&$ 6b}\\
\small{6}& \small{-}         & \small{J20291617+4057371} & \small{O9.7 III$^{2}$            }& \small{O8.5 III           }& \small{-             }& \small{6a}& \small{6a $\&$ 6b}\\
\small{7}& \small{-}         & \small{J20293480+4120089} & \small{O9.5 V$^{5}$              }& \small{O9.5 IV            }& \small{-             }& \small{6a}& \small{6a $\&$ 6b}\\
\small{8}& \small{-}         & \small{J20293563+4024315} & \small{O8 IIIz$^{2}$             }& \small{O8 IIIz            }& \small{-             }& \small{5a}& \small{5a $\&$ 5b}\\
\small{9}& \small{A42}       & \small{J20295701+4109538} & \small{O9.7 III$^{3}$            }& \small{O9.7 III           }& \small{-             }& \small{5a}& \small{5a $\&$ 5b}\\
\small{10}& \small{A18}       & \small{J20300788+4123504} & \small{O8 V$^{11}$                }& \small{O9.7 III(n)        }& \small{-             }& \small{6a}& \small{6a $\&$ 6b}\\
\small{11}& \small{-}         & \small{J20301838+4053466} & \small{O9 V$^{5}$                }&  \small{- }& \small{SB2 }& \small{-}& \small{-}\\
\small{12}& \small{B17}& \small{J20302730+4113253} & \small{O6 Iaf+O9: Ia:$^{4}$        }&  \small{- }& \small{SB2 }& \small{-}& \small{-}\\
\small{13}& \small{MT91-5}  & \small{J20303981+4136507} & \small{O6 V$^{7}$                }&  \small{O6.5 V((f))z       }& \small{-             }& \small{6a}& \small{6a $\&$ 6b}\\    
\small{14}& \small{A26}       & \small{J20305772+4109575} & \small{O9.5 V $^{9}$             }&  \small{O9.5 IV            }& \small{-             }& \small{6a}& \small{6a $\&$ 6b}\\ 
\small{15}& \small{A46}       & \small{J20310019+4049497} & \small{O7 V((f))$^{8}$           }& \small{O7 V((f))          }& \small{-             }& \small{6a}& \small{6a $\&$ 6b}\\
 \small{16}&\small{A41}       & \small{J20310838+4202422} & \small{O9.7 II$^{8}$             }& \small{O9.7 II(n)         }& \small{-             }& \small{6a}& \small{6a $\&$ 6b}\\
\small{17}& \small{Cyg OB2~$\#$1A } & \small{J20311055+4131535} & \small{O8 IV(n)f$^{1}$       }& \small{O8 IV(n)f      }& \small{SB1             }& \small{2}& \small{2 $\&$ 6b}\\
\small{18}& \small{MT91-70}  & \small{J20311833+4121216} & \small{O9 IV $^{1}$               }& \small{O8.5 III           }& \small{SB1             }& \small{6a}& \small{6a $\&$ 6b}\\
\small{19}& \small{A15}       & \small{J20313690+4059092} & \small{O7 Ibf$^{9}$              }& \small{O7 Ibf             }& \small{-             }& \small{6a}& \small{6a $\&$ 6b}\\
\small{20}& \small{Cyg OB2~$\#$3A} & \small{J20313749+4113210} & \small{O8.5 Iab(f) + O6 III:$^{1}$ }& \small{-       }& \small{SB2             }& \small{-}& \small{-}\\
 \small{21}&\small{MT91-138}  & \small{J20314540+4118267} & \small{O8 I$^{7}$                }& \small{O8 III             }& \small{-             }& \small{6a}& \small{6a $\&$ 6b}\\
 \small{22}&\small{Cyg OB2~$\#$20} & \small{J20314965+4128265} & \small{O9.7 IV$^{3}$             }& \small{O9.7 IV            }& \small{SB1             }& \small{5a}& \small{5a $\&$ 5b}\\
 \small{23}&\small{-}         & \small{J20315961+4114504} & \small{O7 V$^{5}$             }& \small{O7.5 Vz            }& \small{-             }& \small{6a}& \small{6a $\&$ 6b}\\ 
 \small{24}&\small{Cyg OB2~$\#$4A } & \small{J20321383+4127120} & \small{O7 IIIf$^{7}$             }& \small{O7 IIIf            }& \small{-             }& \small{4b}& \small{5a $\&$ 5b}\\
 \small{25}&\small{Cyg OB2~$\#$14} & \small{J20321656+4125357} & \small{O9 V$^{7}$             }& \small{O9.5 IV            }& \small{-             }& \small{2}& \small{2 $\&$ 6b}\\
\small{26}& \small{Cyg OB2~$\#$5A} & \small{J20322242+4118190} & \small{O6.5: Iafe + O7 Iafe$^{1}$ }& \small{-           }& \small{SB2             }& \small{-}& \small{-}\\
\small{27}& \small{Cyg OB2~$\#$5B} & \small{J20322248+4118190} & \small{O7 Ib(f)p var?$^{1}$          }& \small{O7 Ib(f)p var?          }& \small{-             }& \small{-}& \small{-}\\ 
 \small{28}&\small{Cyg OB2~$\#$15} & \small{J20322766+4126220} & \small{O8 III$^{1}$              }& \small{O8 III             }& \small{SB1 }& \small{2}& \small{2 $\&$ 6b}\\
 \small{29}&\small{Cyg OB2~$\#$A32}       & \small{J20323033+4034332} & \small{O9.5 IV$^{8}$             }& \small{- }& \small{SB2 }& \small{- }& \small{-}\\  
 \small{30}&\small{A11}  & \small{J20323154+4114082} & \small{O7 Ib(f)$^{4}$             }& \small{O7.5 Ib(f)         }& \small{SB1             }& \small{1}& \small{1 $\&$ 6b}\\
 \small{31}&\small{A38}       & \small{J20323486+4056174} & \small{O8 V$^{9}$                }& \small{O8 Vz(n)           }& \small{-             }& \small{6a}& \small{6a $\&$ 6b}\\
\small{32}& \small{A25}       & \small{J20323843+4040445} & \small{O8 III$^{9}$              }& \small{O8 III((f))        }& \small{-             }& \small{6a}& \small{6a $\&$ 6b}\\   
 \small{33}&\small{Cyg OB2~$\#$16} & \small{J20323857+4125137} & \small{O7.5 IV(n)$^{4}$          }& \small{O7.5: IV(n)         }& \small{- }& \small{5b}& \small{5a $\&$ 5b}\\
 \small{34}&\small{Cyg OB2~$\#$6 } & \small{J20324545+4125374} & \small{O8.5 V(n)$^{4}$           }& \small{O8 V(n)            }& \small{-             }& \small{4b}& \small{2 $\&$ 6b}\\
 \small{35}&\small{Cyg OB2~$\#$17} & \small{J20325002+4123446} & \small{O8 V$^{7}$                }& \small{O8: V:           }& \small{SB1       }& \small{2}& \small{2 $\&$ 6b}\\ 
 \small{36}&\small{MT91-376}  & \small{J20325919+4124254} & \small{O8 V$^{7}$                }& \small{O8 V               }& \small{-             }& \small{2}& \small{2 $\&$ 6b}\\  
 \small{37}&\small{MT91-378}  & \small{J20325964+4115146} & \small{O9.7 III(n)$^{4}$         }& \small{O9.7 IV:(n)        }& \small{SB1 }& \small{1}& \small{1 $\&$ 6b}\\  
 \small{38}&\small{A20}       & \small{J20330292+4047254} & \small{O8 II(f)$^{5}$            }& \small{O7 III((f))      }& \small{-             }& \small{6a}& \small{6a $\&$ 6b}\\
 \small{39}&\small{MT91-390}  & \small{J20330292+4117431} & \small{O7.5 V((f))$^{1}$         }& \small{O8: V(n)            }& \small{SB1 }& \small{6a}& \small{6a $\&$ 6b}\\
 \small{40}&\small{Cyg OB2~$\#$22A}& \small{J20330776+4113186} & \small{O3 If*$^{6}$              }& \small{O3 If*             }& \small{-             }& \small{4a}& \small{1 $\&$ 4cd}\\
 \small{41}&\small{Cyg OB2~$\#$22B}& \small{J20330883+4113174} & \small{O6 V((f))$^{6}$           }& \small{O6 IV((f))         }& \small{SB1 }& \small{4b}& \small{1 $\&$ 4cd}\\
 \small{42}&\small{Cyg OB2~$\#$22E}& \small{J20330944+4112583} & \small{O9.7: V:$^{12}$           }& \small{O9.7 V:             }& \small{-         }& \small{-}& \small{-}\\ 
 \small{43}&\small{Cyg OB2~$\#$22C}& \small{J20330960+4113005} & \small{O9.5 IIIn$^{6}$           }& \small{O9.5 IV:n          }& \small{SB1 }& \small{2}& \small{2 $\&$ 6b}\\
 \small{44}&\small{Cyg OB2~$\#$22D}& \small{J20331011+4113101} & \small{O9.5 Vn$^{4}$             }& \small{O9.5 Vn            }& \small{-             }& \small{2}& \small{-}\\
 \small{45}&\small{Cyg OB2~$\#$9 } & \small{J20331074+4115081} & \small{O4If+O5.5III(f)$^{1}$     }& \small{-       }& \small{SB2        }& \small{-}& \small{-}\\
 \small{46}&\small{MT91-448}  & \small{J20331326+4113287} & \small{O6.5: V$^{1}$             }& \small{O6.5 V((f))        }& \small{SB1 }& \small{2}& \small{2 $\&$ 6b}\\
 \small{47}&\small{MT41-455}  & \small{J20331369+4113057} & \small{O8 V$^{5}$             }& \small{O7 Vz((f))         }& \small{-             }& \small{2}& \small{2 $\&$ 6b}\\
 \small{48}&\small{Cyg OB2~$\#$7 } & \small{J20331411+4120218} & \small{O3 If*$^{7}$              }& \small{O3 If*             }& \small{-             }& \small{4b}& \small{4a,b,c,d}\\
 \small{49}&\small{Cyg OB2~$\#$8B} & \small{J20331476+4118416} & \small{O6 II(f)$^{6}$            }& \small{O6 II(f)           }& \small{-             }& \small{1}& \small{1 $\&$ 6b}\\ 		
 \small{50}&\small{Cyg OB2~$\#$8A} & \small{J20331508+4118504} & \small{O6 Ib(fc) + O4.5 III(fc)$^{1}$}& \small{-          }& \small{SB2 }& \small{-}& \small{-}\\
 \small{51}&\small{Cyg OB2~$\#$23} & \small{J20331571+4120172} & \small{O9.5 V$^{7}$             }& \small{O9.5 V             }& \small{-             }& \small{5a}& \small{5a $\&$ 5b}\\
 \small{52}&\small{Cyg OB2~$\#$8D} & \small{J20331634+4119017} & \small{O8.5 V(n)$^{13}$           }& \small{O8.5 V(n)          }& \small{SB1             }& \small{4b}& \small{1 $\&$ 6b}\\
 \small{53}&\small{Cyg OB2~$\#$24} & \small{J20331748+4117093} & \small{O8 V(n)$^{6}$             }& \small{O8 V(n)            }& \small{-             }& \small{1}& \small{1 $\&$ 6b}\\
 \small{54}&\small{Cyg OB2~$\#$8C} & \small{J20331798+4118311} & \small{O4.5 (fc)pvar$^{13}$       }& \small{O5 I(fc)           }& \small{-             }& \small{4b}& \small{5a $\&$ 5b}\\
\small{55}& \small{MT91-485}  & \small{J20331803+4121366} & \small{O8 V$^{4}$               }& \small{O8 V               }& \small{SB1 }& \small{2}& \small{2 $\&$ 6b}\\
 \small{56}&\small{MT91-507}  & \small{J20332101+4117401} & \small{O9 Vn$^{5}$              }& \small{O9.5 V(n)          }& \small{-             }& \small{1}& \small{1 $\&$ 6b}\\
\small{57}& \small{MT91-516} & \small{J20332346+4109130} & \small{O6 IVf$^{4}$               }& \small{O6 IVf             }& \small{-             }& \small{5a}& \small{5a $\&$ 5b}\\
 \small{58}&\small{Cyg OB2~$\#$25A}& \small{J20332556+4133269} & \small{O8.5 Vz$^{4}$              }& \small{O8.5 V             }& \small{-             }& \small{1}& \small{1 $\&$ 6b}\\
 \small{59}&\small{MT91-534}  & \small{J20332674+4110595} & \small{O8.5 Vz$^{4}$              }& \small{O8.5 V             }& \small{-             }& \small{1}& \small{1 $\&$ 6b}\\  
 \small{60}&\small{Cyg OB2~$\#$74}  & \small{J20333030+4135578} & \small{O8 V$^{10}$                 }& \small{O8 V((f))          }& \small{SB1             }& \small{6a}& \small{6a $\&$ 6b}\\
 \small{61}&\small{Cyg OB2~$\#$70}  & \small{J20333700+4116113} & \small{O9.5 IVn$^{1}$             }& \small{O9.5 IV:n           }& \small{SB1 }& \small{1}& \small{1 $\&$ 6b}\\   
 \small{62}&\small{MT91-611}  & \small{J20334086+4130189} & \small{O7 V$^{7}$                 }& \small{O7 V((f))          }& \small{-             }& \small{6a}& \small{6a $\&$ 6b}\\
\small{63}& \small{Cyg OB2~$\#$10} & \small{J20334610+4133010} & \small{O9.7 Iab$^{3}$             }& \small{O9.7 Iab           }& \small{-             }& \small{5a}& \small{5a $\&$ 5b}\\
 \small{64}&\small{-}         & \small{J20335842+4019411} & \small{O9:$^{5}$                 }& \small{O9.5 Vn            }& \small{-             }& \small{6a}& \small{6a $\&$ 6b}\\
 \small{65}&\small{Cyg OB2~$\#$27A} & \small{J20335952+4117354} & \small{O9.7 V(n) + O9.7: V$^{4}$     }& \small{-                 }& \small{SB2 }& \small{-}& \small{-}\\
\small{66}& \small{MT91-716}  & \small{J20340486+4105129} & \small{O9 V $^{7}$                }& \small{O9 V               }& \small{-             }& \small{5a}& \small{5a $\&$ 5b}\\
 \small{67}&\small{MT91-720}  & \small{J20340601+4108090} & \small{O9.5 V + B1-2 V$^{12}$       }& \small{-                 }& \small{SB2 }& \small{-}& \small{-}\\
 \small{68}&\small{Cyg OB2~$\#$11} & \small{J20340850+4136592} & \small{O5.5Ifc$^{6}$            }& \small{O5.5 Ifc           }& \small{SB1 }& \small{5a}& \small{5a $\&$ 5b}\\
 \small{69}&\small{MT91-736}  & \small{J20340951+4134136} & \small{O9.5 IV:$^{4}$            }& \small{O9.5 IV            }& \small{-             }& \small{2}& \small{2 $\&$ 6b}\\ 
 \small{70}&\small{Cyg OB2~$\#$29} & \small{J20341350+4135027} & \small{O7.5 V((f))(n)z$^{4}$     }& \small{O7.5 V((f))(n)z    }& \small{SB1             }& \small{2}& \small{2 $\&$ 6b}\\
\small{71}& \small{A28}       & \small{J20341604+4102196} & \small{O9.5 V(n)$^{5}$           }& \small{O9.5 V(n)          }& \small{-             }& \small{6a}& \small{6a $\&$ 6b}\\
 \small{72}&\small{S73}       & \small{J20342193+4117016} & \small{O8 Vz + O8 Vz$^{4}$        }& \small{-                 }& \small{SB2 }& \small{-}& \small{-}\\
 \small{73}&\small{MT91-771}  & \small{J20342959+4131455} & \small{O7 V((f)) + O7 IV((f))$^{1}$ }& \small{-    }& \small{SB2             }& \small{-}& \small{-}\\
\small{74}& \small{A24}       & \small{J20344410+4051584} & \small{O6.5 III(f)$^{9}$         }& \small{O6.5 III(f)        }& \small{-             }& \small{1}& \small{1 $\&$ 6b}\\
 \small{75}&\small{A29}  & \small{J20345605+4038180} & \small{O9.7 Iab$^{8}$            }& \small{O9.7 Iab           }& \small{-             }& \small{6a}& \small{6a $\&$ 6b}\\
\small{76}& \small{B18}& \small{J20345785+4143542} & \small{O7: Ib$^{2}$              }& \small{O7.5 IV:(f)        }& \small{$a$           }& \small{6a}& \small{6a $\&$ 6b}\\
 \small{77}&\small{Cyg OB2~$\#$A36}       & \small{J20345878+4136174} & \small{O9.5 II + O8.5 III$^{12}$           }& \small{ -         }& \small{SB2             }& \small{-}& \small{-}\\ 
 \small{78}&\small{A37}       & \small{J20360451+4056129} & \small{O5 V((f))$^{8}$           }& \small{O5 V((f))          }& \small{-             }& \small{6a}& \small{6a $\&$ 6b}\\

\hline  
\multicolumn{8}{l}{\footnotesize  SpT sources from previous works: (1) \cite{maiz19}, (2) \cite{berlanas18a}, (3) \cite{berlanas18b},} \\ 
 \multicolumn{8}{l}{\footnotesize (4) \cite{maiz16}, (5) \cite{com12}, (6)\cite{sota11}, (7) \cite{kiminki07}, (8) \cite{hanson03}.}\\
 \multicolumn{8}{l}{\footnotesize (9) \cite{neg08}, (10)\cite{mt91}, (11) \cite{morgan54}, (12) Ma\'iz-Apell\'aniz, private comment.}\\ 
 \multicolumn{8}{l}{\footnotesize  (13) \cite{sota14}. Other notes: (a) Recently classified as O8 III by \cite{roman-lopes19}}   \\    
\end{longtable}
%
 

\begin{longtable}{p{0.5cm}|c|cccc|c}
\caption{\small{Spectroscopic stellar parameters and rotational velocities of the final O-type sample obtained using the \texttt{iacob-gbat} and \texttt{iacob-broad} tools. Uncertainties for $vsini$  are in the order of 10 -- 20$\%$. }}
\label{table_param}\\
\hline
\hline
 \small{ID}& \small{SpT} &\small{$T_{\rm eff}$ } & \small{log $g$ }& \small{-log $Q$ } & \small{Y(He)}& \small{$vsini$} \\
 \small{}& \small{} &\small{[kK]} & \small{[dex] }& \small{[dex]} & \small{[10$^{-2}$dex]}& \small{[km s$^{-1}$]} \\
 \hline\\[-1.5ex] 
\endfirsthead	

\caption{\small{continued.}} \\
\hline
\hline
 \small{ID}& \small{SpT} & \small{$T_{\rm eff}$ } & \small{log $g$ }& \small{-log $Q$ } & \small{Y(He)}& \small{$vsini$} \\
  \small{}& \small{} &\small{[kK]} & \small{[dex] }& \small{[dex]} & \small{[10$^{-2}$dex]}& \small{[km s$^{-1}$]} \\
 \hline\\[-1.5ex] 
\endhead

 \small{ 1  }   & \small{B0 IV              }  & \small{   31.0 $\pm$  0.5}       & \small{  3.99 $\pm$  0.05}        & \small{$>$ 13.5         }       & \small{   10.1 $\pm$  2.5}        & \small{   20.}    \\
 \small{ 2  }   & \small{O8 Vz              }  & \small{   35.8 $\pm$  0.6}       & \small{  3.88 $\pm$  0.06}        & \small{$>$ 13.0         }       & \small{   10.0 $\pm$  2.5}        & \small{  125.}    \\
 \small{ 3  }   & \small{O9.5 II            }  & \small{   29.3 $\pm$  0.8}       & \small{  3.14 $\pm$  0.07}        & \small{  12.8 $\pm$  0.2}       & \small{   11.3 $\pm$  2.8}        & \small{  190.}    \\
 \small{ 4  }   & \small{OC9.7 Iab          }  & \small{   28.2 $\pm$  1.3}       & \small{  3.12 $\pm$  0.13}        & \small{  12.5 $\pm$  0.1}       & \small{  $<$ 8.3         }        & \small{   90.}    \\
 \small{ 5  }   & \small{O7 Ib(f)           }  & \small{   35.5 $\pm$  1.1}       & \small{  3.47 $\pm$  0.12}        & \small{  12.5 $\pm$  0.2}       & \small{   14.3 $\pm$  4.3}        & \small{  180.}    \\
 \small{ 6  }   & \small{O8.5 III           }  & \small{   35.5 $\pm$  1.5}       & \small{$<$ 4.20          }        & \small{  13.4 $\pm$  0.5}       & \small{  $<$ 6.9         }        & \small{   110.}    \\
 \small{ 7  }   & \small{O9.5 IV            }  & \small{   32.6 $\pm$  1.0}       & \small{  4.10 $\pm$  0.15}        & \small{  13.5 $\pm$  0.5}       & \small{    9.9 $\pm$  3.3}        & \small{  105.}    \\
 \small{ 8  }   & \small{O8 IIIz            }  & \small{   35.1 $\pm$  1.0}       & \small{  3.61 $\pm$  0.10}        & \small{ $>$ 13.1        }       & \small{  10.4  $\pm$  2.5}        & \small{   55.}    \\
 \small{ 9  }   & \small{O9.7 III           }  & \small{   31.5 $\pm$  0.8}       & \small{  3.56 $\pm$  0.16}        & \small{$>$ 13.3         }       & \small{   11.4 $\pm$  2.7}        & \small{   85.}    \\
 \small{ 10 }   & \small{O9.7 III(n)        }  & \small{   30.8 $\pm$  1.8}       & \small{  3.48 $\pm$  0.18}        & \small{$>$ 13.0         }       & \small{ $<$ 15.5         }        & \small{  185.}    \\
 \small{ 13 }   & \small{O6.5 V((f))z       }  & \small{   38.0 $\pm$  0.7}       & \small{  3.80 $\pm$  0.07}        & \small{  13.3 $\pm$  0.5}       & \small{    11.0 $\pm$  2.5}       & \small{   70.}    \\
 \small{ 14 }   & \small{O9.5 IV            }  & \small{   33.0 $\pm$  1.2}       & \small{  3.87 $\pm$  0.19}        & \small{$>$ 13.3         }       & \small{   10.0 $\pm$  2.5}        & \small{   85.}    \\
 \small{ 15 }   & \small{O7 V((f))          }  & \small{   36.8 $\pm$  1.0}       & \small{  3.77 $\pm$  0.11}        & \small{  13.6 $\pm$  0.7}       & \small{  $<$ 8.0         }        & \small{   75.}    \\
 \small{ 16 }   & \small{O9.7 II(n)         }  & \small{   29.6 $\pm$  0.7}       & \small{  3.31 $\pm$  0.12}        & \small{  12.9 $\pm$  0.2}       & \small{  $<$ 8.7         }        & \small{   80.}    \\
 \small{ 17 }   & \small{O8 IV(n)f      }  & \small{   35.0 $\pm$  0.5}       & \small{  3.69 $\pm$  0.05}        & \small{  13.0 $\pm$  0.2}       & \small{  $<$ 6.6         }        & \small{  185.}    \\
 \small{ 18 }   & \small{O8.5 III           }  & \small{   31.8 $\pm$  1.2}       & \small{  3.39 $\pm$  0.12}        & \small{  13.5 $\pm$  0.5}       & \small{   10.0 $\pm$  2.5}        & \small{   85.}    \\
 \small{ 19 }   & \small{O7 Ibf             }  & \small{   35.7 $\pm$  1.2}       & \small{  3.53 $\pm$  0.16}        & \small{  12.3 $\pm$  0.2}       & \small{   10.6 $\pm$  2.6}        & \small{  270.}    \\
 \small{ 21 }   & \small{O8 III             }  & \small{   33.3 $\pm$  0.6}       & \small{  3.52 $\pm$  0.07}        & \small{  12.9 $\pm$  0.2}       & \small{  $<$ 6.6         }        & \small{   85.}    \\
 \small{ 22 }   & \small{O9.7 IV            }  & \small{   33.0 $\pm$  0.5}       & \small{  3.91 $\pm$  0.06}        & \small{  13.6 $\pm$  0.3}       & \small{    9.7 $\pm$  2.5}        & \small{   30.}    \\
 \small{ 23 }   & \small{O7.5 Vz            }  & \small{   38.6 $\pm$  1.6}       & \small{  4.10 $\pm$  0.20}        & \small{$>$ 12.9         }       & \small{ $<$ 11.8         }        & \small{  130.}    \\
 \small{ 24 }   & \small{O7 IIIf            }  & \small{   35.1 $\pm$  0.5}       & \small{  3.41 $\pm$  0.05}        & \small{  12.9 $\pm$  0.2}       & \small{   9.9 $\pm$  2.5}        & \small{  105.}   \\
 \small{ 25 }   & \small{O9.5 IV            }  & \small{   31.7 $\pm$  0.7}       & \small{  3.78 $\pm$  0.08}        & \small{$>$ 13.4         }       & \small{    9.8 $\pm$  2.5}        & \small{  190.}   \\
 \small{ 28 }   & \small{O8 III             }  & \small{   35.0 $\pm$  0.5}       & \small{  3.80 $\pm$  0.05}        & \small{  13.5 $\pm$  0.4}       & \small{  $<$ 6.9         }        & \small{  200.}    \\
 \small{ 30 }   & \small{O7.5 Ib(f)         }  & \small{   35.8 $\pm$  0.9}       & \small{  3.60 $\pm$  0.11}        & \small{  12.7 $\pm$  0.1}       & \small{  $<$ 6.9         }        & \small{  135.}    \\
 \small{ 31 }   & \small{O8 Vz(n)           }  & \small{   37.1 $\pm$  0.9}       & \small{  4.03 $\pm$  0.16}        & \small{$>$ 13.2         }       & \small{   10.5 $\pm$  2.5}        & \small{  120.}    \\
 \small{ 32 }   & \small{O8 III((f))        }  & \small{   35.1 $\pm$  1.2}       & \small{  3.63 $\pm$  0.14}        & \small{  13.1 $\pm$  0.3}       & \small{  $<$ 7.3         }        & \small{  100.}    \\
 \small{ 33 }   & \small{O7.5: IV(n)        }  & \small{   35.8 $\pm$  0.6}       & \small{  3.68 $\pm$  0.05}        & \small{  13.5 $\pm$  0.5}       & \small{   10.1 $\pm$  2.5}        & \small{  210.}    \\
 \small{ 34 }   & \small{O8 V(n)            }  & \small{   35.0 $\pm$  0.5}       & \small{  3.84 $\pm$  0.08}        & \small{  13.5 $\pm$  0.3}       & \small{  $<$ 6.6         }        & \small{  190.}    \\
 \small{ 35 }   & \small{O8 V:              }  & \small{   35.5 $\pm$  0.7}       & \small{  3.86 $\pm$  0.08}        & \small{  13.5 $\pm$  0.5}       & \small{  $<$ 7.3         }        & \small{  190.}    \\
 \small{ 36 }   & \small{O8 V               }  & \small{   35.8 $\pm$  0.6}       & \small{  3.80 $\pm$  0.05}        & \small{  13.1 $\pm$  0.2}       & \small{    9.7 $\pm$  2.5}        & \small{  195.}    \\
 \small{ 37 }   & \small{O9.7 IV:(n)        }  & \small{   30.3 $\pm$  0.9}       & \small{  3.77 $\pm$  0.14}        & \small{  13.3 $\pm$  0.3}       & \small{   10.5 $\pm$  2.5}        & \small{  225.}    \\
 \small{ 38 }   & \small{O7 III((f))      }  & \small{   38.2 $\pm$  0.6}       & \small{  3.85 $\pm$  0.11}        & \small{  12.7 $\pm$  0.1}       & \small{  $<$ 7.1         }        & \small{   80.}    \\
 \small{ 39 }   & \small{O8 V(n)            }  & \small{   34.9 $\pm$  1.2}       & \small{  3.55 $\pm$  0.16}        & \small{  13.8 $\pm$  0.9}       & \small{ $<$ 12.3         }        & \small{  130.}    \\
 \small{ 40 }   & \small{O3 If*             }  & \small{   44.4 $\pm$  0.9}       & \small{  3.83 $\pm$  0.07}        & \small{  12.5 $\pm$  0.1}       & \small{    9.9 $\pm$  2.5}        & \small{   100.}    \\
 \small{ 41 }   & \small{O6 IV((f))         }  & \small{   37.3 $\pm$  0.6}       & \small{  3.63 $\pm$  0.06}        & \small{  13.0 $\pm$  0.2}       & \small{  $<$ 6.6         }        & \small{  145.}    \\
 \small{ 43 }   & \small{O9.5 IV:n          }  & \small{   31.9 $\pm$  0.6}       & \small{  3.91 $\pm$  0.15}        & \small{  13.5 $\pm$  0.3}       & \small{  $<$ 6.6         }        & \small{  275.}    \\
 \small{ 46 }   & \small{O6.5 V((f))        }  & \small{   37.0 $\pm$  1.3}       & \small{  3.77 $\pm$  0.15}        & \small{  13.2 $\pm$  0.5}       & \small{   13.5 $\pm$  4.1}        & \small{  195.}    \\
 \small{ 47 }   & \small{O7 Vz((f))         }  & \small{   35.8 $\pm$  0.5}       & \small{  3.87 $\pm$  0.07}        & \small{  13.3 $\pm$  0.4}       & \small{  $<$ 6.9         }        & \small{  170.}   \\
 \small{ 48 }   & \small{O3 If*             }  & \small{   44.0 $\pm$  2.0}       & \small{  3.72 $\pm$  0.11}        & \small{  12.1 $\pm$  0.1}       & \small{   16.0 $\pm$  4.0}        & \small{   80.}    \\
 \small{ 49 }   & \small{O6 II(f)           }  & \small{   36.8 $\pm$  0.5}       & \small{  3.59 $\pm$  0.05}        & \small{  12.6 $\pm$  0.1}       & \small{  $<$ 6.6         }        & \small{  110.}    \\
 \small{ 51 }   & \small{O9.5 V             }  & \small{   32.3 $\pm$  1.1}       & \small{  3.84 $\pm$  0.14}        & \small{$>$ 13.4         }       & \small{   13.3 $\pm$  3.4}        & \small{  190.}    \\
 \small{ 52 }   & \small{O8.5 V(n)          }  & \small{   34.4 $\pm$  0.6}       & \small{  3.83 $\pm$  0.08}        & \small{$>$ 13.5         }       & \small{   10.0 $\pm$  2.5}        & \small{  170.}    \\

 \small{ 53 }   & \small{O8 V(n)            }  & \small{   35.6 $\pm$  0.6}       & \small{  3.81 $\pm$  0.11}        & \small{  13.0 $\pm$  0.2}       & \small{  $<$ 6.4         }        & \small{  230.}    \\
 \small{ 54 }   & \small{O5 I(fc)           }  & \small{   38.2 $\pm$  0.5}       & \small{  3.69 $\pm$  0.08}        & \small{  12.5 $\pm$  0.1}       & \small{  $<$ 7.1         }        & \small{  150.}    \\
 \small{ 55 }   & \small{O8 V               }  & \small{   37.3 $\pm$  0.5}       & \small{  4.10 $\pm$  0.07}        & \small{  13.5 $\pm$  0.4}       & \small{   9.9 $\pm$  2.5}        & \small{  130.}    \\
 \small{ 56 }   & \small{O9.5 V(n)          }  & \small{   33.0 $\pm$  0.5}       & \small{  3.80 $\pm$  0.05}        & \small{$>$ 13.7         }       & \small{   14.3 $\pm$  2.5}        & \small{  195.}    \\
 \small{ 57 }   & \small{O6 IVf             }  & \small{   38.0 $\pm$  0.5}       & \small{  3.70 $\pm$  0.05}        & \small{  12.7 $\pm$  0.1}       & \small{  $<$ 6.6         }        & \small{  85.}    \\
 \small{ 58 }   & \small{O8.5 V             }  & \small{   36.0 $\pm$  0.5}       & \small{  3.80 $\pm$  0.05}        & \small{  13.0 $\pm$  0.2}       & \small{  $<$ 7.1         }        & \small{   85.}    \\
 \small{ 59 }   & \small{O8.5 V             }  & \small{   36.0 $\pm$  0.5}       & \small{  3.80 $\pm$  0.05}        & \small{  13.5 $\pm$  0.5}       & \small{ 9.9 $\pm$  2.5 }          & \small{   95.}    \\
 \small{ 60 }   & \small{O8 V((f))          }  & \small{   35.9 $\pm$  1.1}       & \small{  3.72 $\pm$  0.13}        & \small{  12.9 $\pm$  0.2}       & \small{  $<$ 6.6         }        & \small{  120.}    \\
 \small{ 61 }   & \small{O9.5 IV:n          }  & \small{   31.0 $\pm$  0.5}       & \small{  3.50 $\pm$  0.05}        & \small{$>$ 13.0         }       & \small{ $<$ 11.1         }        & \small{  225.}    \\
 \small{ 62 }   & \small{O7 V((f))          }  & \small{   37.0 $\pm$  1.0}       & \small{  3.82 $\pm$  0.13}        & \small{$>$ 13.0         }       & \small{   11.3 $\pm$  2.5}        & \small{   70.}    \\
 \small{ 63 }   & \small{O9.7 Iab           }  & \small{   28.4 $\pm$  0.6}       & \small{  3.03 $\pm$  0.06}        & \small{  12.7 $\pm$  0.1}       & \small{    9.4 $\pm$  2.6}        & \small{   55.}    \\
 \small{ 64 }   & \small{O9.5 Vn            }  & \small{   33.6 $\pm$  1.7}       & \small{  3.76 $\pm$  0.25}        & \small{  13.4 $\pm$  0.4}       & \small{ $<$ 18.4         }        & \small{  265.}    \\
 \small{ 66 }   & \small{O9 V               }  & \small{   34.6 $\pm$  0.7}       & \small{  4.05 $\pm$  0.10}        & \small{  13.8 $\pm$  0.8}       & \small{    9.9 $\pm$  2.5}        & \small{   55.}    \\
 \small{ 68 }   & \small{O5.5 Ifc           }  & \small{   36.3 $\pm$  0.5}       & \small{  3.50 $\pm$  0.05}        & \small{  12.3 $\pm$  0.1}       & \small{   10.1 $\pm$  2.5}        & \small{   95.}   \\
 \small{ 69 }   & \small{O9.5 IV            }  & \small{   33.9 $\pm$  0.6}       & \small{  3.96 $\pm$  0.10}        & \small{$>$ 13.0         }       & \small{    9.2 $\pm$  2.5}        & \small{  165.}   \\
 \small{ 70 }   & \small{O7.5 V((f))(n)z    }  & \small{   37.0 $\pm$  0.5}       & \small{  3.80 $\pm$  0.05}        & \small{  13.5 $\pm$  0.3}       & \small{   9.7 $\pm$  2.5 }        & \small{  170.}    \\
 \small{ 71 }   & \small{O9.5 V(n)          }  & \small{   31.2 $\pm$  1.2}       & \small{  3.86 $\pm$  0.20}        & \small{$>$ 13.0         }       & \small{   15.7 $\pm$  6.0}        & \small{  155.}   \\
 \small{ 74 }   & \small{O6.5 III(f)        }  & \small{   36.6 $\pm$  0.6}       & \small{  3.58 $\pm$  0.05}        & \small{  12.9 $\pm$  0.2}       & \small{   12.0 $\pm$  2.6}        & \small{  150.}    \\
 \small{ 75 }   & \small{O9.7 Iab           }  & \small{   27.7 $\pm$  0.6}       & \small{  2.98 $\pm$  0.07}        & \small{  13.0 $\pm$  0.2}       & \small{    9.8 $\pm$  2.5}        & \small{   90.}    \\
 \small{ 76 }   & \small{O7.5 IV:(f)        }  & \small{   36.3 $\pm$  1.2}       & \small{  3.92 $\pm$  0.14}        & \small{  12.8 $\pm$  0.3}       & \small{   10.3 $\pm$  2.5}        & \small{   95.}    \\
 \small{ 78 }   & \small{O5 V((f))          }  & \small{   39.8 $\pm$  1.7}       & \small{  3.63 $\pm$  0.12}        & \small{  12.6 $\pm$  0.2}       & \small{   10.5 $\pm$  2.7}        & \small{  225.}    \\

\hline  
\end{longtable}\vspace{1cm}

\begin{longtable}{p{0.5cm}|ccccc|ccc}
\caption{\small{ O-type sample with reliable astrometry (RUWE $\leq$ 1.4). Visual extinctions and absolute visual magnitudes have been calculated using the extinction law derived by \cite{rieke85}, distances derived by \cite{bailer-jones18} and photometry from Table~\ref{table_fotom}. Membership groups are from results obtained by \cite{berlanas19} (Group 1: stars at $\sim$1.35 kpc; Group 2: main Cygnus OB2 group at $\sim$1.76 kpc; Group 0: objects lying between Groups 1 and 2 that could not be reliably assigned to any of them; Group 3: foreground or background contaminants). Radii, luminosities and masses derived using the \texttt{iacob-gbat} tool. We include the \texttt{iacob-gbat} formal uncertainties and errors related to $M_{V}$.}}
\label{table_gaia}\\
\hline
\hline
 \small{ID}& \small{Gaia source} &\small{Distance [pc]} & \small{A$_{V}$ [mag]}& \small{M$_{V}$ [mag]} & \small{Group}&\small{R  [R$_{\odot}$]} & \small{log L/L$_{\odot}$ [dex]}& \small{M$_{sp}$ [M$_{\odot}$]} \\ 	 	 
 \hline\\[-1.5ex] 
\endfirsthead	

\caption{\small{continued.}} \\
\hline
\hline
 \small{ID}& \small{Gaia source} &\small{Distance [pc]} & \small{A$_{V}$ [mag]}& \small{M$_{V}$ [mag]} & \small{Group}&\small{R  [R$_{\odot}$]} & \small{log L/L$_{\odot}$ [dex]}& \small{M$_{sp}$ [M$_{\odot}$]}\\ 	 	   	 	 	 	 
 \hline\\[-1.5ex] 
\endhead

\small{1 }  & \small{2067642233192106624 }    & \small{2634$^{ + 208}_{ - 181}$}    & \small{4.01 }           & \small{-4.55$^{  - 0.17}_{ + 0.15}$}      & \small{3 }  & \small{  10.10 $^{+0.70}_{-0.79}$ }  & \small{   4.92$^{+0.06}_{-0.07}$  }       & \small{ 37.0$^{+6.5 }_{- 7.0 }$   }   \\ 
\small{2 }  & \small{2067398416488472704 }    & \small{1652$^{ + 77}_{  - 71 }$}    & \small{2.43 }           & \small{-3.95$^{  - 0.10}_{ + 0.10}$}      & \small{2 }  & \small{  6.80  $^{+0.33}_{-0.33}$ }  & \small{   4.84$^{+0.04}_{-0.04}$  }       & \small{ 13.4$^{+2.4 }_{- 2.4 }$   }   \\   
\small{3}   & \small{2068074620437883520 }    & \small{1489$^{ + 118}_{ - 103}$}    & \small{5.87 }           & \small{-5.06$^{  - 0.17}_{ + 0.15} $}     & \small{2 }  & \small{  13.20 $^{+0.96}_{-1.07}$ }  & \small{   5.06$^{+0.07}_{-0.07}$  }       & \small{ 8.1 $^{+2.3 }_{- 2.4 }$   }   \\   
\small{4 }  & \small{2067625637438421248 }    & \small{1757$^{ + 126}_{ - 111}$}    & \small{6.84 }           & \small{-6.68$^{  - 0.15}_{ + 0.14}$}      & \small{2 }  & \small{  28.70 $^{+2.01}_{-2.13}$ }  & \small{   5.68$^{+0.07}_{-0.07}$  }       & \small{ 41.2$^{+9.6 }_{- 9.8 }$   }   \\   
\small{5 }  & \small{2067430718942052224 }    & \small{1324$^{ + 76}_{  - 69 }$}    & \small{3.59 }           & \small{-5.61$^{  - 0.12}_{ + 0.11}$}      & \small{1 }  & \small{  14.80 $^{+0.81}_{-0.87}$ }  & \small{   5.48$^{+0.05}_{-0.06}$  }       & \small{ 24.7$^{+5.6 }_{- 5.7 }$   }   \\   
\small{6}   & \small{2067807885789820160 }    & \small{1777$^{ + 130}_{ - 114}$}    & \small{6.76 }           & \small{-4.95$^{  - 0.15}_{ + 0.15}$}      & \small{2 }  & \small{  10.90 $^{+0.78}_{-0.78}$ }  & \small{   5.21$^{+0.06}_{-0.07}$  }       & \small{ 40.0$^{+15.9}_{- 15.9}$   }  \\   
\small{7}   & \small{2068007034832335104 }    & \small{1755$^{ + 118}_{ - 104}$}    & \small{6.60 }           & \small{-4.30$^{  - 0.14}_{ + 0.13}$}      & \small{2 }  & \small{  8.60  $^{+0.55}_{-0.59}$ }  & \small{   4.89$^{+0.06}_{-0.06}$  }       & \small{ 36.8$^{+11.3}_{- 11.5}$   }  \\   
\small{9 }  & \small{2067816681882846464 }    & \small{1623$^{ + 65}_{  - 60 }$}    & \small{4.64 }           & \small{-3.95$^{  - 0.08}_{ + 0.08}$}      & \small{2 }  & \small{  7.50  $^{+0.29}_{-0.29}$ }  & \small{   4.68$^{+0.04}_{-0.04}$  }       & \small{ 8.1 $^{+2.3 }_{- 2.3 }$   }   \\   
\small{10}  & \small{2068008164405551104 }    & \small{1763$^{ + 220}_{ - 177}$}    & \small{6.87 }           & \small{-4.46$^{  - 0.26}_{ + 0.23}$}      & \small{2 }  & \small{  9.60  $^{+1.09}_{-1.21}$ }  & \small{   4.88$^{+0.10}_{-0.11}$  }       & \small{ 11.1$^{+4.0 }_{- 4.2 }$   }   \\   
\small{15}  & \small{2067788713055642880 }    & \small{1561$^{ + 65}_{  - 60 }$}    & \small{4.42 }           & \small{-4.50$^{  - 0.08}_{ + 0.09}$}      & \small{2 }  & \small{  8.60  $^{+0.37}_{-0.33}$ }  & \small{   5.08$^{+0.05}_{-0.04}$  }       & \small{ 16.3$^{+2.4 }_{- 2.3 }$   }   \\   
\small{16}  & \small{2068155125305302144 }    & \small{1297$^{ + 100}_{ - 87 }$}    & \small{5.88 }           & \small{-5.02$^{  - 0.16}_{ + 0.15} $}     & \small{1 }  & \small{  12.80 $^{+0.89}_{-0.95}$ }  & \small{   5.06$^{+0.07}_{-0.07}$  }       & \small{ 12.9$^{+2.7 }_{- 2.7 }$   }   \\   
\small{18}  & \small{2067840149584105344 }    & \small{1595$^{ + 101}_{ - 89 }$}    & \small{6.15 }           & \small{-4.80$^{  - 0.13}_{ + 0.13}$}      & \small{2 }  & \small{  10.90 $^{+0.66}_{-0.66}$ }  & \small{   5.04$^{+0.06}_{-0.06}$  }       & \small{ 10.9$^{+1.9 }_{- 1.9 }$   }   \\   
\small{19}  & \small{2067796787591714048 }    & \small{1694$^{ + 216}_{ - 174}$}    & \small{7.96 }           & \small{-6.07$^{  - 0.27}_{ + 0.23}$}      & \small{2 }  & \small{  18.30 $^{+1.98}_{-2.31}$ }  & \small{   5.69$^{+0.09}_{-0.11}$  }       & \small{ 43.6$^{+14.1}_{- 15.2}$   }   \\   
\small{22}  & \small{2067847502568383744 }    & \small{1603$^{ + 88}_{  - 79 }$}    & \small{4.13 }           & \small{-3.68$^{  - 0.12}_{ + 0.11}$}      & \small{2 }  & \small{  6.40  $^{+0.34}_{-0.37}$ }  & \small{   4.64$^{+0.05}_{-0.05}$  }       & \small{ 12.9$^{+1.9 }_{- 2.1 }$   }   \\   
\small{23}  & \small{2067827466541470080 }    & \small{1760$^{ + 78}_{  - 71 }$}    & \small{6.16 }           & \small{-4.45$^{  - 0.09}_{ + 0.09} $}     & \small{2 }  & \small{  8.10  $^{+0.39}_{-0.39}$ }  & \small{   5.13$^{+0.04}_{-0.04}$  }       & \small{ 31.6$^{+12.0}_{- 12.0}$   }   \\   
\small{24}  & \small{2067835682818357376 }    & \small{1431$^{ + 80}_{  - 72 }$}    & \small{4.31 }           & \small{-5.01$^{  - 0.11}_{ + 0.12} $}     & \small{0 }  & \small{  11.30 $^{+0.63}_{-0.58}$ }  & \small{   5.24$^{+0.05}_{-0.05}$  }       & \small{ 12.2$^{+2.0 }_{- 1.9 }$   }   \\   
\small{25}  & \small{2067835614098871040 }    & \small{1645$^{ + 83}_{  - 75 }$}    & \small{4.23 }           & \small{-4.20$^{  - 0.11}_{ + 0.10}$}      & \small{2 }  & \small{  8.30  $^{+0.39}_{-0.43}$ }  & \small{   4.81$^{+0.04}_{-0.05}$  }       & \small{ 16.3$^{+2.5 }_{- 2.6 }$   }   \\   
\small{28}  & \small{2067834926904094848 }    & \small{1519$^{ + 73}_{  - 67 }$}    & \small{4.43 }           & \small{-4.23$^{  - 0.11}_{ + 0.10}$}      & \small{2 }  & \small{  7.90  $^{+0.38}_{-0.41}$ }  & \small{   4.92$^{+0.04}_{-0.05}$  }       & \small{ 14.5$^{+2.1 }_{- 2.2 }$   }   \\   
\small{30}  & \small{2067829596846026880 }    & \small{1309$^{ + 110}_{ - 95 }$}    & \small{7.53 }           & \small{-5.60$^{  - 0.18}_{ + 0.16} $}     & \small{1 }  & \small{  14.70 $^{+1.12}_{-1.25}$ }  & \small{   5.50$^{+0.06}_{-0.07}$  }       & \small{ 32.3$^{+7.7 }_{- 8.1 }$   }   \\   
\small{31}  & \small{2067769609040987648 }    & \small{1723$^{ + 99}_{  - 89 }$}    & \small{6.14 }           & \small{-4.14$^{  - 0.13}_{ + 0.11} $}     & \small{2 }  & \small{  7.20  $^{+0.38}_{-0.44}$ }  & \small{   4.94$^{+0.05}_{-0.06}$  }       & \small{ 21.7$^{+5.9 }_{- 6.1 }$   }  \\    
\small{33}  & \small{2067833204620693632 }    & \small{1514$^{ + 85}_{  - 77 }$}    & \small{4.33 }           & \small{-4.53$^{  - 0.12}_{ + 0.11} $}     & \small{2 }  & \small{  8.80  $^{+0.46}_{-0.49}$ }  & \small{   5.07$^{+0.05}_{-0.05}$  }       & \small{ 14.2$^{+2.5 }_{- 2.5 }$   }   \\   
\small{34}  & \small{2067833243277076864 }    & \small{1487$^{ + 83}_{  - 75 }$}    & \small{4.40 }           & \small{-4.77$^{  - 0.12 }_{+ 0.11}$}      & \small{0 }  & \small{  10.00 $^{+0.51}_{-0.56}$ }  & \small{   5.13$^{+0.05}_{-0.05}$  }       & \small{ 26.0$^{+4.8 }_{- 4.9 }$   }   \\   
\small{35}  & \small{2067832968398974208 }    & \small{1391$^{ + 78}_{  - 70 }$}    & \small{4.61 }           & \small{-4.09$^{  - 0.12}_{ + 0.11} $}     & \small{1 }  & \small{  7.20  $^{+0.38}_{-0.41}$ }  & \small{   4.90$^{+0.05}_{-0.05}$  }       & \small{ 14.9$^{+2.4 }_{- 2.5 }$   }  \\   
\small{36}  & \small{2067832624801783040 }    & \small{1625$^{ + 104}_{ - 93 }$}    & \small{4.61 }           & \small{-4.10$^{  - 0.13}_{ + 0.13}$}      & \small{2 }  & \small{  7.30  $^{+0.45}_{-0.45}$ }  & \small{   4.90$^{+0.06}_{-0.05}$  }       & \small{ 12.5$^{+2.1 }_{- 2.1 }$   }    \\
\small{37}  & \small{2067782936320586240 }    & \small{1742$^{ + 128}_{ - 112}$}    & \small{6.48 }           & \small{-4.62$^{  - 0.15}_{ + 0.15} $}     & \small{2 }  & \small{  10.60 $^{+0.76}_{-0.76}$ }  & \small{   4.93$^{+0.07}_{-0.07}$  }       & \small{ 25.9$^{+5.7 }_{- 5.7 }$   }   \\
\small{39}  & \small{2067784516868550016 }    & \small{1573$^{ + 110}_{ - 97 }$}    & \small{6.35 }           & \small{-4.66$^{  - 0.15}_{ + 0.14} $}     & \small{2 }  & \small{  9.70  $^{+0.65}_{-0.69}$ }  & \small{   5.09$^{+0.06}_{-0.07}$  }       & \small{ 13.2$^{+2.6 }_{- 2.7 }$   }   \\
\small{40}  & \small{2067781905528395264 }    & \small{1534$^{ + 180}_{ - 147}$}    & \small{7.25 }           & \small{-5.89$^{  - 0.24}_{ + 0.22}$}      & \small{2 }  & \small{  14.80 $^{+1.44}_{-1.65}$ }  & \small{   5.87$^{+0.09}_{-0.10}$  }       & \small{ 54.8$^{+13.6}_{- 14.8}$   }  \\
\small{43}  & \small{2067781905528395776 }    & \small{1475$^{ + 84}_{  - 75 }$}    & \small{6.84 }           & \small{-4.68$^{  - 0.11}_{ + 0.12} $}     & \small{2 }  & \small{  10.40 $^{+0.58}_{-0.53}$ }  & \small{   5.01$^{+0.05}_{-0.05}$  }       & \small{ 33.6$^{+12.5}_{- 12.4}$   }    \\
\small{46}  & \small{2067781939888133248 }    & \small{1670$^{ + 136}_{ - 117}$}    & \small{6.69 }           & \small{-4.71$^{  - 0.17}_{ + 0.16}$}      & \small{2 }  & \small{  9.40  $^{+0.67}_{-0.74}$ }  & \small{   5.17$^{+0.07}_{-0.07}$  }       & \small{ 21.4$^{+5.9 }_{- 6.0 }$   }   \\
\small{47}  & \small{2067781871173898624 }    & \small{1627$^{ + 87}_{  - 79 }$ }   & \small{5.68 }           & \small{-4.27$^{  - 0.12}_{ + 0.11}$}      & \small{2 }  & \small{  7.90  $^{+0.41}_{-0.45}$ }  & \small{   4.97$^{+0.05}_{-0.05}$  }       & \small{ 16.9$^{+3.1 }_{- 3.2 }$   }  \\
\small{48}  & \small{2067785070923663104 }    & \small{1530$^{ + 73}_{  - 67 }$}    & \small{5.26 }           & \small{-5.78$^{  - 0.10}_{ + 0.10} $}     & \small{2 }  & \small{  13.90 $^{+0.81}_{-0.81}$ }  & \small{   5.83$^{+0.06}_{-0.06}$  }       & \small{ 46.5$^{+9.1 }_{- 9.1 }$   }    \\
\small{49}  & \small{2067784619950644480 }    & \small{1531$^{ + 104}_{ - 91 } $}   & \small{5.20 }           & \small{-5.80$^{  - 0.14}_{ + 0.14} $}     & \small{2 }  & \small{  15.80 $^{+1.04}_{-1.04}$ }  & \small{   5.61$^{+0.06}_{-0.06}$  }       & \small{ 33.5$^{+5.2 }_{- 5.2 }$   }    \\
\small{51}  & \small{2067785070923661440 }    & \small{1607$^{ + 72}_{  - 66 }$}    & \small{4.34 }           & \small{-3.62$^{  - 0.10}_{ + 0.09}$}      & \small{2 }  & \small{  6.30  $^{+0.28}_{-0.31}$ }  & \small{   4.59$^{+0.05}_{-0.05}$  }       & \small{ 10.1$^{+2.6 }_{- 2.7 }$   }   \\
\small{52}  & \small{2067785002204178688 }    & \small{1517$^{ + 78}_{  - 71 }$}    & \small{4.48 }           & \small{-4.01$^{  - 0.11}_{ + 0.11} $}     & \small{2 }  & \small{  7.20  $^{+0.38}_{-0.38}$ }  & \small{   4.79$^{+0.05}_{-0.05}$  }       & \small{ 13.6$^{+1.7 }_{- 1.7 }$   }  \\
\small{53}  & \small{2067783799613328128 }    & \small{1751$^{ + 137}_{ - 119}$}    & \small{5.38 }           & \small{-5.01$^{  - 0.16 }_{+ 0.15}$}      & \small{2 }  & \small{  11.10 $^{+0.77}_{-0.82}$ }  & \small{   5.26$^{+0.06}_{-0.07}$  }       & \small{ 29.9$^{+6.8 }_{- 6.9 }$   }    \\
\small{54}  & \small{2067784246289931776 }    & \small{1713$^{ + 97}_{  - 87 }$}    & \small{4.62 }           & \small{-5.98$^{  - 0.11}_{ + 0.12}$}      & \small{2 }  & \small{  16.90 $^{+0.95}_{-0.88}$ }  & \small{   5.72$^{+0.05}_{-0.05}$  }       & \small{ 51.2$^{+10.6}_{- 10.3}$   }   \\
\small{55}  & \small{2067785208362826112 }    & \small{1571$^{ + 115}_{ - 100}$}    & \small{4.68 }           & \small{-4.23$^{  - 0.15}_{ + 0.14}$}      & \small{2 }  & \small{  7.50  $^{+0.49}_{-0.53}$ }  & \small{   4.98$^{+0.06}_{-0.06}$  }       & \small{ 25.9$^{+5.2 }_{- 5.3 }$   }    \\
\small{56}  & \small{2067784173274044928 }    & \small{1611$^{ + 76}_{  - 70 }$ }   & \small{4.97 }           & \small{-3.75$^{  - 0.10}_{ + 0.10} $}     & \small{2 }  & \small{  6.50  $^{+0.31}_{-0.31}$ }  & \small{   4.66$^{+0.04}_{-0.04}$  }       & \small{ 11.1$^{+1.8 }_{- 1.8 }$   }  \\
\small{58}  & \small{2067928110513529216 }    & \small{2025$^{ + 195}_{ - 165}$}    & \small{5.53 }           & \small{-5.47$^{  - 0.20}_{ + 0.19} $}     & \small{2 }  & \small{  13.80 $^{+1.21}_{-1.27}$ }  & \small{   5.46$^{+0.08}_{-0.08}$  }       & \small{ 44.4$^{+8.8 }_{- 9.2 }$   }   \\
\small{59}  & \small{2067780054401820544 }    & \small{1731$^{ + 96}_{  - 86 }$}    & \small{5.90 }           & \small{-4.53$^{  - 0.11}_{ + 0.11}$}      & \small{2 }  & \small{  8.80  $^{+0.46}_{-0.46}$ }  & \small{   5.07$^{+0.04}_{-0.04}$  }       & \small{ 18.0$^{+2.6 }_{- 2.5 }$   }  \\
\small{60}  & \small{2067929794140714752 }    & \small{1108$^{ + 72}_{  - 63 }$}    & \small{6.02 }           & \small{-4.17$^{  - 0.14}_{ + 0.13} $}     & \small{1 }  & \small{  7.50  $^{+0.46}_{-0.49}$ }  & \small{   4.93$^{+0.06}_{-0.06}$  }       & \small{ 11.7$^{+2.4 }_{- 2.5 }$   } \\
\small{61}  & \small{2067781016474500864 }    & \small{1542$^{ + 80}_{  - 73 }$}    & \small{5.50 }           & \small{-4.46$^{  - 0.11}_{ + 0.11}$}      & \small{2 }  & \small{  9.50  $^{+0.49}_{-0.49}$ }  & \small{   4.87$^{+0.05}_{-0.05}$  }       & \small{ 10.9$^{+3.0 }_{- 3.0 }$   }  \\
\small{62}  & \small{2067881548771533312 }    & \small{1734$^{ + 95}_{  - 86 }$}    & \small{5.09 }           & \small{-4.01$^{  - 0.12}_{ + 0.11}$}      & \small{2 }  & \small{  6.80  $^{+0.36}_{-0.39}$ }  & \small{   4.89$^{+0.05}_{-0.06}$  }       & \small{ 12.1$^{+2.3 }_{- 2.4 }$   }  \\ 
\small{66}  & \small{2067766379225921152 }    & \small{1556$^{ + 70}_{  - 65 }$ }   & \small{5.58 }           & \small{-3.59$^{  - 0.09}_{ + 0.09} $}     & \small{2 }  & \small{  5.90  $^{+0.26}_{-0.26}$ }  & \small{   4.66$^{+0.04}_{-0.04}$  }       & \small{ 14.2$^{+2.2 }_{- 2.2 }$   }  \\  
\small{68}  & \small{2067888218857234304 }    & \small{1639$^{ + 75}_{  - 69 }$}    & \small{5.23 }           & \small{-6.53$^{  - 0.10}_{ + 0.10} $}     & \small{2 }  & \small{  22.60 $^{+1.06}_{-1.06}$ }  & \small{   5.90$^{+0.04}_{-0.04}$  }       & \small{ 57.4$^{+7.3 }_{- 7.3 }$   }  \\   
\small{69}  & \small{2067887802243913216 }    & \small{1549$^{ + 69}_{  - 63 }$}    & \small{5.50 }           & \small{-3.75$^{  - 0.09}_{ + 0.09} $}     & \small{2 }  & \small{  6.40  $^{+0.28}_{-0.28}$ }  & \small{   4.70$^{+0.04}_{-0.04}$  }       & \small{ 14.2$^{+2.8 }_{- 2.8 }$   }  \\   
\small{70}  & \small{2067887840900094848 }    & \small{1597$^{ + 90}_{  - 81 }$}    & \small{5.56 }           & \small{-4.58$^{  - 0.12}_{ + 0.12}$}      & \small{2 }  & \small{  8.90  $^{+0.50}_{-0.50}$ }  & \small{   5.12$^{+0.05}_{-0.05}$  }       & \small{ 18.4$^{+2.7 }_{- 2.7 }$   }  \\  
\small{71}  & \small{2067763080691028736 }    & \small{1660$^{ + 91}_{  - 82 }$}    & \small{6.30 }           & \small{-4.12$^{  - 0.11}_{ + 0.11}$}      & \small{2 }  & \small{  8.20  $^{+0.43}_{-0.43}$ }  & \small{   4.75$^{+0.05}_{-0.05}$  }       & \small{ 15.5$^{+5.2 }_{- 5.2 }$   }  \\ 
\small{74}  & \small{2064757698797394688 }    & \small{1674$^{ + 109}_{ - 97 }$}    & \small{6.92 }           & \small{-5.31$^{  - 0.13}_{ + 0.13}$}      & \small{2 }  & \small{  12.60 $^{+0.76}_{-0.76}$ }  & \small{   5.38$^{+0.05}_{-0.05}$  }       & \small{ 22.4$^{+3.8 }_{- 3.8 }$   }   \\    
\small{75}  & \small{2064739041458261120 }    & \small{1499$^{ + 87}_{  - 78 }$}    & \small{6.64 }           & \small{-5.90$^{  - 0.12}_{ + 0.12}$}      & \small{2 }  & \small{  20.30 $^{+1.16}_{-1.16}$ }  & \small{   5.35$^{+0.05}_{-0.05}$  }       & \small{ 15.6$^{+2.3 }_{- 2.3 }$   }     \\   
\small{76}  & \small{2067913164027256960 }    & \small{1835$^{ + 169}_{ - 143}$}    & \small{7.51 }           & \small{-5.60$^{  - 0.19}_{ + 0.18}$}      & \small{2 }  & \small{  14.50 $^{+1.22}_{-1.28}$ }  & \small{   5.50$^{+0.08}_{-0.08}$  }       & \small{ 61.4$^{+18.1}_{- 18.4}$   } \\  
\small{78}  & \small{2064838375463800448 }    & \small{1704$^{ + 128}_{ - 112}$}    & \small{6.09 }           & \small{-5.03$^{  - 0.16}_{ + 0.14} $}     & \small{2 }  & \small{  10.60 $^{+0.71}_{-0.80}$ }  & \small{   5.37$^{+0.06}_{-0.07}$  }       & \small{ 19.0$^{+4.6 }_{- 4.8 }$   }   \\
                                                                            
\hline

\end{longtable}

\newpage

\section{Best-fitting models}\label{appB}
 In this appendix, we show an example of the online material available at the Zenodo open-access repository (\url{http://doi.org/10.5281/zenodo.3984973}). We present the FASTWIND best-fitting model to the observed spectra for one star of the Cygnus OB2 O-type population analyzed in this work (Cyg OB2 $\#$15). The observed spectra (in black) is overplotted to the best-fitting model (in red) resulting from the \texttt{iacob-gbat} analysis. He\,I, He\,II, and H lines are indicated with solid, dashed, and dotted short vertical lines, respectively.

\begin{figure*}[h!]
\centering
\includegraphics[width=20cm,height=15cm,angle=270]{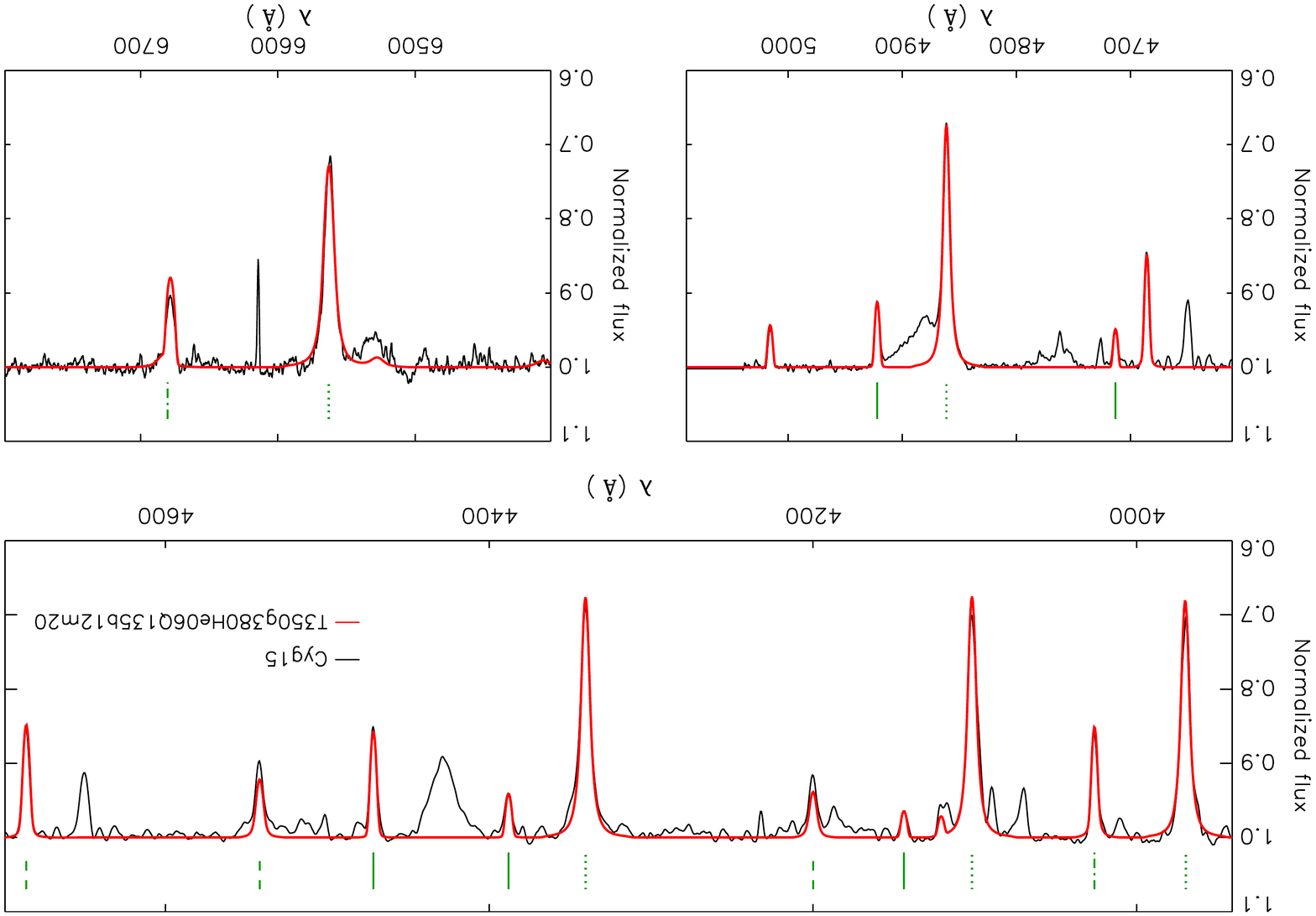}
\label{ALS15111}
\end{figure*}

\end{appendix}

\end{document}